\documentclass[preprintnumbers,superscriptaddress,amsmath,amssymb,pre,aps,10pt]{revtex4-1}

\usepackage{graphicx,easymat,chemarrow}
\usepackage{dcolumn}
\usepackage{bm}
\usepackage{color}
\usepackage{multirow}
\usepackage{listings}
\usepackage{CJK}

\begin{document}

\preprint{Fractal/MFXWT}

\title{Multifractal cross wavelet analysis}

\author{Zhi-Qiang Jiang}
 \affiliation{Department of Finance, East China University of Science
   and Technology, Shanghai 200237, China}
 \affiliation{Research Center for Econophysics, East China University of
   Science and Technology, Shanghai 200237, China}
 \affiliation{Department of Physics and Center for Polymer Studies,
   Boston  University, Boston, MA 02215, USA}

\author{Xing-Lu Gao}
 \affiliation{Department of Finance, East China University of Science
   and Technology, Shanghai 200237, China}

\author{Wei-Xing Zhou}
 \email{wxzhou@ecust.edu.cn}
 \affiliation{Department of Finance, East China University of Science
   and Technology, Shanghai 200237, China}
 \affiliation{Research Center for Econophysics, East China University of
   Science and Technology, Shanghai 200237, China}
 \affiliation{Department of Mathematics, East China University of
   Science and Technology, Shanghai 200237, China}

 \author{H. Eugene Stanley}
 \affiliation{Department of Physics and Center for Polymer Studies,
   Boston  University, Boston, MA 02215, USA}

\date{\today}

\begin{abstract}

Complex systems are composed of mutually interacting components and the
output values of these components usually exhibit long-range
cross-correlations. Using wavelet analysis, we propose a method of
characterizing the joint multifractal nature of these long-range cross
correlations, a method we call multifractal cross wavelet analysis
(MFXWT). We assess the performance of the MFXWT method by performing
extensive numerical experiments on the dual binomial measures with
multifractal cross correlations and the bivariate fractional Brownian
motions (bFBMs) with monofractal cross correlations. For binomial
multifractal measures, we find the empirical joint multifractality of
MFXWT to be in approximate agreement with the theoretical formula. For
bFBMs, MFXWT may provide spurious multifractality because of the wide
spanning range of the multifractal spectrum. We also apply the MFXWT
method to stock market indices, and in pairs of index returns and
volatilities we find an intriguing joint multifractal behavior. The tests on surrogate series also reveal that the cross correlation behavior, particularly the cross correlation with zero lag, is the main origin of cross multifractality.


\end{abstract}


\maketitle


\section{Introduction}
\label{sec:Introduction}

In recent years a series of multifractal cross-correlation analysis
methods have been developed and applied to a number of different fields.
The goal has been to unveil possible multifractal long-range cross
correlations between two time series.
Such long-range cross correlations in pairs of series have widely applied in financial markets, ranging from uncovering the facts of cross multifractal nature \cite{Wang-Xie-2013-ND, Wang-Suo-Yu-Lei-2013-PA, Ma-Wei-Huang-2013-PA} in different markets to building trading strategies to get excess returns \cite{Zhou-Chen-2016-PA}, from improving the estimation of hedge ratio \cite{Wang-Xie-He-Chen-2014-PA} to incorporating the copula-multifractality into the calculation of volatilities \cite{Chen-Wei-Lang_Lin-Liu-2014-PA}.
An early method, joint
multifractal analysis, was invented in 1990 to study the relationship
between the dissipation rates of kinetic energy and passive scalar
fluctuations in fully developed turbulence and to handle the joint
partition function of two multifractal measures
\cite{Meneveau-Sreenivasan-Kailasnath-Fan-1990-PRA}. This method is also
referred to as the multifractal cross-correlation analysis based on the
partition function approach (MFXPF)
\cite{Xie-Jiang-Gu-Xiong-Zhou-2015-NJP}. A special case of MFXPF,
multifractal statistical moment cross-correlation analysis (MFSMXA), was
independently invented in 2012 to study volatility time series in
finance \cite{Wang-Shang-Ge-2012-Fractals}. In 2015, the main properties
of the joint multifractal nature of binomial measures were derived and
numerically validated \cite{Xie-Jiang-Gu-Xiong-Zhou-2015-NJP}.

Another multifractal cross-correlation analysis method is multifractal
height cross-correlation analysis (MF-HXA) \cite{Kristoufek-2011-EPL},
which is a bivariate generalization of height-height correlation
analysis \cite{Barabasi-Vicsek-1991-PRA}. The MF-HXA method also has its
origin in turbulence and is an extension of the cross-correlation
analysis of the structure functions of temperature and velocity
dissipation fields in a heated turbulent jet
\cite{Antonia-VanAtta-1975-JFM}. Hence it is also a multifractal
cross-correlation analysis based on structure function (MFXSF).

Other multifractal cross-correlation analysis methods include
multifractal detrended cross-correlation analysis based on detrended
fluctuation analysis (MFXDFA) \cite{Zhou-2008-PRE}, which is a
multifractal version of detrended cross-correlation analysis (DCCA)
\cite{Podobnik-Stanley-2008-PRL}, multifractal detrended
cross-correlation analysis based on detrending moving-average analysis
(MFXDMA) \cite{Jiang-Zhou-2011-PRE} based on multifractal detrending
moving-average analysis (MF-DMA) \cite{Gu-Zhou-2010-PRE} and detrending
moving-average analysis (DMA)
\cite{Alessio-Carbone-Castelli-Frappietro-2002-EPJB,Carbone-Castelli-2003-SPIE,Carbone-Castelli-Stanley-2004-PRE,Arianos-Carbone-2007-PA,Carbone-2007-PRE,Carbone-Kiyono-2016-PRE,Tsujimoto-Miki-Shimatani-Kiyono-2016-PRE,Kiyono-Tsujimoto-2016-PRE},
multifractal cross-correlation analysis (MFCCA)
\cite{Oswiecimka-Drozdz-Forczek-Jadach-Kwapien-2014-PRE,Kwapien-Oswiecimka-Drozdz-2015-PRE},
and multifractal detrended partial correlation analysis (MFDPXA)
\cite{Qian-Liu-Jiang-Podobnik-Zhou-Stanley-2015-PRE}.

Wavelet transform has long been applied to the study of fractals and
multifractals
\cite{Holschneider-1988-JSP,Arneodo-Grasseau-Holschneider-1988-PRL} and
a partition function approach based on wavelet transform has been
proposed \cite{Muzy-Bacry-Arneodo-1991-PRL}. Here we generalize
multifractal wavelet analysis to the bivariate case and propose a new
joint multifractal analysis based on the wavelet transform of two time
series, which is a multifractal generalization of the cross wavelet
transform
\cite{Hudgins-Friehe-Mayer-1993-PRL,Maraun-Kurths-2004-NPG,AguiarConraria-Soares-2014-JES}. We
thus can also call it multifractal cross wavelet analysis
(MFXWT). Similar to when we use the MFXPF method, we introduce two
orders in MFXWT. We test the validity of the method by conducting
numerical experiments with two mathematical models and gain explicit
analytical results. Finally we apply the method to an empirical time
series.

The rest of paper is organized as follows. In Section 2, we present a framework of the MFXWT approach. Extensive numerical experiments using binomial measures and bivariate fractional Brownian motions with known analytical multifracal expressions are conducted in Section 3 to check the validity of MFXWT approach. In Section 4 we apply the MFXWT algorithms to the pair of daily returns, as well as the pair of daily volatilities. Statistical tests indicate that the MFXWT method has the ability to detect the cross multifractality in pairs of financial series. Section 5 concludes.

\section{Methods}
\label{S1:Algo:MFXWT}

Following
Refs.~\cite{Kantelhardt-Zschiegner-KoscielnyBunde-Havlin-Bunde-Stanley-2002-PA,Oswiecimka-Kwapien-Drozdz-2006-PRE}, we define the wavelet transform
of a given time series $x(t)$ as
\begin{equation}
w(s,i) = \frac{1}{s} \sum_{t = 1}^n x(t) \psi [(t-i)/s],
~~~~~~i=1,\cdots,n, \label{Eq:MFXWT:WTTransform}
\end{equation}
where $\psi(x)$ is the wavelet kernel shifted by $i$, $s$ is the scale,
and $n$ is the length of $x(t)$. We use the wavelet transform to
decompose the signals in the time-scale plane. The resulting wavelet
coefficients are an indicator of the singular behavior of signals when
the wavelet kernel is $\int x^{m+1} \psi(s) {\rm{d}} x = 0$
\cite{Bacry-Muzy-Arneodo-1993-JSP}, and from this we approximate the
signal trends by polynomials up to order $m$. A good choice of $\psi(x)$
is derivative $m$ of a Gaussian, $\psi^m(x) = {\rm{d}}^m
(e^{-x^2/2})/{\rm{d}} x^m$. Here we use the ``Mexican hat'' $m=2$.

Using the wavelet-based scaling (or multiscaling) estimator
\cite{Audit-Bacry-Muzy-Arneodo-2002-IEEEtit,
  Muzy-Bacry-Arneodo-1994-IJBC} and cross correlation (or multifractal)
analysis \cite{Podobnik-Stanley-2008-PRL, Zhou-2008-PRE,
  Jiang-Zhou-2011-PRE, Kristoufek-2011-EPL, Wang-Shang-Ge-2012-Fractals,
  Xie-Jiang-Gu-Xiong-Zhou-2015-NJP}, we propose a new method for
detecting the multifractal cross correlations in a pair of series $x(t)$
and $y(t)$ based on wavelet analysis, i.e., a multifractal cross wavelet
analysis with two moment orders (MFXWT $(p, q)$).

We first perform a wavelet transform of the two time series and obtain
the wavelet coefficients $w_x(s,i)$ and $w_y(s,i)$. We then define the
joint partition function with moments $p$ and $q$ based on the obtained
wavelet coefficients,
\begin{equation}
  \chi_{xy}(p,q,s) = \sum_{i=1}^n
  |w_x(s,i)|^{p/2}|w_y(s,i)|^{q/2}. \label{Eq:MFXWT:Chi}
\end{equation}

Because some wavelet coefficients approach 0, the partition function
diverges for $p < 0$ or $q < 0$. When $w_x = w_y$ and $p=q$, we use
wavelet analysis to recover the traditional partition function.
The part $|w_x(s,i)|^{p/2} |w_y(s,i)|^{q/2}$ is a generalized cross wavelet spectrum and it recovers the traditional cross wavelet spectrum when $p=4$ and $q=4$ \cite{Grinsted-Moore-Jevrejeva-2004-NPG}, as the wavelet coefficients are real number. The cross wavelet spectrum can be used to calculate the wavelet coherency, which is able to uncover the co-movement between two series in the time-frequency domain \cite{Rua-Nunes-2009-JEF, Vacha-Barunik-2012-EE}. The definition of the partition function allows us to uncover the more intricate relationship between the coherency and the scale under different scopes, which corresponds to the cross multifractal behaviors.
If the underlying processes are jointly multifractal, the result is a scaling
behavior,
\begin{equation}
\chi_{xy}(p,q,s) \sim s^{{\cal{T}}_{xy}(p, q)}.\label{Eq:MFXWT:Chi:Scaling}
\end{equation}
where ${\cal{T}}_{xy}(p, q)$ is the joint mass exponent function. Note
that we can estimate ${\cal{T}}_{xy}(p, q)$ by regressing $\ln \chi_{xy}
(p,q,s)$ against $\ln s$ in the scaling range for a given pair $(p,q)$.

Analogous to the double Legendre transforms in the joint multifractal
analysis based on the partition function approach MFXPF$(p,q)$
\cite{Xie-Jiang-Gu-Xiong-Zhou-2015-NJP}, we define the joint singularity
strength function $h_x$ and $h_y$
\begin{eqnarray}
 h_x(p,q) & = & 2\partial {\cal{T}}_{xy}(p, q) / \partial p, \label{Eq:MFXWT:hx}\\
 h_y(p,q) & = & 2\partial {\cal{T}}_{xy}(p, q) / \partial q, \label{Eq:MFXWT:hy}
\end{eqnarray}
and the multifractal spectrum $D_{xy}(h_x, h_y)$
\begin{equation}
 D_{xy}(h_x, h_y) = p h_x /2 + q h_y /2  - {\cal{T}}_{xy}. \label{Eq:MFXWT:Dh}
\end{equation}

The values $h_x(p,q)$, $h_y(p,q)$, and $D_{xy}(h_x, h_y)$ from the
MFXWT$(p,q)$ method differ from the joint singularity strengths
$\alpha_x(p,q)$, $\alpha_y(p,q)$ and the joint multifractal spectrum
$f_{xy}(\alpha_x, \alpha_y)$ obtained from the MFXPF$(p,q)$ method. For
example, when $p = 0$ and $q = 0$ the joint partition function in
Eq.~(\ref{Eq:MFXWT:Chi}) is equal to the number of wavelet coefficients
and corresponds to the total number of data points in the original
series, which means that ${\cal{T}}_{xy}(0, 0) = 0$ and that in the
MFXPF$(p, q)$ method $\tau_{xy} (0,0)=-1$. We also find that all the
estimated $h_x$ and $h_y$ are less than 0. Although this violates our
intuition that the singularity strength should be positive, these
differences in value do not mean that our method is useless because the
joint multifractal quantities obtained from both methods still share the
same physical meanings and geometric features, which allows us to
determine the cross correlations in the time series pairs. Following the
usual numerical experiments and empirical analysis in which the
multifractal analysis based wavelet estimators are performed on integral
series, we also test our MFXWT$(p, q)$ method on integral series. The
obtained results are not easy to explain, however, and it is very
difficult to link to the theoretical values in the $p$-model
\cite{Meneveau-Sreenivasan-1987-PRL}. Thus we focus our investigations
on non-cumulative series, e.g., stock returns rather than stock prices.

Reference~\cite{Muzy-Bacry-Arneodo-1991-PRL} found errors in the
estimation of joint singularity strength and joint multifractal spectrum
based on the Legendre transform. They proposed an alternative method of
computing $h$ and $D(h)$ using the direct estimation canonical
method. Similarly we directly estimate the joint singularity strength
$h_x$ and $h_y$ and the joint multifractal spectrum $D_{xy}(p,q)$ using
\begin{eqnarray}
 h_x(p,q) & = & \lim_{s \rightarrow 0} \frac{1}{\ln s} \sum_i
 \mu_{xy}(p,q,s,i) \ln |w_x(s,i)|, \label{Eq:MFXWT:DE:hx}\\
 h_y(p,q) & = & \lim_{s \rightarrow 0} \frac{1}{\ln s} \sum_i
 \mu_{xy}(p,q,s,i) \ln |w_y(s,i)|, \label{Eq:MFXWT:DE:hy} \\
 D_{xy}(p,q) &=& \lim_{s \rightarrow 0} \frac{1}{\ln s} \sum_i
 \mu_{xy}(p,q,s,i) \ln \mu_{xy}(p,q,s,i). \label{Eq:MFXWT:DE:Dxy}
\end{eqnarray}
where
\begin{equation}
\mu_{xy} (p,q,s,i) = |w_x(s,t)|^{p/2} |w_y(s,i)|^{q/2} /
\chi_{xy}(p,q,s).\nonumber
\end{equation}
Thus we can directly determine the joint singularity strength functions
$h_x(p,q)$ and $h_y(p,q)$ and the joint multifractal function
$D_{xy}(p,q)$ from Eqs.~(\ref{Eq:MFXWT:DE:hx}--\ref{Eq:MFXWT:DE:Dxy}).

\section{Numerical experiments}
\label{S1:NumSim}

To test the validity and performance of the proposed MFXWT $(p,q)$
method we conduct two binomial measurements generated (i) from the
multiplicative $p$-model \cite{Meneveau-Sreenivasan-1987-PRL} and from
(ii) bivariate fractional Brownian motions (bFBMs)
\cite{Lavancier-Philippe-Surgailis-2009-SPL,Coeurjolly-Amblard-Achard-2010-EUSIPCO,Amblard-Coeurjolly-Lavancier-Philippe-2013-BSMF}.

\subsection{Binomial measures}

We first numerically test the validity of the MFXWT$(p,q)$ method using
two binomial measures from the $p$-model with known analytic
multifractal properties, (i) $\{x(i):i = 1, 2, \cdots, 2^k\}$ and (ii)
$\{y(i): i = 1, 2, \cdots, 2^k\}$
\cite{Meneveau-Sreenivasan-1987-PRL}. Each binomial measure is generated
iteratively. We start with the zeroth iteration $k = 0$, where the data
set $z(i)$ consists of one value, $z^{(0)}(1)= 1 $. In iteration $k$,
the data set $\{z^{(k)}(i): i = 1, 2, \cdots, 2^k\}$ is obtained from
\begin{equation}
  \begin{array}{l}
    z^{(k)}(2i-1)= p_z z^{(k-1)}(i)\\
    z^{(k)}(2i)  = (1-p_z)z^{(k-1)}(i)
\end{array}
  \label{Eq:pModel}
\end{equation}
for $i = 1, 2, \cdots, 2^{k-1}$. When $k\to\infty$, $z^{(k)}(i)$
approaches a binomial measure with a scaling exponent function
$H_{zz}(q)$ and a mass exponent function $ \tau_{zz}(q)$ that have an
analytic form
\cite{Halsey-Jensen-Kadanoff-Procaccia-Shraiman-1986-PRA,Meneveau-Sreenivasan-1987-PRL}
\begin{eqnarray}
 H_{zz}(q) & = & 1/q-\log_2[p_z^q+(1-p_z)^q]/q, \label{Eq:pModel:Hzz}\\
\tau_{zz}(q) & = & -\log_2[p_z^q+(1-p_z)^q]. \label{Eq:pModel:Tauzz}
\end{eqnarray}
In our numerical experiment, the parameters of the two binomial measures
from the $p$-model are set at $p_x =0.3$ for $x(i)$ and $p_y = 0.4$ for
$y(i)$ with an iterative step $k=16$. The analytic scaling exponent
functions $H_{xx}(q)$ and $H_{yy}(q)$ of $x$ and $y$ are shown in
Eq.~(\ref{Eq:pModel:Hzz}). Because the two series are generated using
the same rule, the two series $x$ and $y$ exhibit a strong correlation
with a coefficient of 0.82.

\begin{figure}[tb]
  \centering
\includegraphics[width=5.5cm]{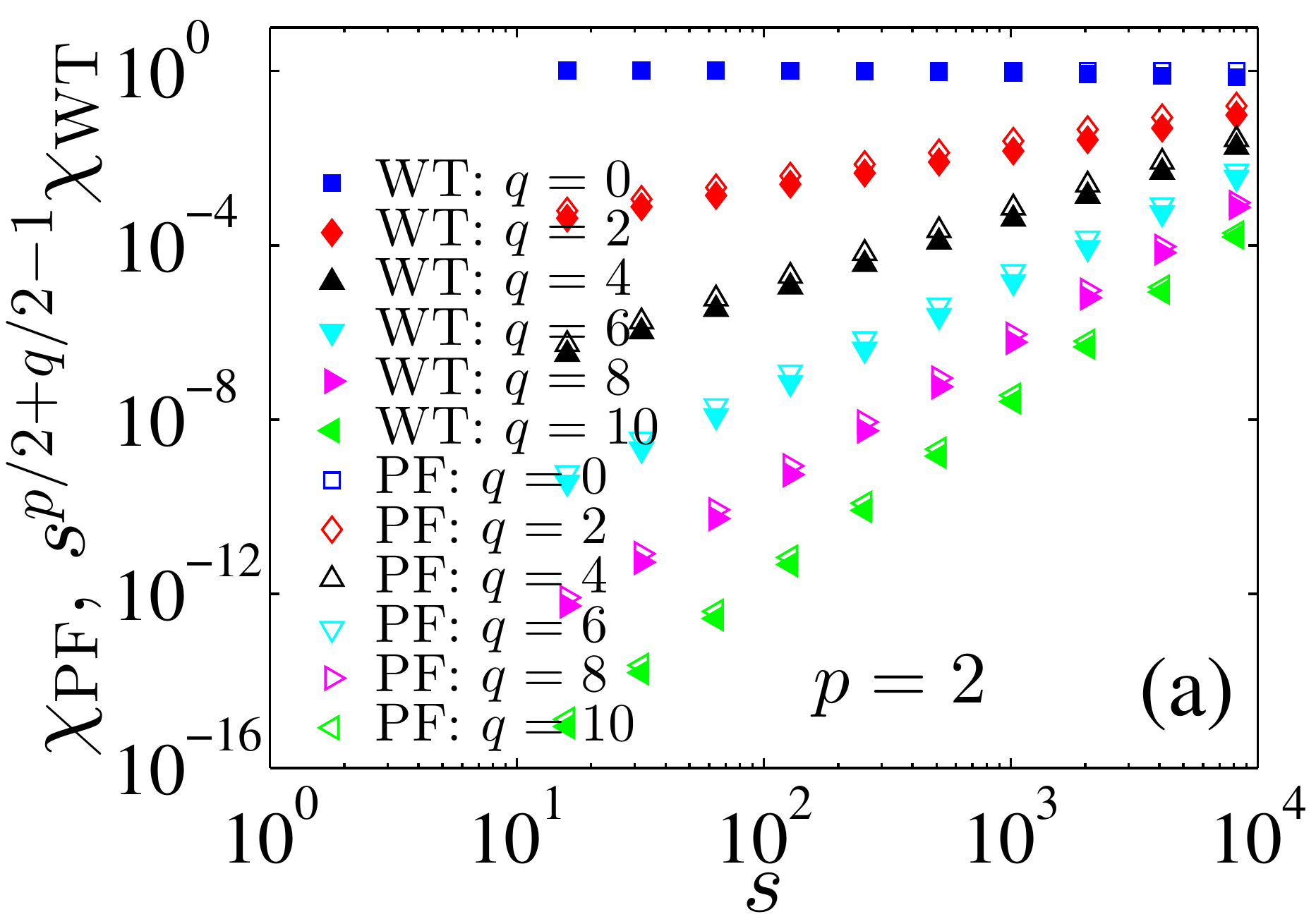}
\includegraphics[width=5.5cm]{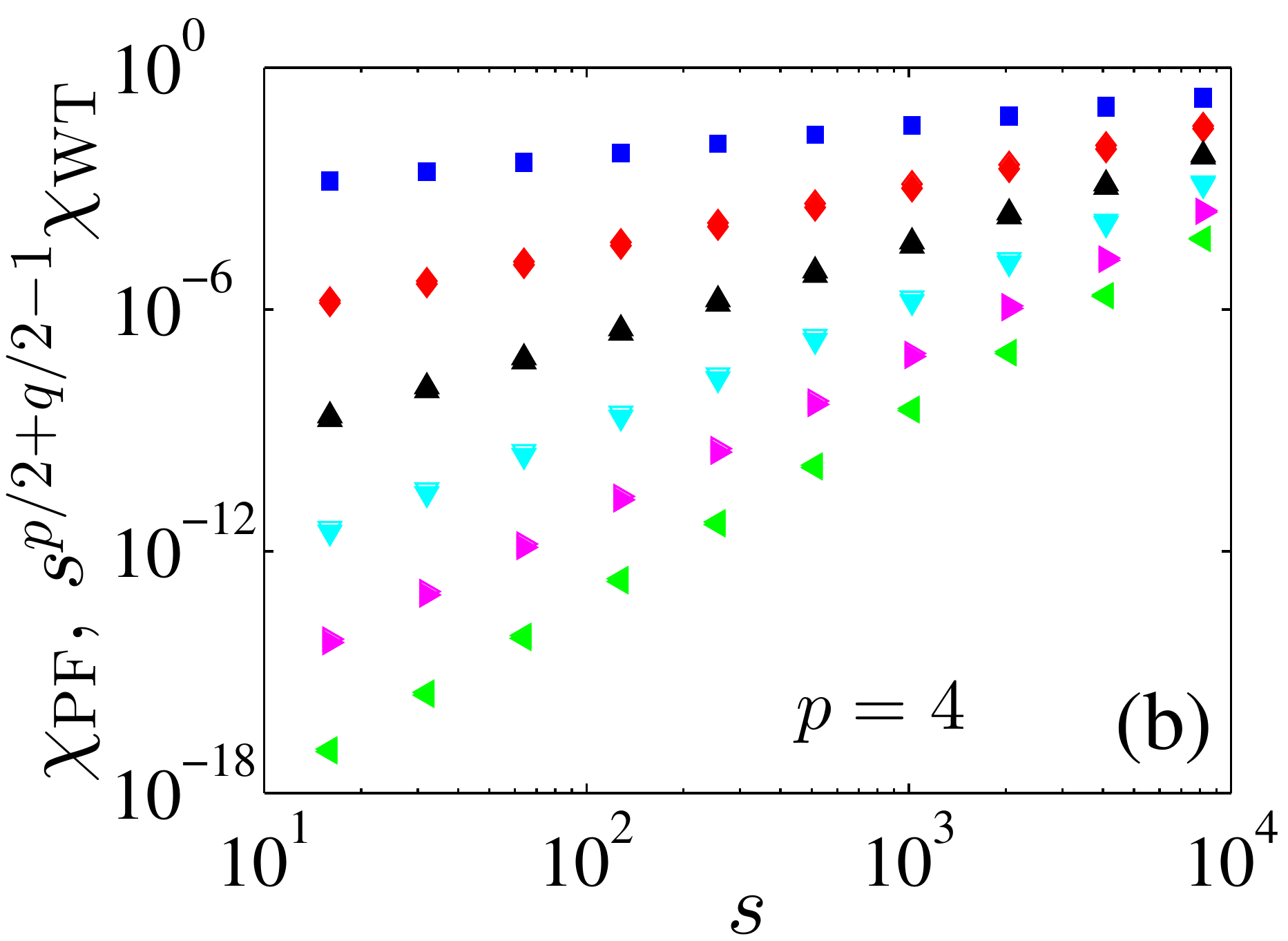}
\includegraphics[width=5.5cm]{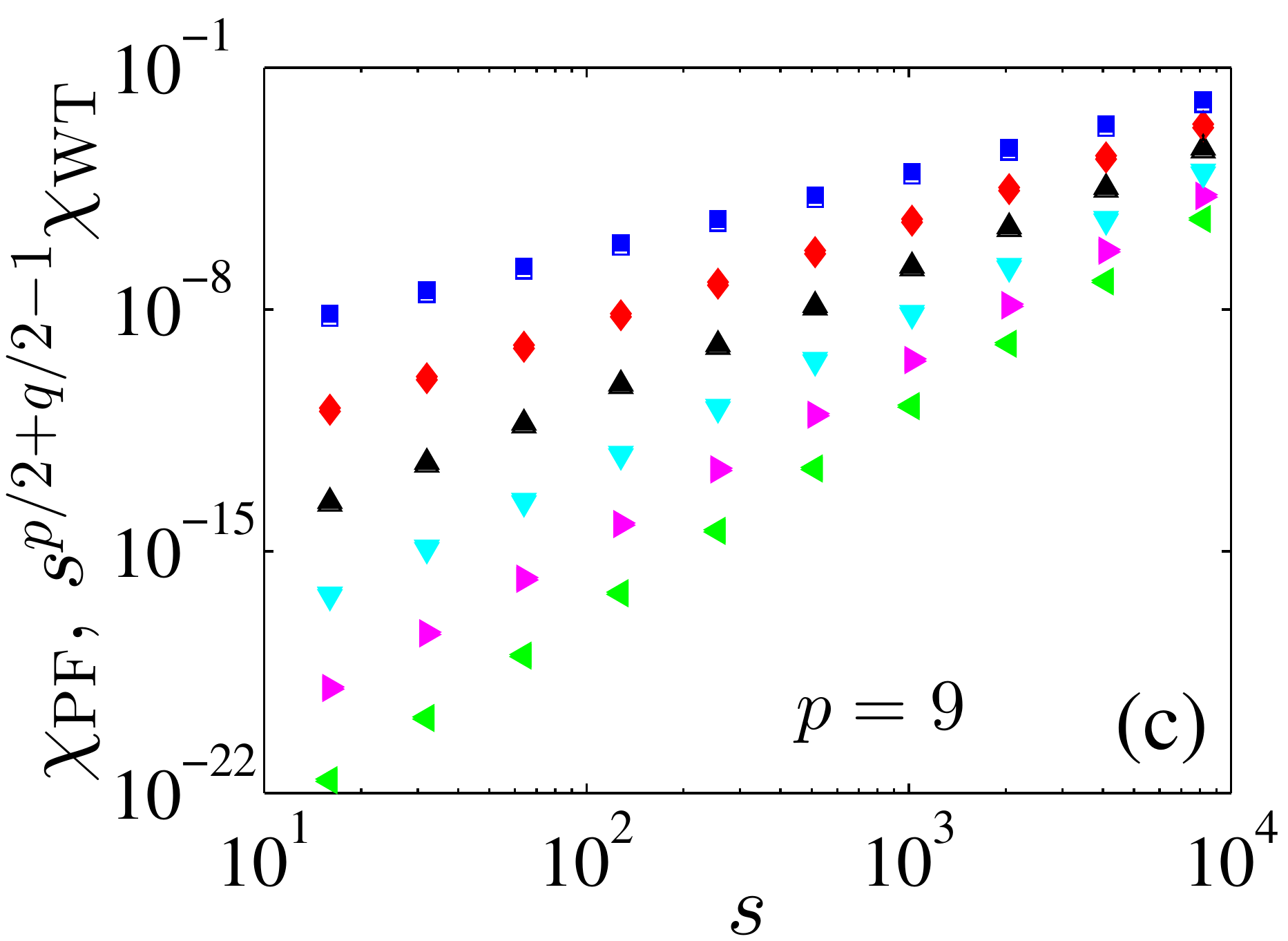}
  \caption{\label{Fig:MFXWT:pmodel:Compare:PFs} (Color online)
    Comparison of the scaling behaviors of two binomial measures for the
    joint partition functions obtained from MFXPF$(p,q)$ and
    MFXWT$(p,q)$ with different values of $p$ and $q$. In the plots, $q$
    varies from 0 to 10 with a step of 2. The joint partition functions
    of MFXWT$(p,q)$ are scaled by a factor of $s^{p/2+q/2-1}$, which
    results in the almost same scaling pattern as the joint partition functions
    of MFXPF$(p,q)$. }
\end{figure}

\begin{figure*}[htb]
\centering
\includegraphics[width=5.5cm]{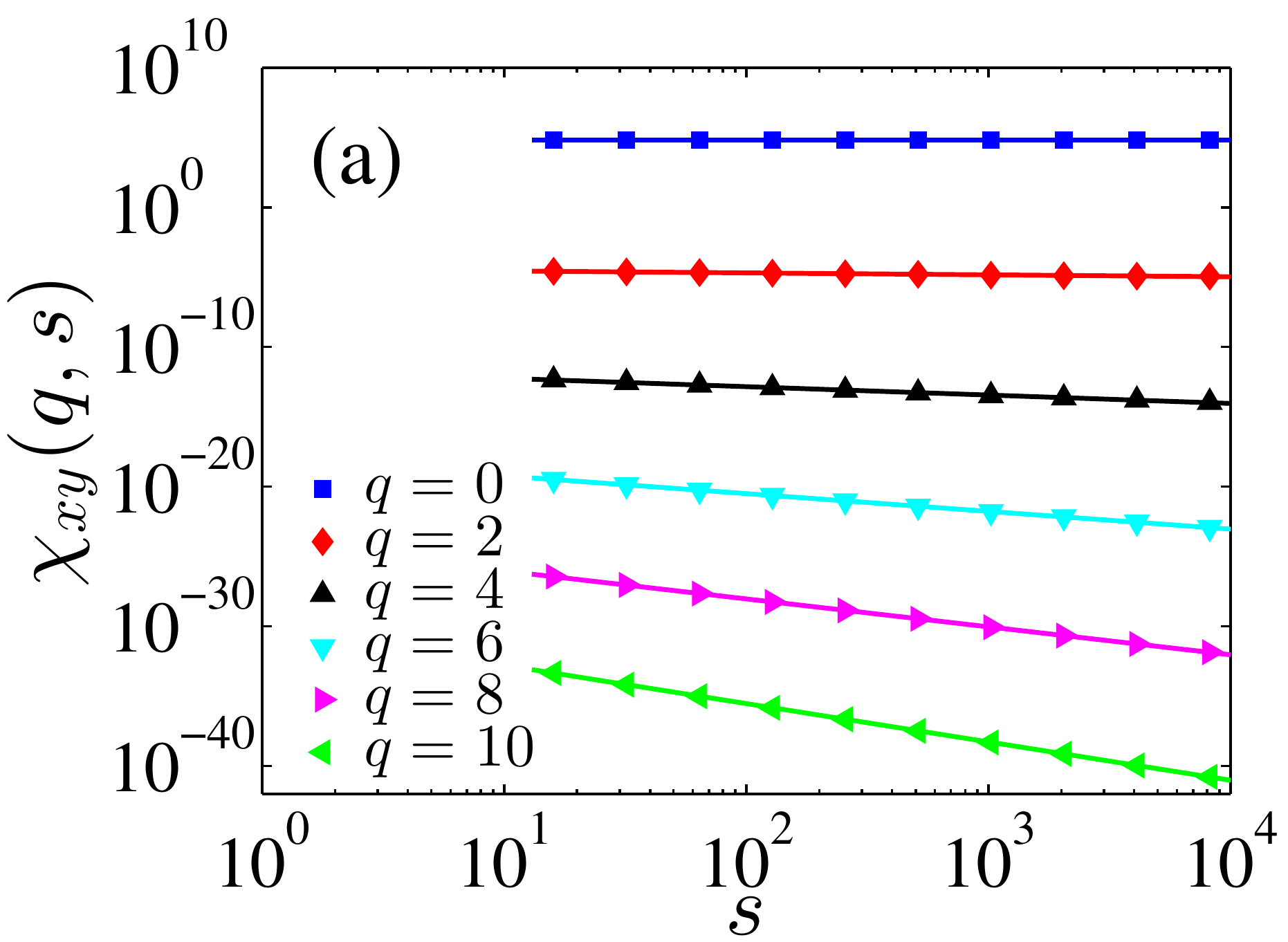}
\includegraphics[width=5.5cm]{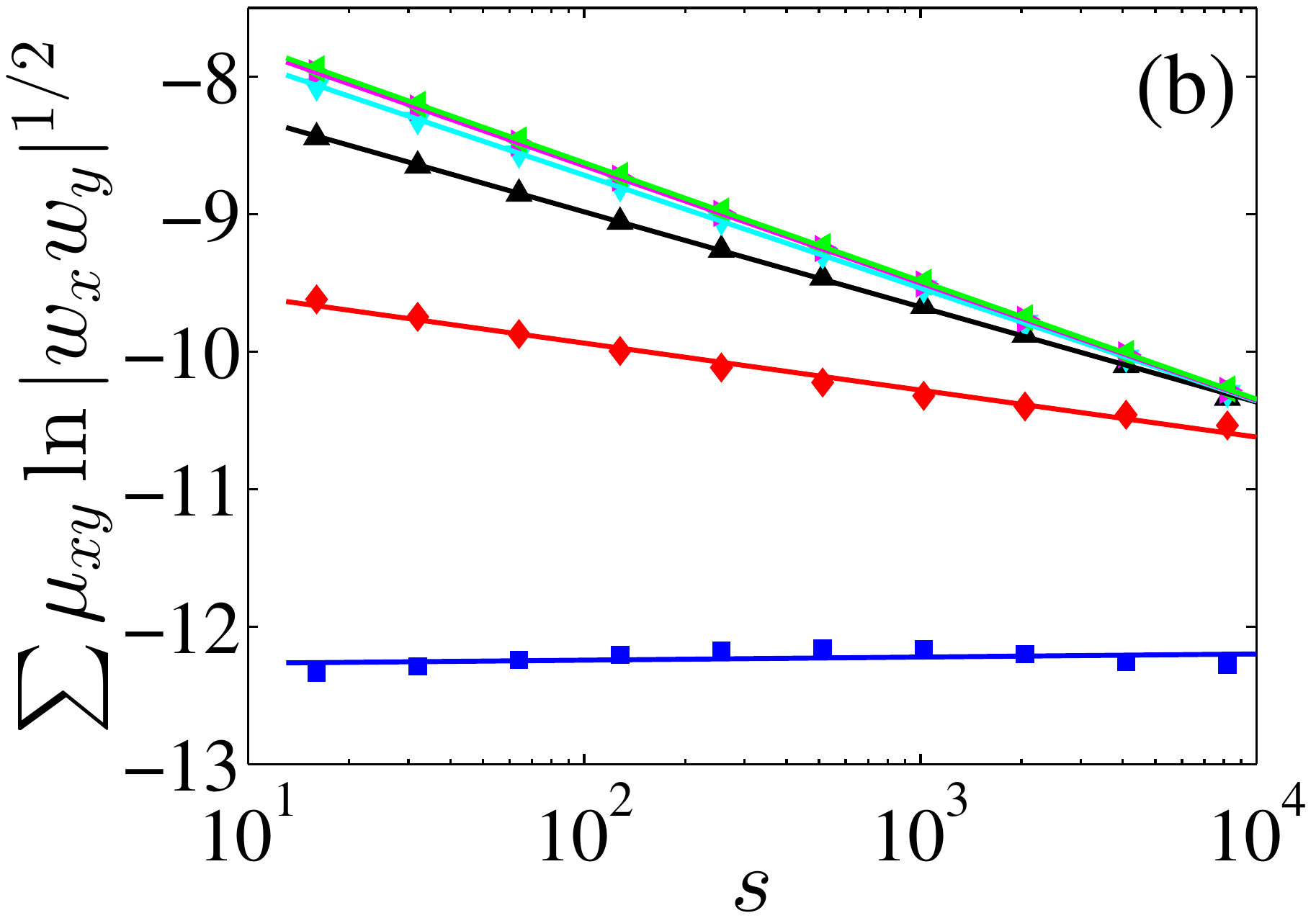}
\includegraphics[width=5.5cm]{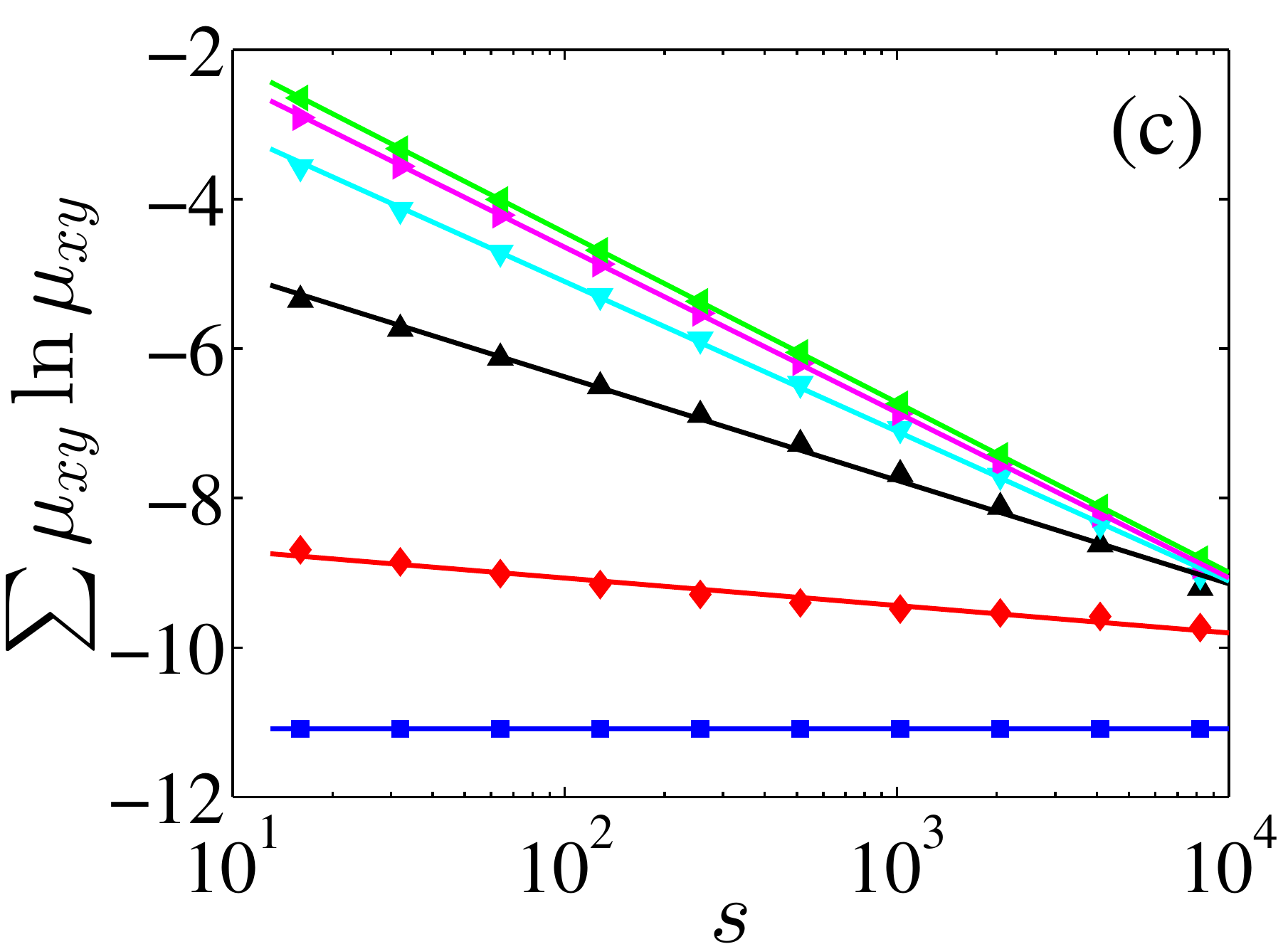}
\includegraphics[width=5.5cm]{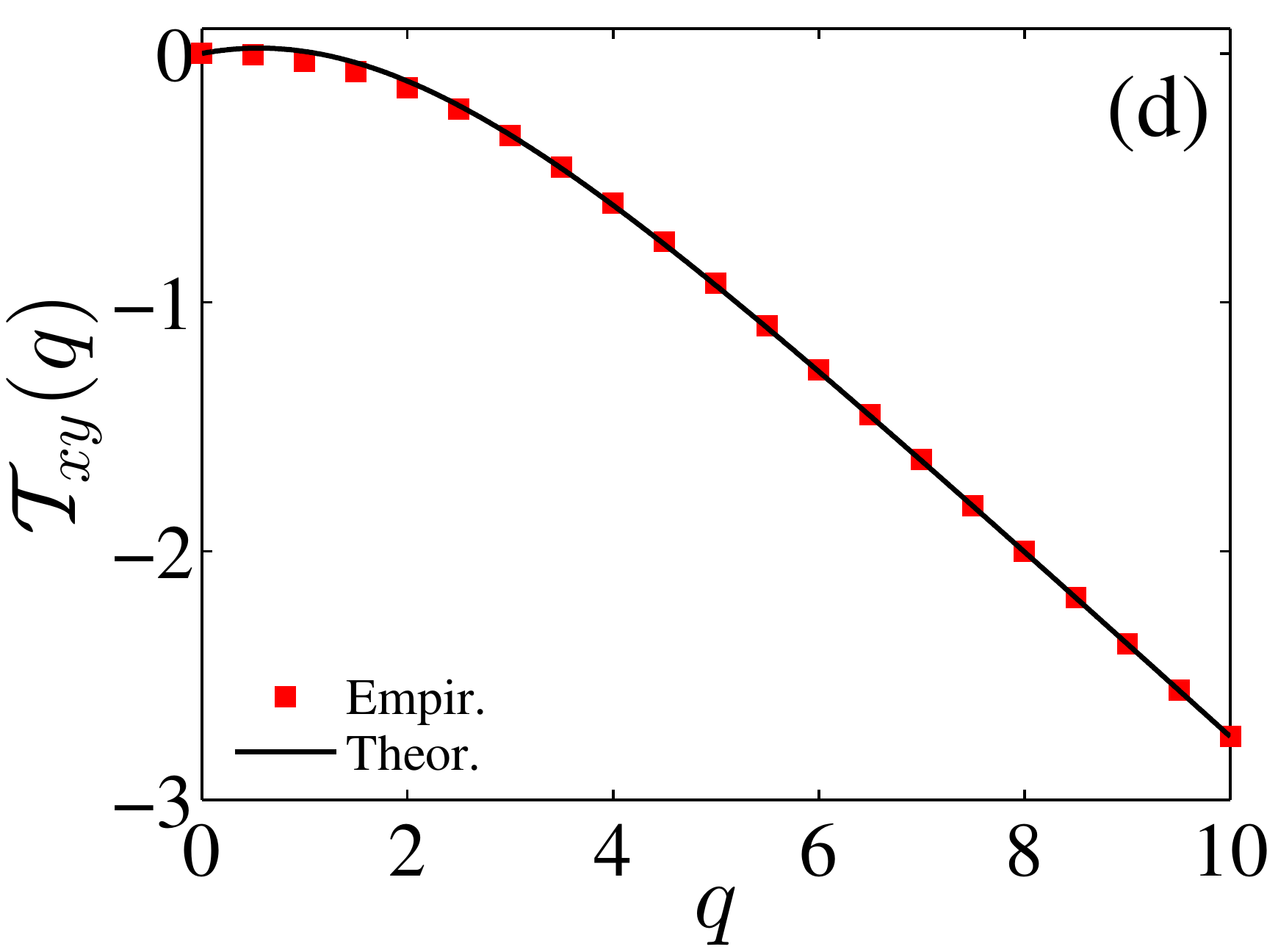}
\includegraphics[width=5.5cm]{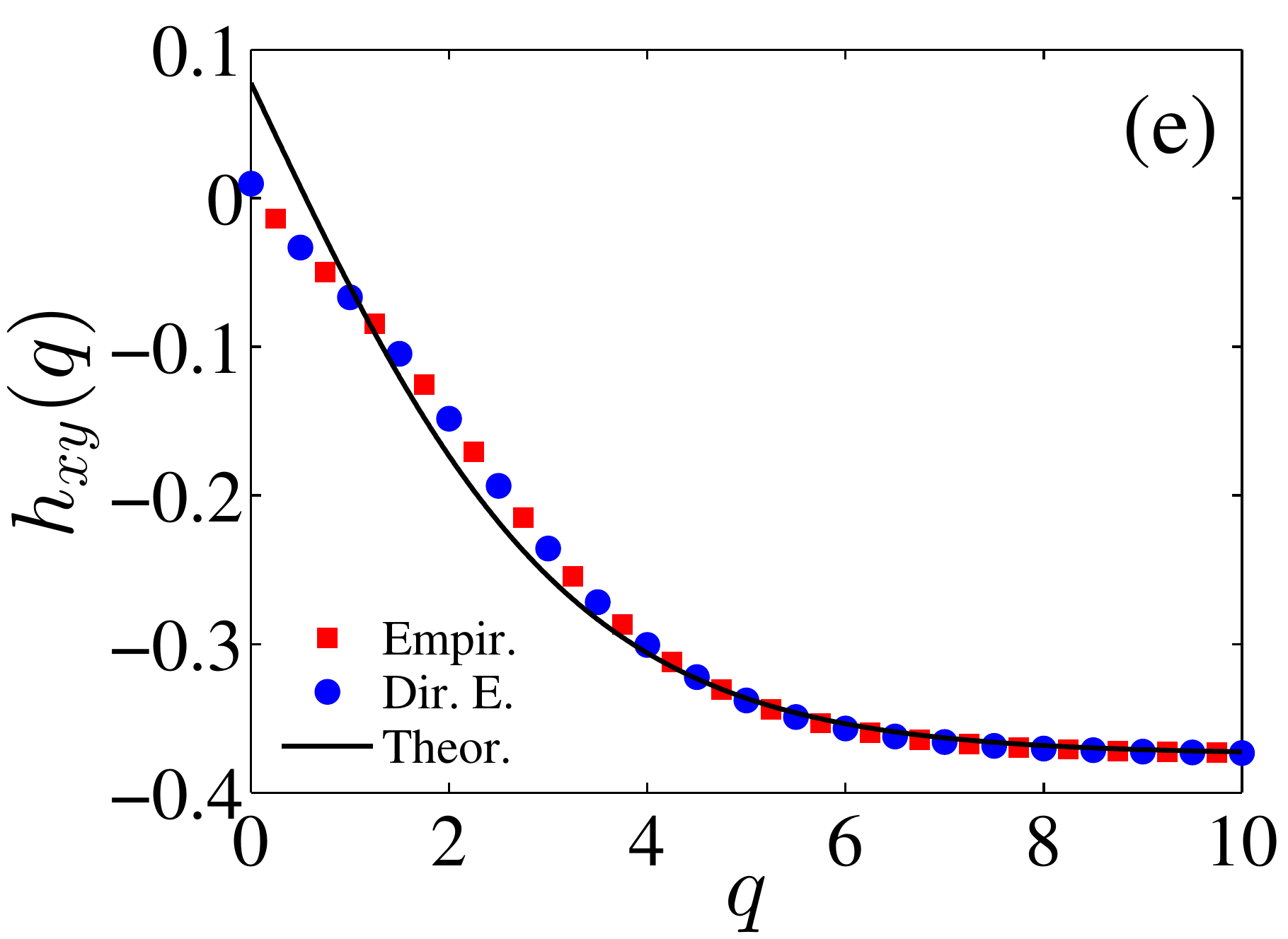}
\includegraphics[width=5.5cm]{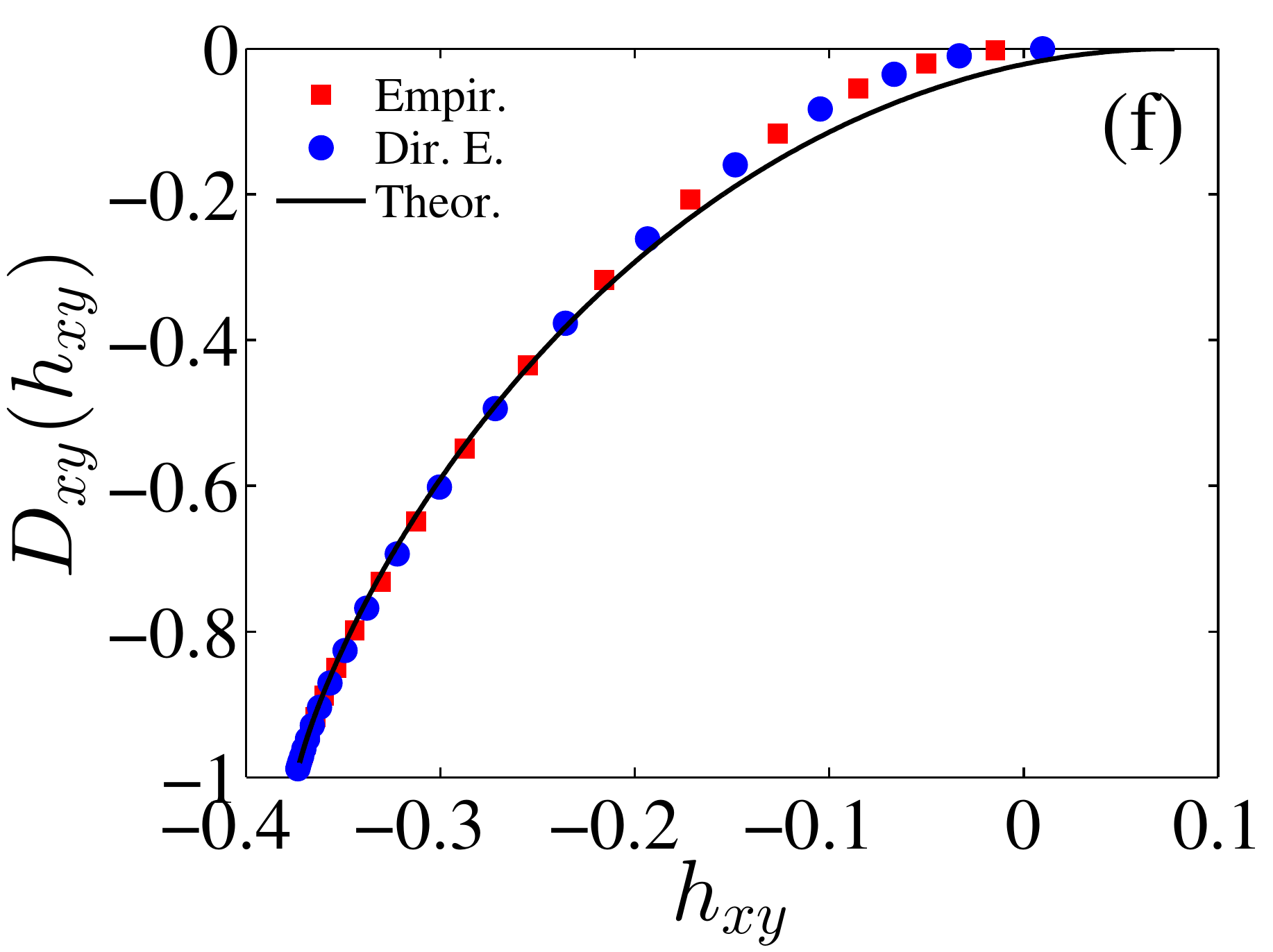}
  \caption{\label{Fig:MFXWT:pmodel:Q} (Color online) Multifractal cross
    wavelet analysis of two binomial measures with $p_x=0.3$ and $p_y =
    0.4$ based on the MFXWT($q$) method.  (a) Power-law behaviors
    between $\chi_{xy}(q,s)$ and the scale $s$ for different $q$. (b)
    Linear relationship of $\sum_t \mu_{xy}(q,s,t) \ln |w_x(s,t)
    w_y(s,t)|^{1/2}$ against $\ln s$. (c) Linear relationship of $\sum_t
    \mu_{xy} (q,s,t) \ln \mu_{xy} (q,s,t)$ with respect to $\ln s$. (d)
    Joint mass exponent function ${\cal{T}}_{xy}(q)$. (e) Joint
    singularity strength function $h_{xy}(q)$. (f) Joint multifractal
    singularity spectrum $D_{xy}(h_{xy})$.}
\end{figure*}

Xie et al. analytically derived the joint multifractal properties for
two binomial measures constructed from the $p$-model
\cite{Xie-Jiang-Gu-Xiong-Zhou-2015-NJP}. The joint mass exponent
function $\tau_{xy}(p, q)$,
\begin{equation}
 \tau_{xy}(p, q)  =  \frac{p\gamma}{2\ln 2} - \frac{\ln \left[ p_y^Q +
     (1-p_y)^Q \right]}{\ln 2},
 \label{Eq:pModel:Tauxypq}
\end{equation}
the two joint singularity strength functions $\alpha_{x}(p, q)$ and
$\alpha_{y}(p,q)$,
\begin{equation}
 \alpha_x(p, q) = \frac{\gamma}{\ln 2} - \frac{\beta}{ \ln 2}
 \frac{p_y^Q \ln p_y + (1-p_y)^Q \ln(1-p_y)}{p_y^Q +
   (1-p_y)^Q},  \label{Eq:pModel:Alphax}
\end{equation}
\begin{equation}
 \alpha_y(p, q) =  - \frac{1}{ \ln 2} \frac{p_y^Q \ln p_y + (1-p_y)^Q
   \ln(1-p_y)}{p_y^Q + (1-p_y)^Q},  \label{Eq:pModel:Alphay}
\end{equation}
and the joint multifractal spectrum $f_{xy}(p,q)$ is expressed as
%
\begin{equation}
 f_{xy}(\alpha_x, \alpha_y)  =   \frac{QZ^Q \ln{Z} + (1+Z^Q) \ln(1 +
   Z^Q)}{\ln2(1+Z^Q)},  \label{Eq:pModel:fxyalpha}
\end{equation}
%

where $\beta = \frac{\ln p_x - \ln(1-p_x)}{ \ln p_y - \ln (1-p_y)}$,
$\gamma = \beta \ln (1-p_y) - \ln (1-p_x)$, $Q = \beta p/2 + q/2$, and
$Z=\frac{1-p_y}{p_y}$. These theoretical formulas have been found to
numerically agree with the empirical results from the MFXPF$(p,q)$
method \cite{Xie-Jiang-Gu-Xiong-Zhou-2015-NJP}, and this allows us to
check whether these theoretical formulas can be employed as a benchmark
test of the performance of the MFXWT$(p,q)$ algorithm when it is applied
to binomial measures. By comparing the scaling behaviors of the joint
partition functions from both methods, we find the theoretical formulas
of the joint mass exponent function ${\cal{T}}_{xy} (p,q)$, the joint
singularity strength functions $h_x(p,q)$ and $h_y(p,q)$, and the joint
multifractal spectrum $D_{xy}(h_x, h_y)$ for MFXWT$(p,q)$.

Figure~\ref{Fig:MFXWT:pmodel:Compare:PFs} shows the scaling behavior of
the joint partition functions obtained from the MFXPF$(p,q)$ and
MFXWT$(p,q)$ methods with different values of $p$ and $q$. The joint
partition functions of MFXWT$(p,q)$ are scaled by a factor of
$s^{p/2+q/2-1}$. We find that there is a slightly difference between the markers of both methods in panel (a) and such differences disappeared in panel (b) and (c).  This indicates almost the same scaling behavior between the scaled joint partition functions of MFXWT$(p,q)$ and the joint partition functions of MFXPF$(p,q)$, which allow us to connect the theoretical joint
multifractal formulas of binomial measures to the empirical joint
multifractal features of MFXWT$(p,q)$ by using
\begin{eqnarray}
 {\cal{T}}_{xy}(p, q)  + p/2 + q/2-1 & = & \tau_{xy}(p,
 q), \label{Eq:pModel:WTTau:PFTau}\\
  h_x(p, q) + 1 & = & \alpha_x(p, q), \label{Eq:pModel:WThx:PFalphax}\\
  h_y(p, q) + 1 & = & \alpha_y(p, q), \label{Eq:pModel:WThy:PFalphay}\\
  D_{xy} (h_x, h_y) + 1 & = & f_{xy}(\alpha_x,
  \alpha_y), \label{Eq:pModel:WTDh:PFfalpha}
\end{eqnarray}
where $\tau_{xy}(p,q)$, $\alpha_x(p,q)$, $\alpha_y(p,q)$, and
$f_{xy}(\alpha_x, \alpha_y)$ are given by
Eqs.~(\ref{Eq:pModel:Tauxypq}--\ref{Eq:pModel:fxyalpha}).

These formulas are an efficient test of the estimation accuracy of the
MFXWT$(p,q)$ method in the joint multifractal analysis of two binomial
measures. Using the partition function approach and wavelet analysis to
detect the multifractal nature of a single time series,
$\tau_{xx}(q)={\cal{T}}_{xx}(q)+q$ and $\alpha_{x}(q)=h_{x}(q)+1$
\cite{Turiel-Parga-2000-NCmp,Kestener-Arneodo-2003-PRL,Turiel-PerezVicente-Grazzini-2006-JComputP}.

We first examine the case of $p=q$. Figure~\ref{Fig:MFXWT:pmodel:Q}(a)
shows the scaling behavior between the joint partition functions
$\chi_{xy}(q,s)$ and the scale $s$. Note that there is a significant
power-law dependence over more than three orders of magnitude. By
estimating the power-law exponents between $\chi_{xy}(q,s)$ and $s$ for
different $q$, we find the joint mass exponent function ${\cal{T}}(q)$
[see the plot in Fig.~\ref{Fig:MFXWT:pmodel:Q}(d)].
Figure~\ref{Fig:MFXWT:pmodel:Q}(d) also shows the theoretical values of
${\cal{T}}(q)$ obtained from Eq.~(\ref{Eq:pModel:WTTau:PFTau}). The two
curves closely match, suggesting that our MFXWT$(p,q)$ algorithm
accurately analyzes the joint multifractal nature in two binomial
measures. As expected, the nonlinear behavior of ${\cal{T}}(q)$ against
$q$ also demonstrates the joint multifractality in binomial measures.

Figures~\ref{Fig:MFXWT:pmodel:Q}(b) and \ref{Fig:MFXWT:pmodel:Q}(c) show
the power-law scaling behaviors of two quantities ($\sum \mu_{xy} \ln
|w_x w_y|^{1/2}$ and $\sum \mu_{xy} \ln \mu_{xy}$) against the scale
$s$, whose power-law exponents are estimates of the joint singularity
strength $h_{xy}$ and the joint multifractal function $D(h_{xy})$.

\begin{figure*}[htb]
\centering
\includegraphics[width=4.2cm]{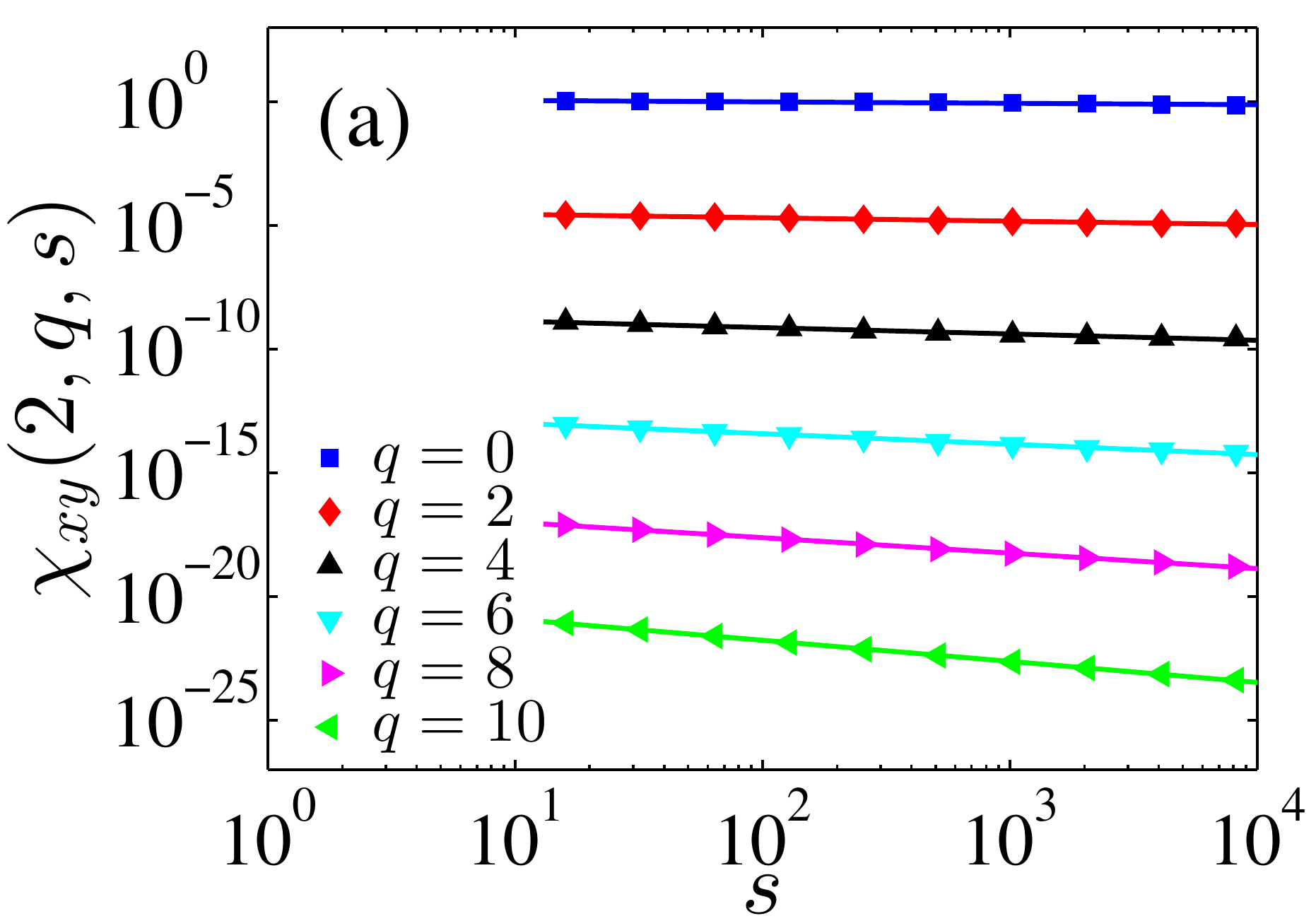}
\includegraphics[width=4.2cm]{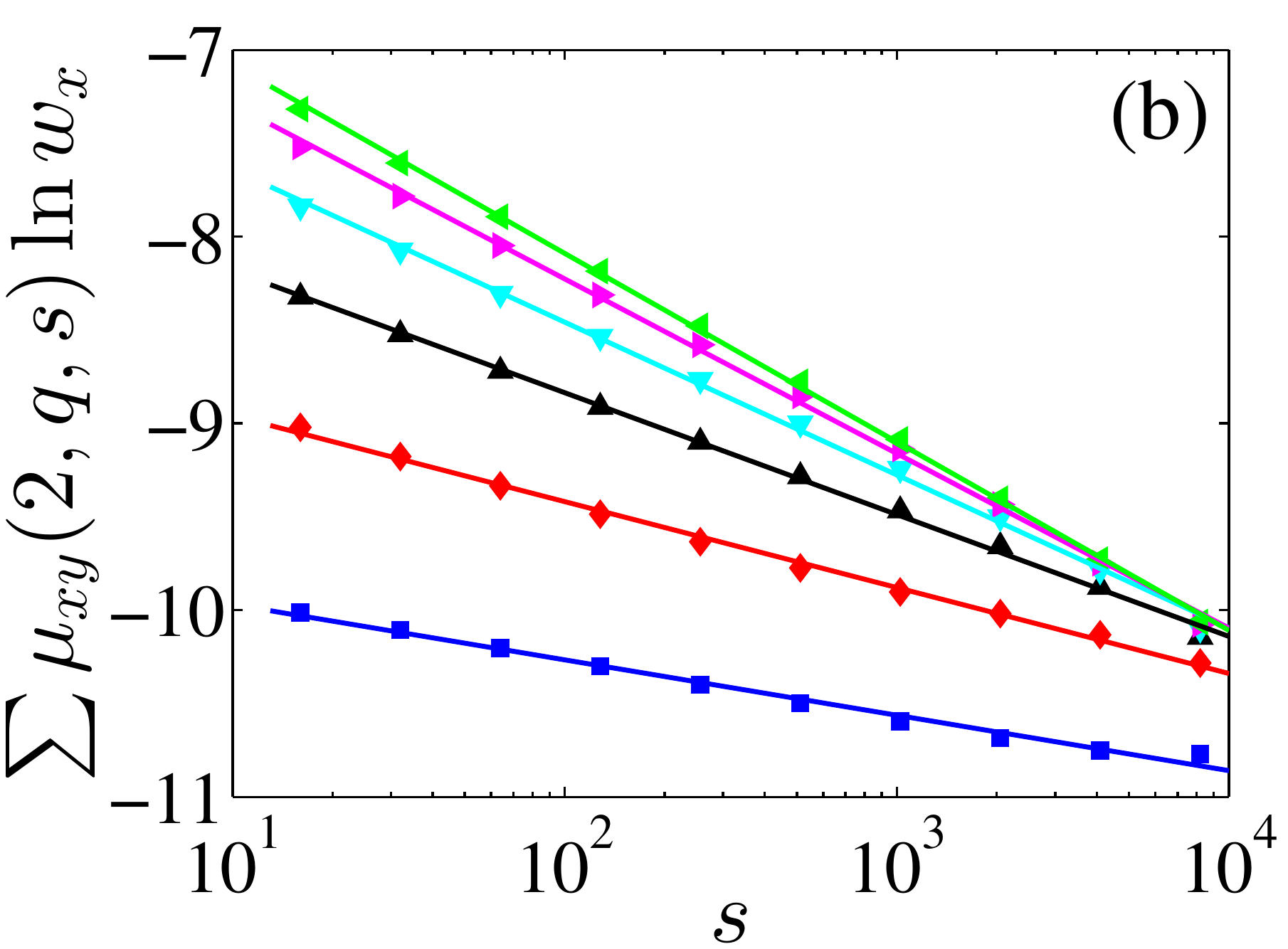}
\includegraphics[width=4.2cm]{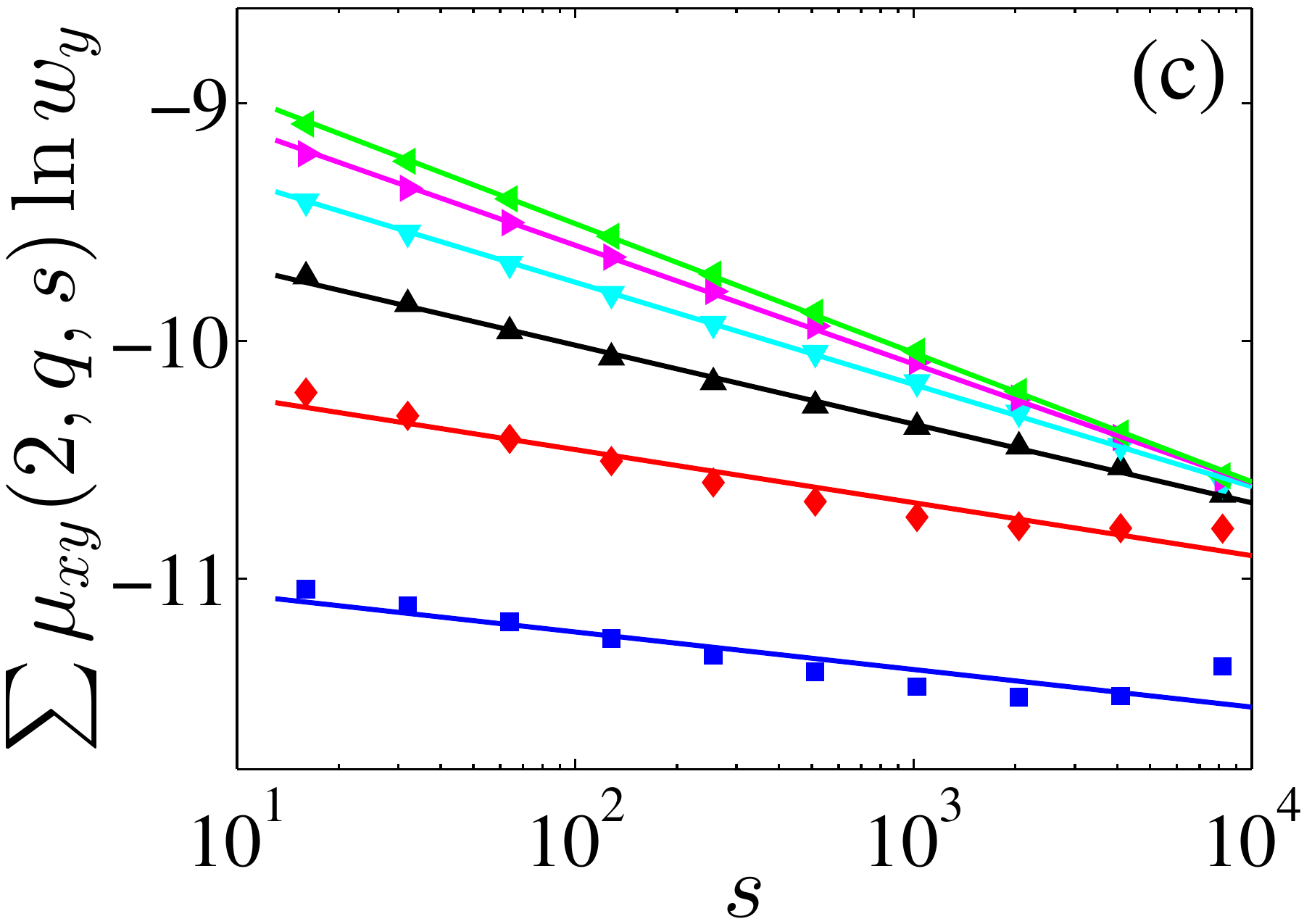}
\includegraphics[width=4.2cm]{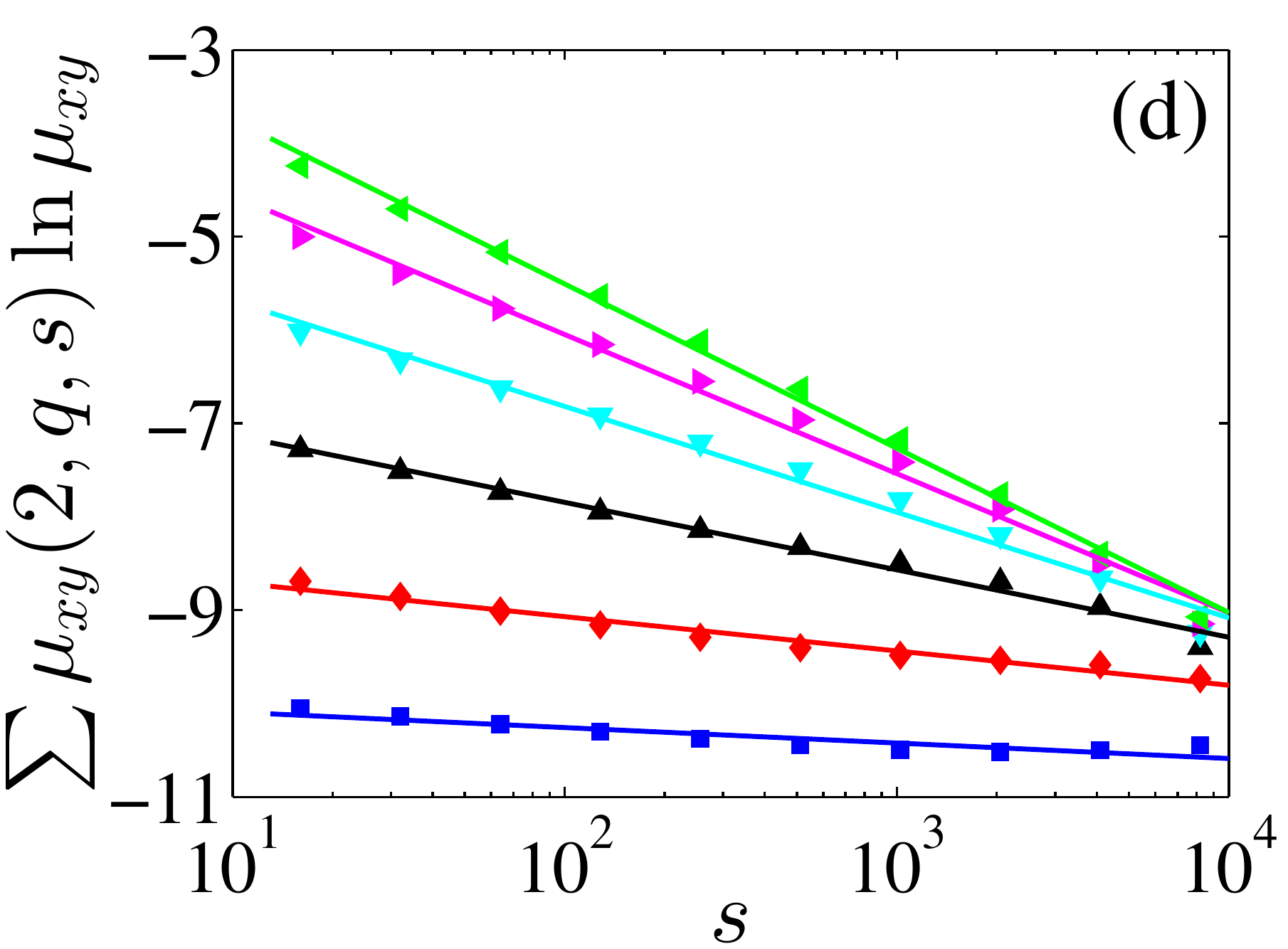}
\includegraphics[width=4.2cm]{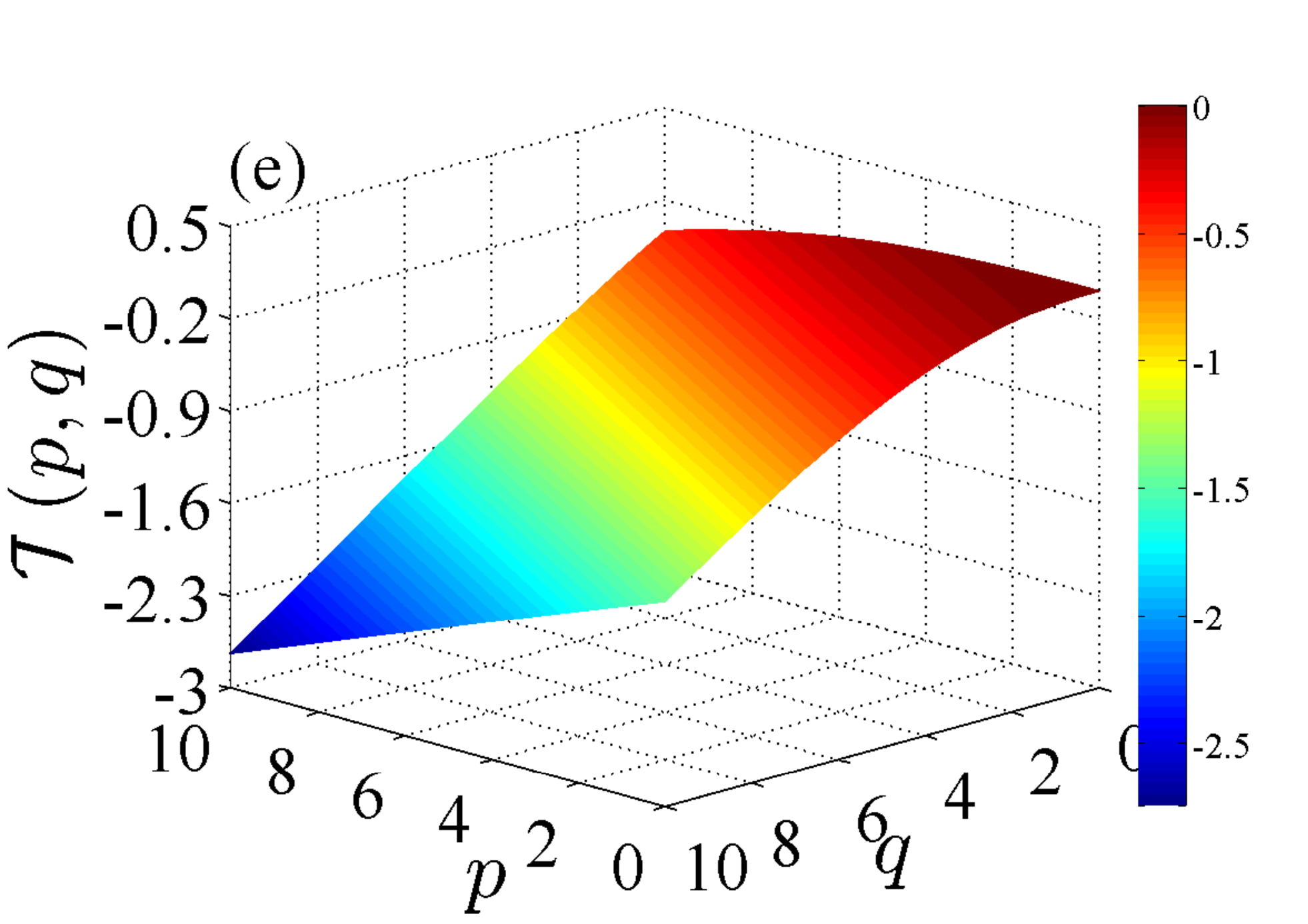}
\includegraphics[width=4.2cm]{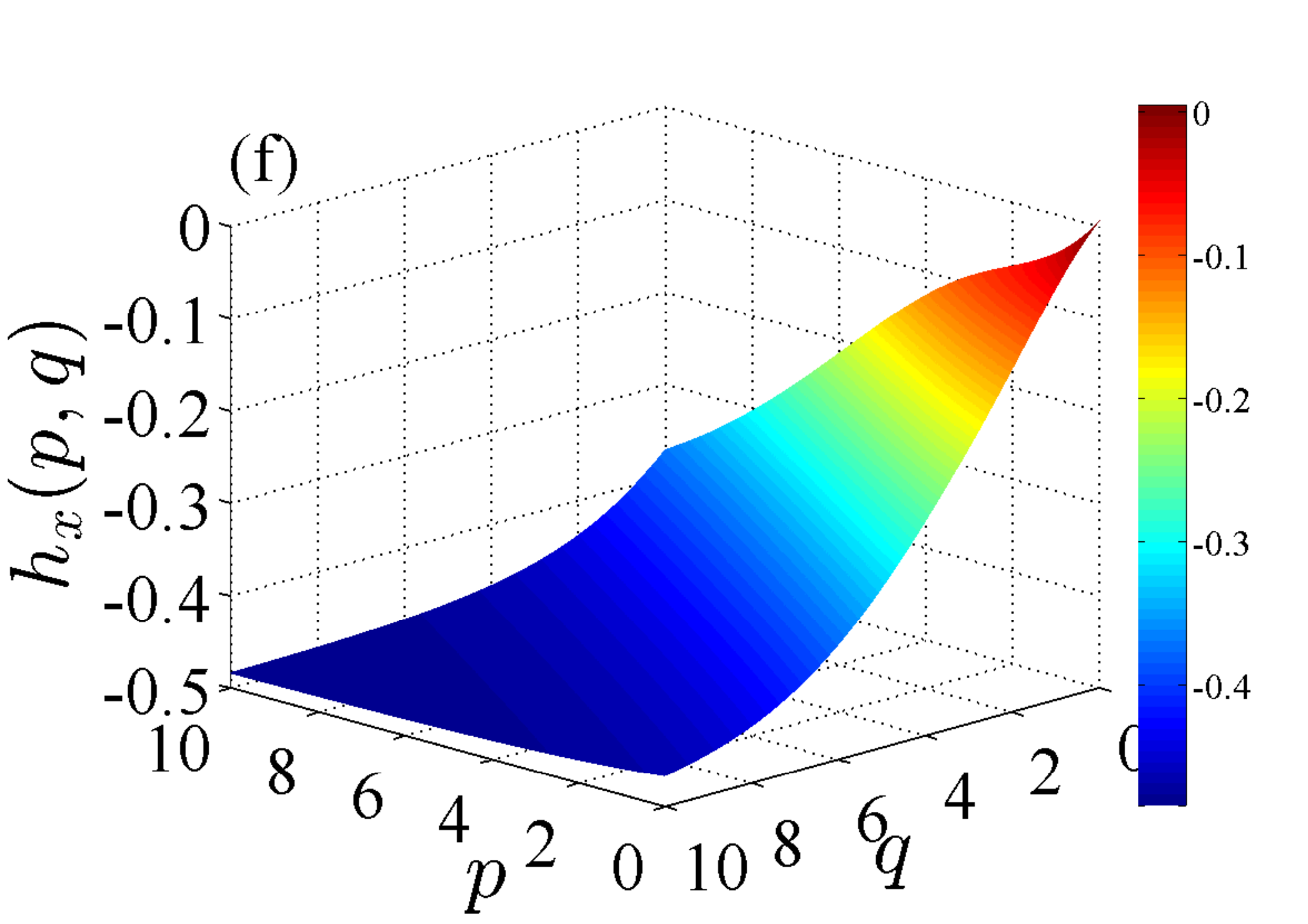}
\includegraphics[width=4.2cm]{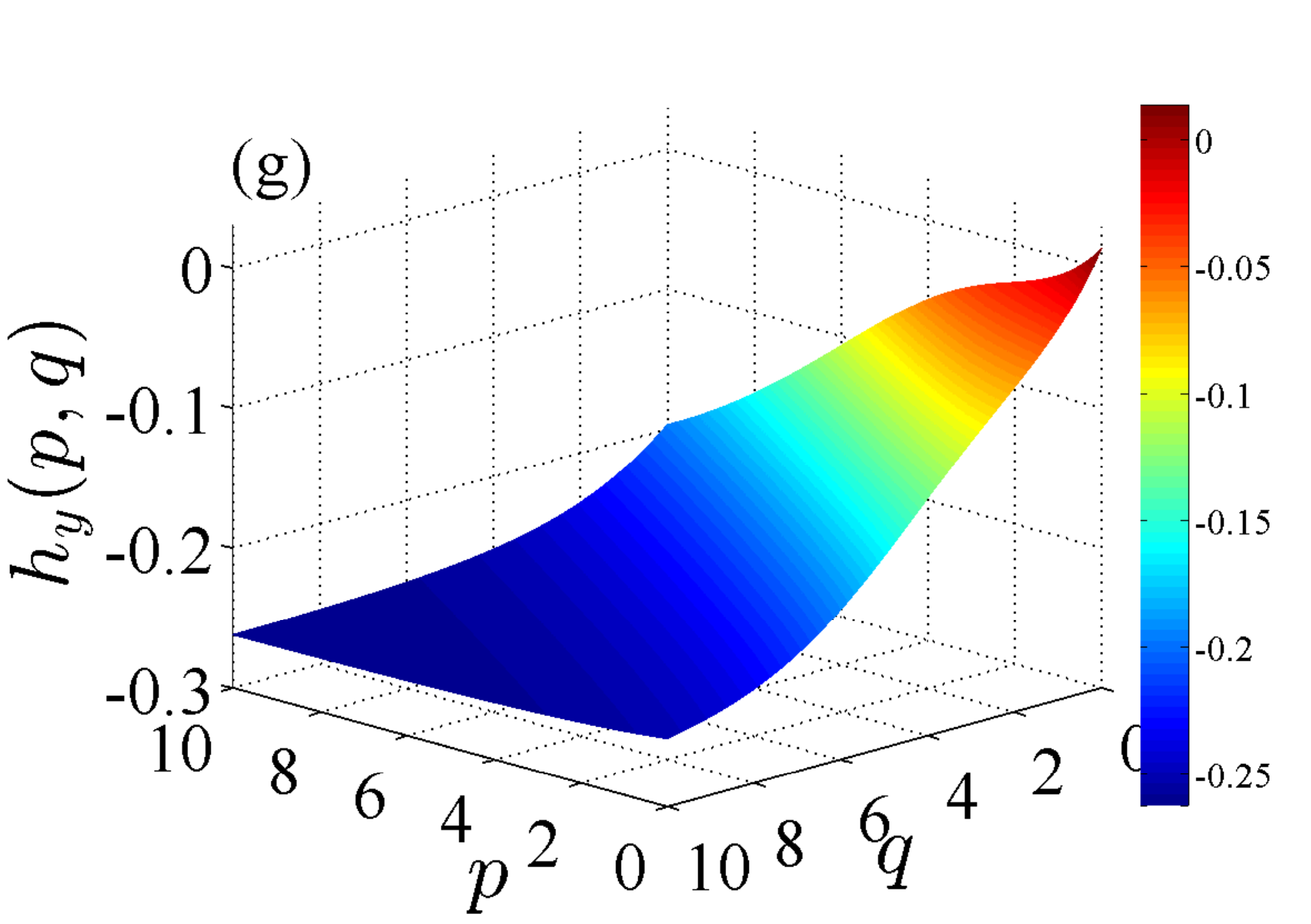}
\includegraphics[width=4.2cm]{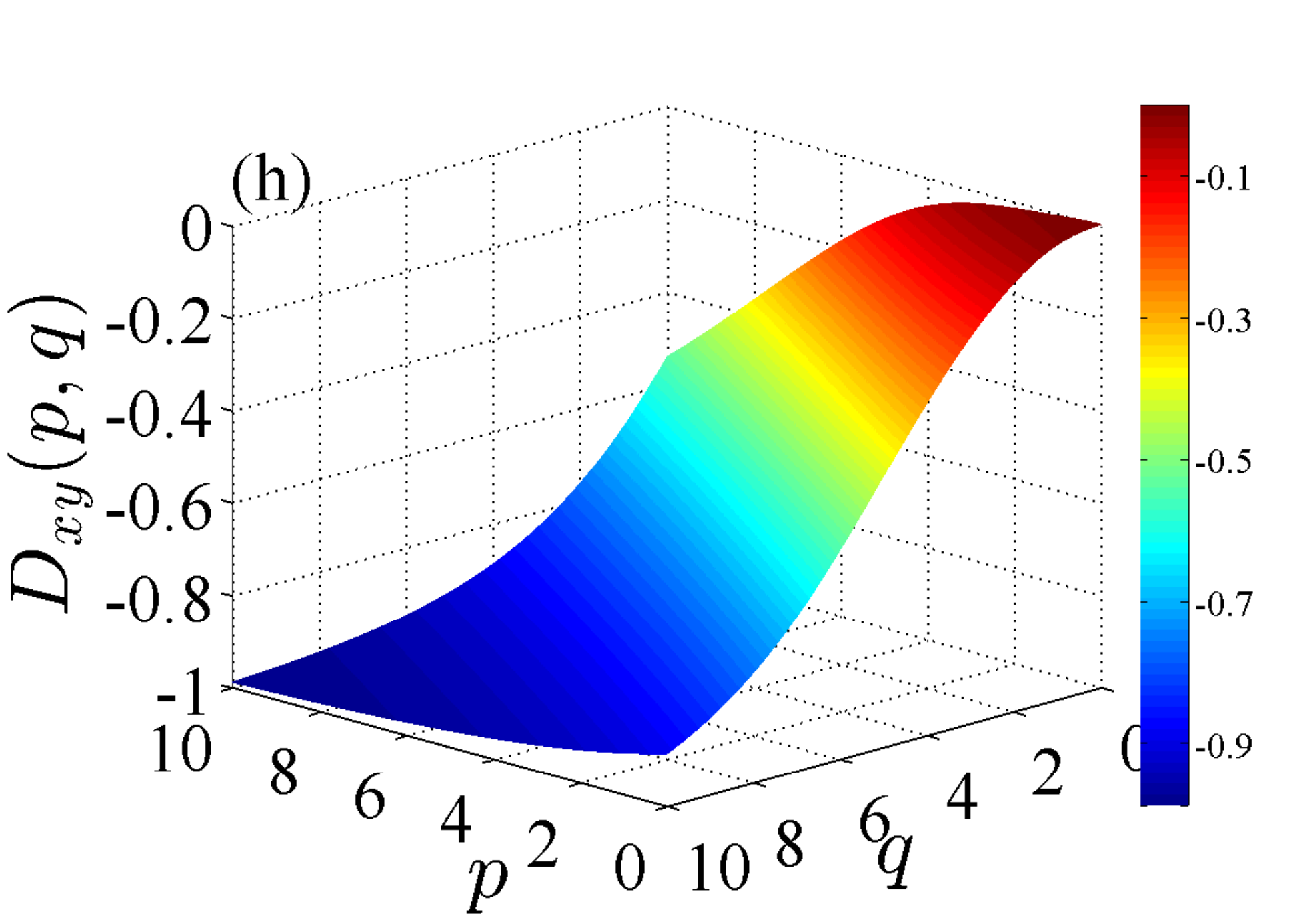}
\includegraphics[width=4.2cm]{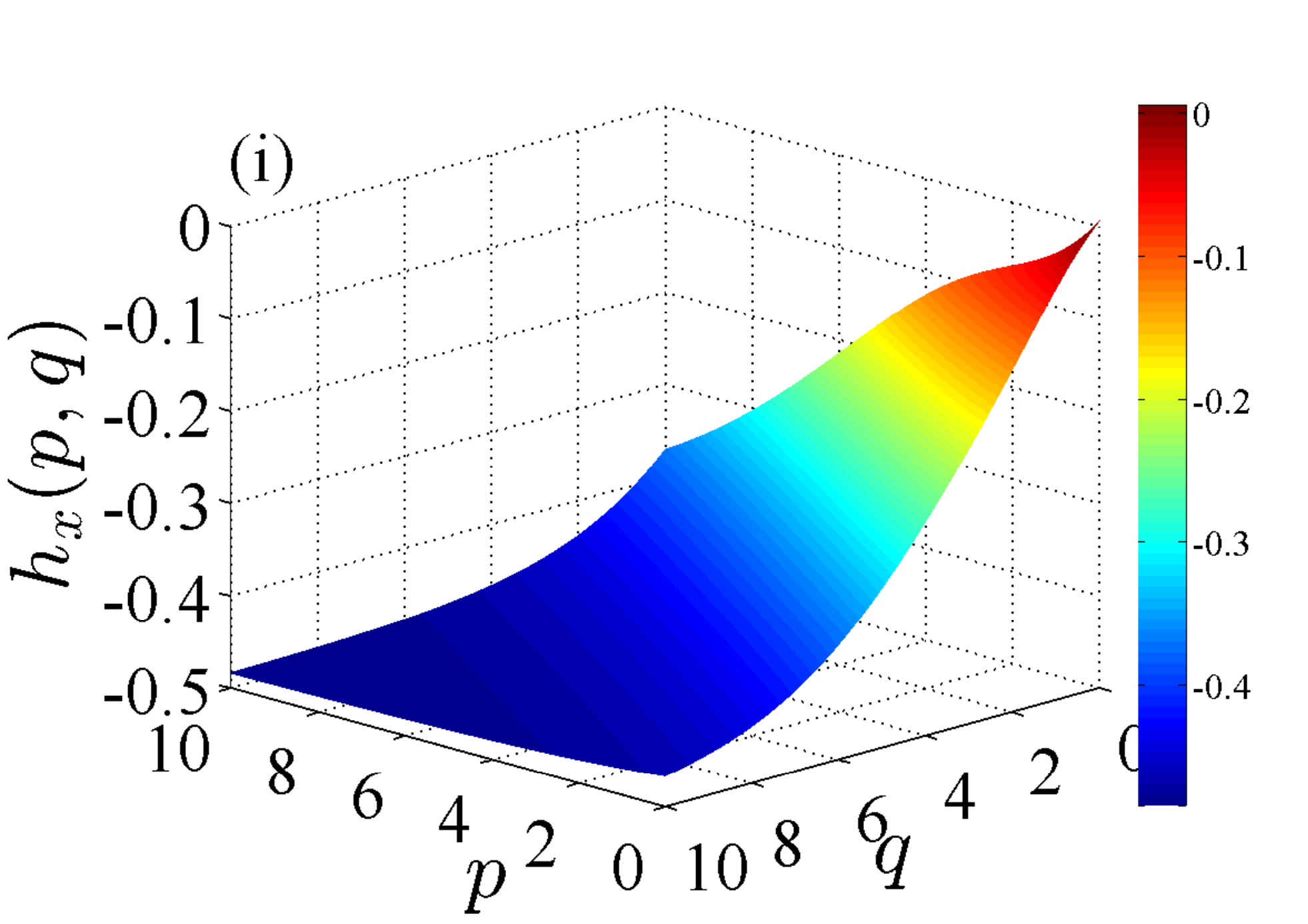}
\includegraphics[width=4.2cm]{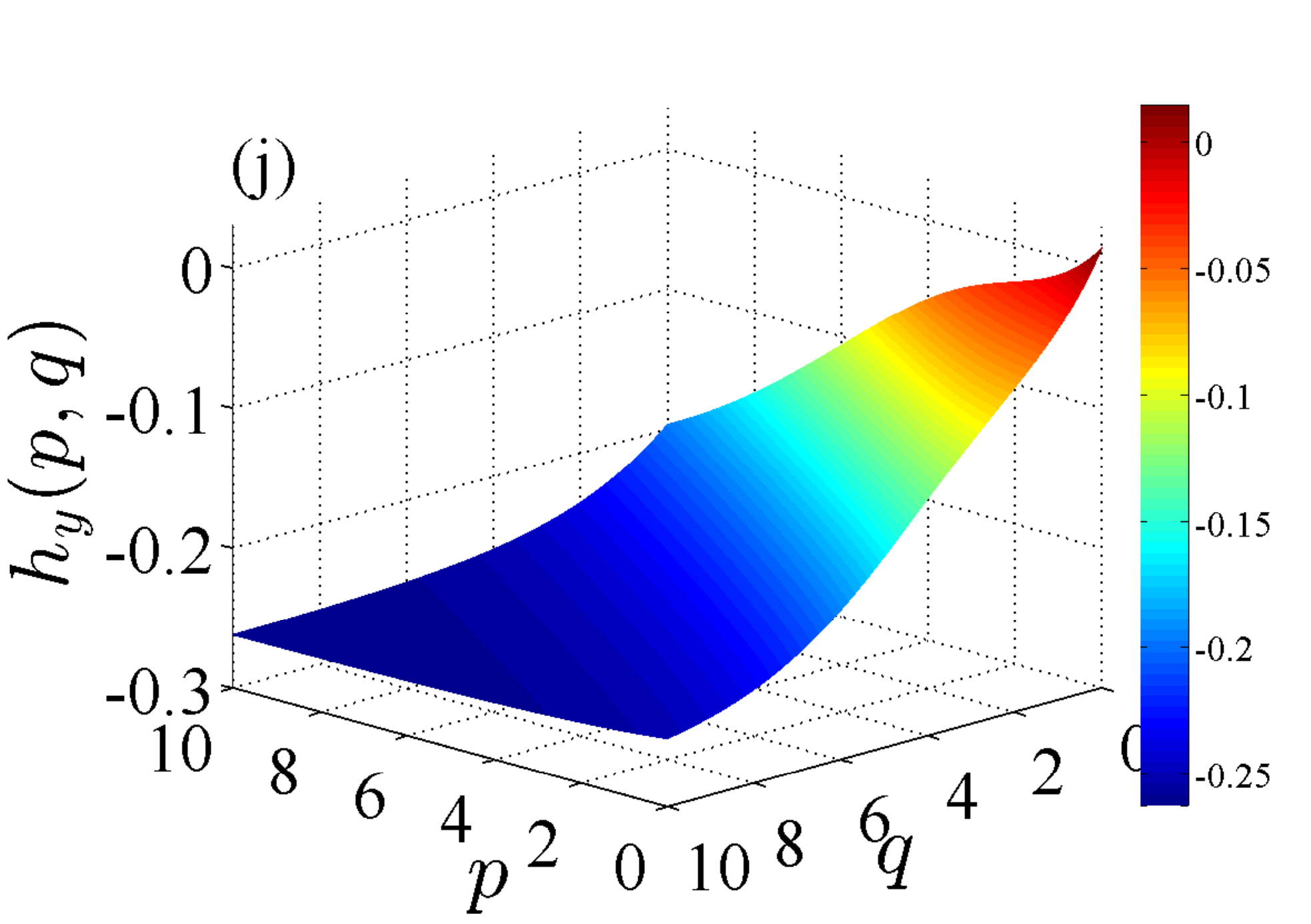}
\includegraphics[width=4.2cm]{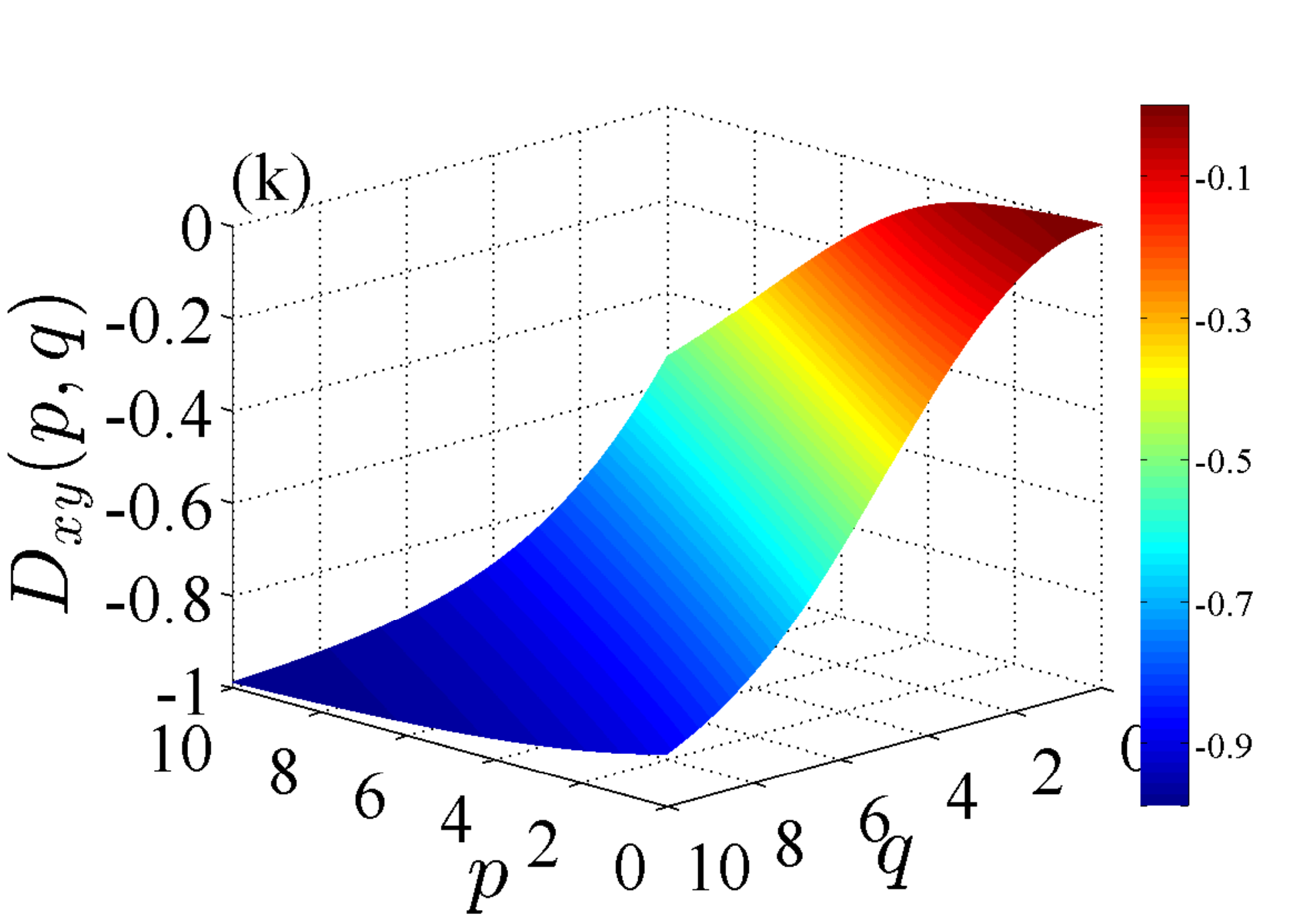}
\includegraphics[width=4.2cm]{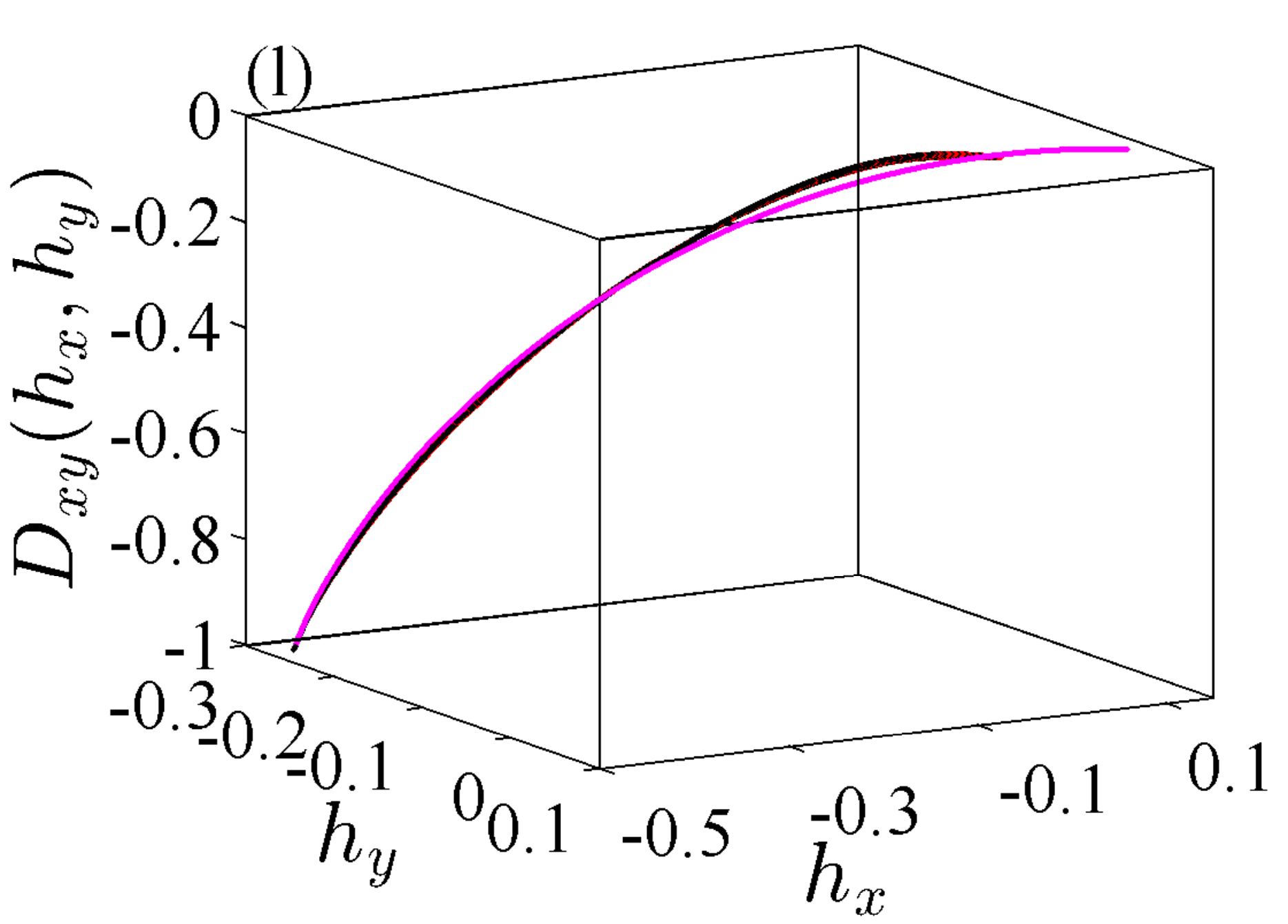}
\caption{\label{Fig:MFXWT:pmodel:PQ} (Color online) Multifractal cross
  wavelet transform analysis of two binomial measures with $p_x = 0.3$
  and $p_y = 0.4$ based on the bi-order MFXWT$(p,q)$ method. (a)
  Power-law plots of $\chi_{xy}(p,q,s)$ with respect to the scale $s$
  for different $q$ with fixed $p=2$. (b) Linear dependence of $\sum_t
  \mu_{xy} (2,q,s,t) \ln |w_x(s,t)|$ against $\ln s$ for different $q$
  with fixed $q=2$. (c) Linear dependence of $\sum_t \mu_{xy} (2,q,s,t)
  \ln |w_y(s,t)|$ against $\ln s$ for different $q$ with fixed
  $q=2$. (d) Linear dependence of $\sum_t \mu_{xy} (2,q,s,t) \ln
  \mu_{xy} (2,q,s,t)$ against $\ln s$ for different $q$ with fixed
  $q=2$. (e)--(h) Joint mass exponent function ${\cal(T)}_{xy} (p,q)$,
  joint singularity functions $h_x(p,q)$ and $h_y(p,q)$, and joint
  multifractal function $D_{xy}(p,q)$ from
  Eqs.~(\ref{Eq:MFXWT:Chi:Scaling}-\ref{Eq:MFXWT:Dh}). (i)--(k) Joint
  singularity functions $h_x(p,q)$ and $h_y(p,q)$ and joint multifractal
  function $D_{xy}(p,q)$ from
  Eqs.~(\ref{Eq:MFXWT:DE:hx}-\ref{Eq:MFXWT:DE:Dxy}). (l) Joint
  multifractal spectrum $D_{xy}(h_x, h_y)$. }
\end{figure*}

Figure~\ref{Fig:MFXWT:pmodel:Q}(e) compares the joint singularity
strength $h_{xy}(q)$ obtained using different methods. The solid line
corresponds to theoretical values. The squares and circles are obtained
from the first derivation of the joint mass exponent ${\cal{T}}_{xy}(q)$
and the direct estimating method, respectively. Note that although the
empirical $h_{xy}(q)$ of both methods of estimating coincide, both
empirical curves agree with the theoretical values only when $q \ge
1$. We see deviations when $q < 1$, and the reason for this is not
clear.

Figure~\ref{Fig:MFXWT:pmodel:Q}(f) shows the corresponding
joint multifractal spectra of binomial measures in both theoretical
values and estimated values. The $D_{xy}(h_{xy})$ values obtained from
Eqs.~(\ref{Eq:MFXWT:Dh}) and (\ref{Eq:MFXWT:DE:Dxy}) agree and collapse
on the theoretical curves. Our results suggest that the accuracy of the
MFXWT$(p,q)$ is acceptable for analyzing the joint multifractality in
binomial measures.

Releasing the $p=q$ restriction in Fig.~\ref{Fig:MFXWT:pmodel:Q},
Fig.~\ref{Fig:MFXWT:pmodel:PQ}(a) shows how the joint partition function
$\chi_{xy}(2,q,s)$ depends on the scale $s$ for different $q$ with fixed
$p=2$. We see power-law behaviors. For each pair of $(p,q)$, the slope
of the straight line in an estimate of the corresponding joint mass
exponent ${\cal{T}}_{xy}(p,q)$. Figure~\ref{Fig:MFXWT:pmodel:PQ}(e)
shows a plot of the joint mass exponent function ${\cal{T}}_{xy}(p,q)$
with respect to $p$ and $q$. Note that again there are nonlinear
features between ${\cal{T}}_{xy}(p,q)$ and $(p,q)$, and this verifies
joint multifractality in the two binomial measures. Following
Eqs.~(\ref{Eq:MFXWT:hx}-\ref{Eq:MFXWT:Dh}), if we have the mass exponent
${\cal{T}}_{xy}(p,q)$ we can numerically compute the joint singularity
strength functions $h_x(p,q)$ and $h_y(p,q)$ and the joint multifractal
function $D_{xy}(p,q)$. Figures~\ref{Fig:MFXWT:pmodel:PQ}(f),
\ref{Fig:MFXWT:pmodel:PQ}(g) and \ref{Fig:MFXWT:pmodel:PQ}(h) show the
corresponding $h_x(p,q)$, $h_y(p,q)$, and $D_{xy}(p,q)$,
respectively. The wide spanning range of $h_x$, $h_y$, and $D_{xy}$
further corroborates the joint multifractality in binomial measures.

The direct estimation method presented in
Eqs.~(\ref{Eq:MFXWT:DE:hx}--\ref{Eq:MFXWT:DE:Dxy}) is an alternative way
to estimate the joint singularity strength $h_x(p,q)$ and $h_y(p,q)$ and
the joint multifractal function $D_{xy}(p,q)$. By estimating the three
quantities $\sum \mu_{xy}\ln|w_x|$, $\sum\mu_{xy}\ln |w_y|$, and
$\sum\mu_{xy}\ln\mu_{xy}$ we find power-law scaling behaviors between
these quantities and the scale $s$, as shown in
Fig.~\ref{Fig:MFXWT:pmodel:PQ}(b--d). Their power-law exponents
correspond to the joint singularity strength function $h_x(p,q)$ in
Fig.~\ref{Fig:MFXWT:pmodel:PQ}(i) and $h_y(p,q)$ in
Fig.~\ref{Fig:MFXWT:pmodel:PQ}(j) and the joint multifractal function
$D_{xy}(p,q)$ in Fig.~\ref{Fig:MFXWT:pmodel:PQ}(k). Note that $h_x(p,q)$
in Fig.~\ref{Fig:MFXWT:pmodel:PQ}(f) and
Fig.~\ref{Fig:MFXWT:pmodel:PQ}(i), $h_y(p,q)$ in
Fig.~\ref{Fig:MFXWT:pmodel:PQ}(g) and Fig.~\ref{Fig:MFXWT:pmodel:PQ}(j),
and $D_{xy}(p,q)$ in Fig.~\ref{Fig:MFXWT:pmodel:PQ}(h) and
Fig.~\ref{Fig:MFXWT:pmodel:PQ}(k) obtained using both methods agree.

\begin{figure*}[htb]
\centering
\includegraphics[width=5.5cm]{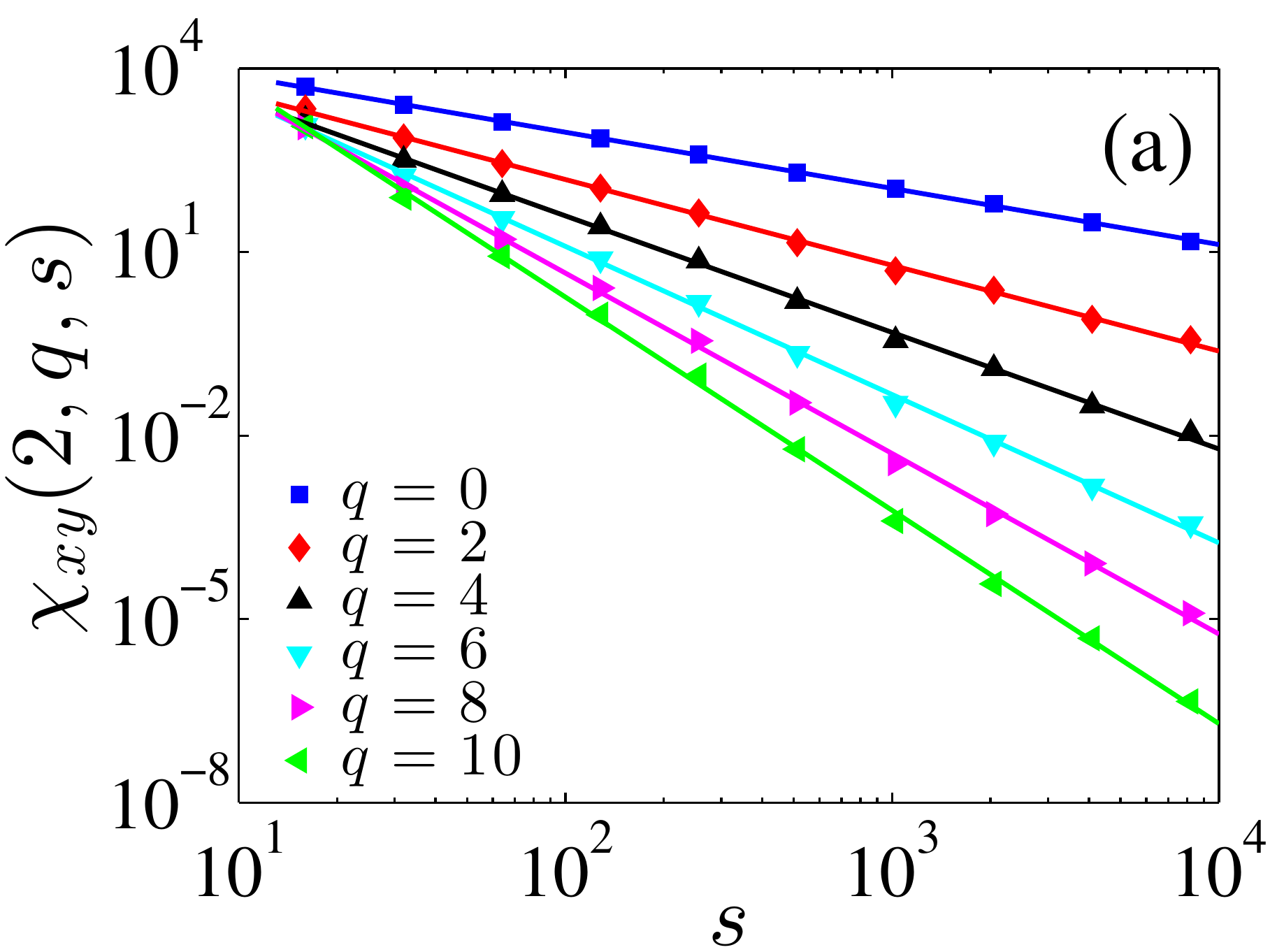}
\includegraphics[width=5.5cm]{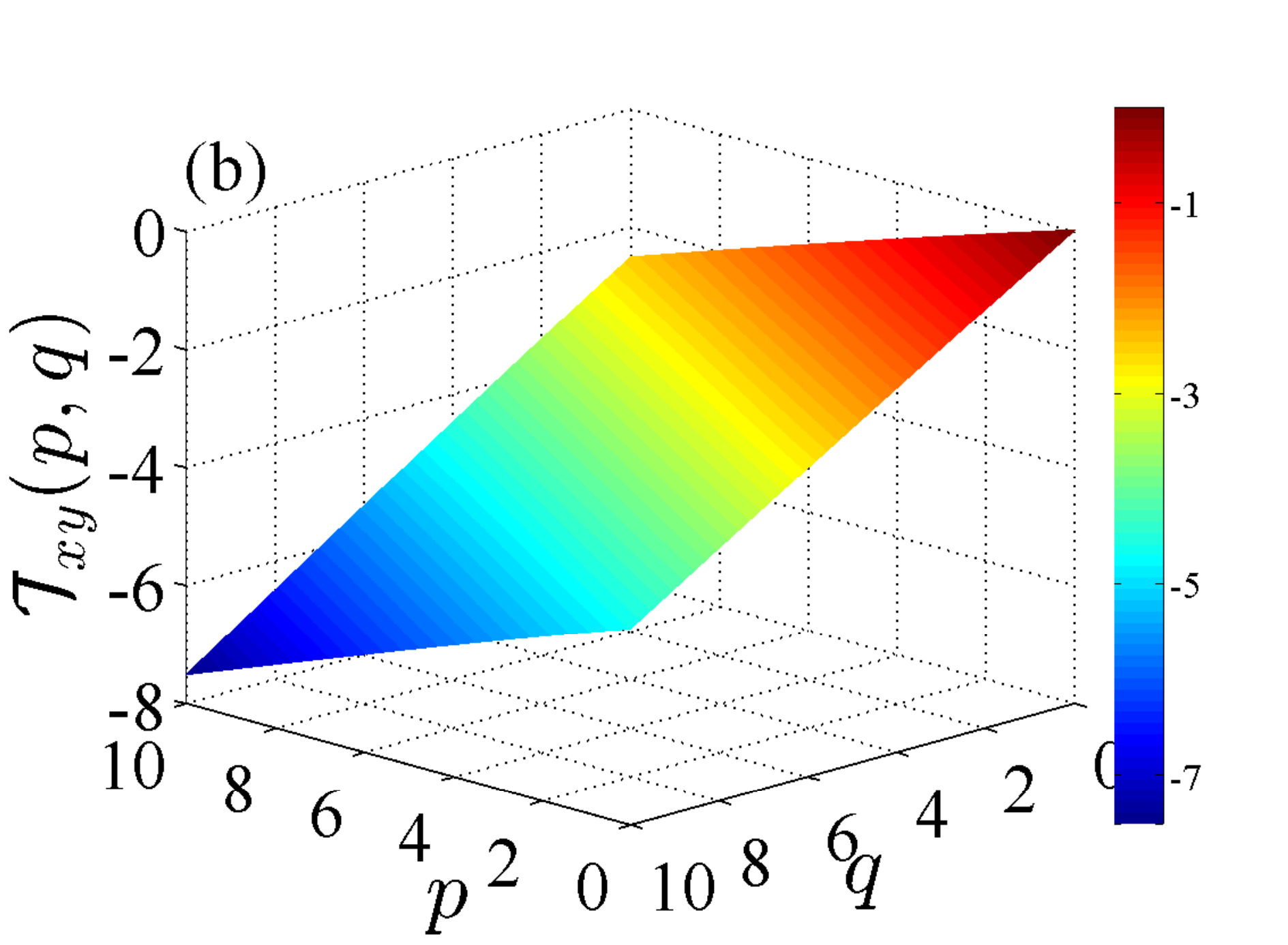}
\includegraphics[width=5.5cm]{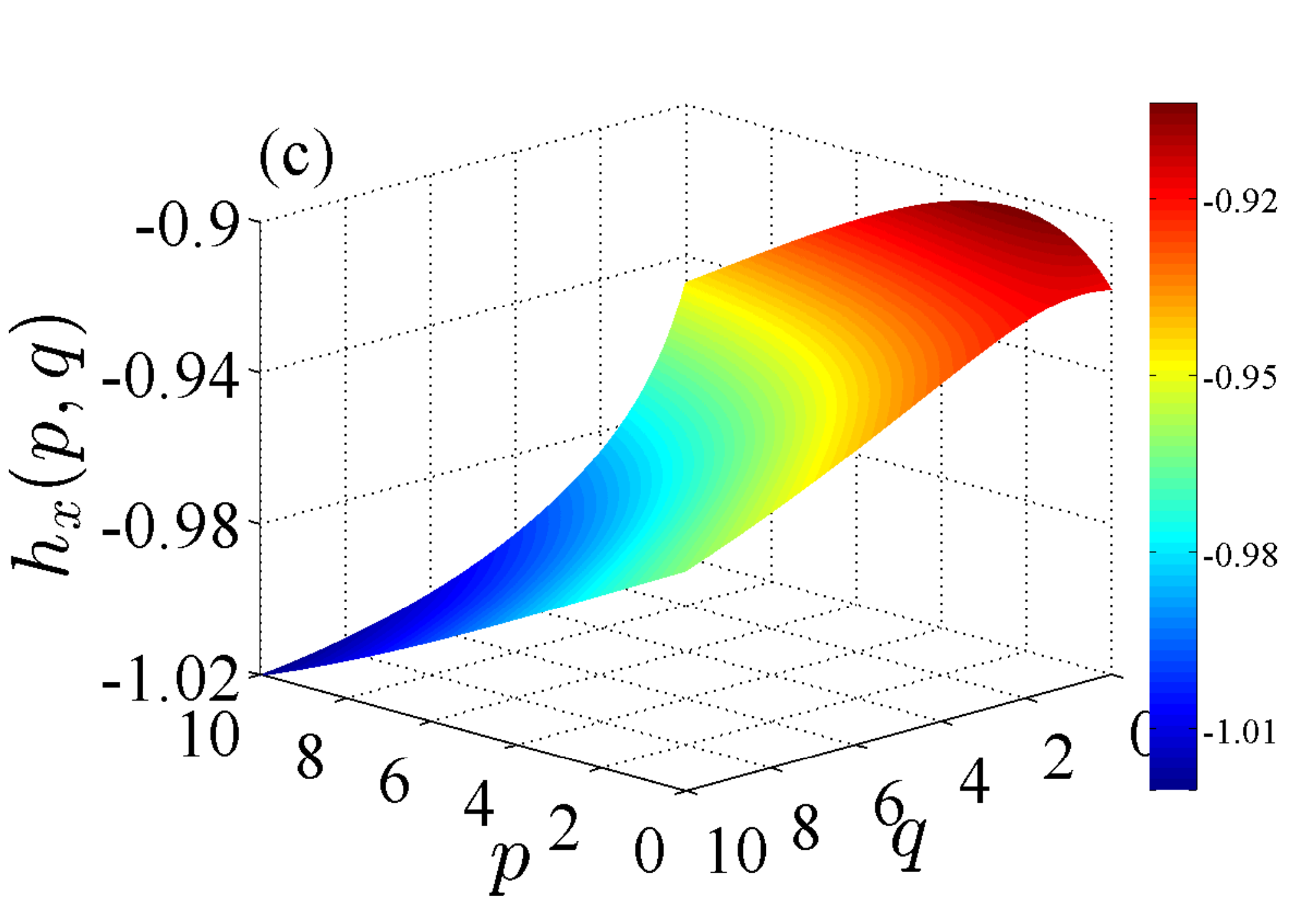}
\includegraphics[width=5.5cm]{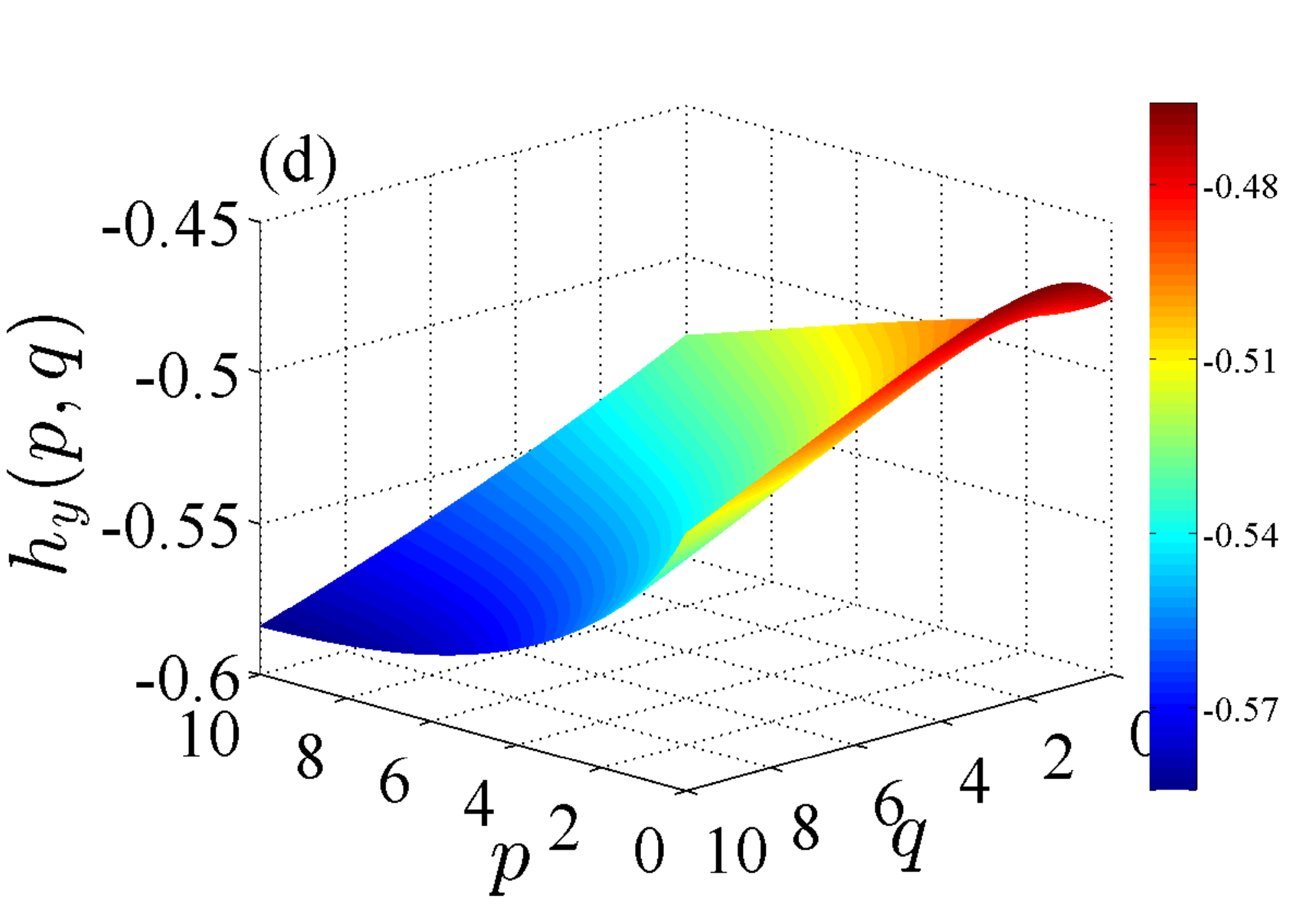}
\includegraphics[width=5.5cm]{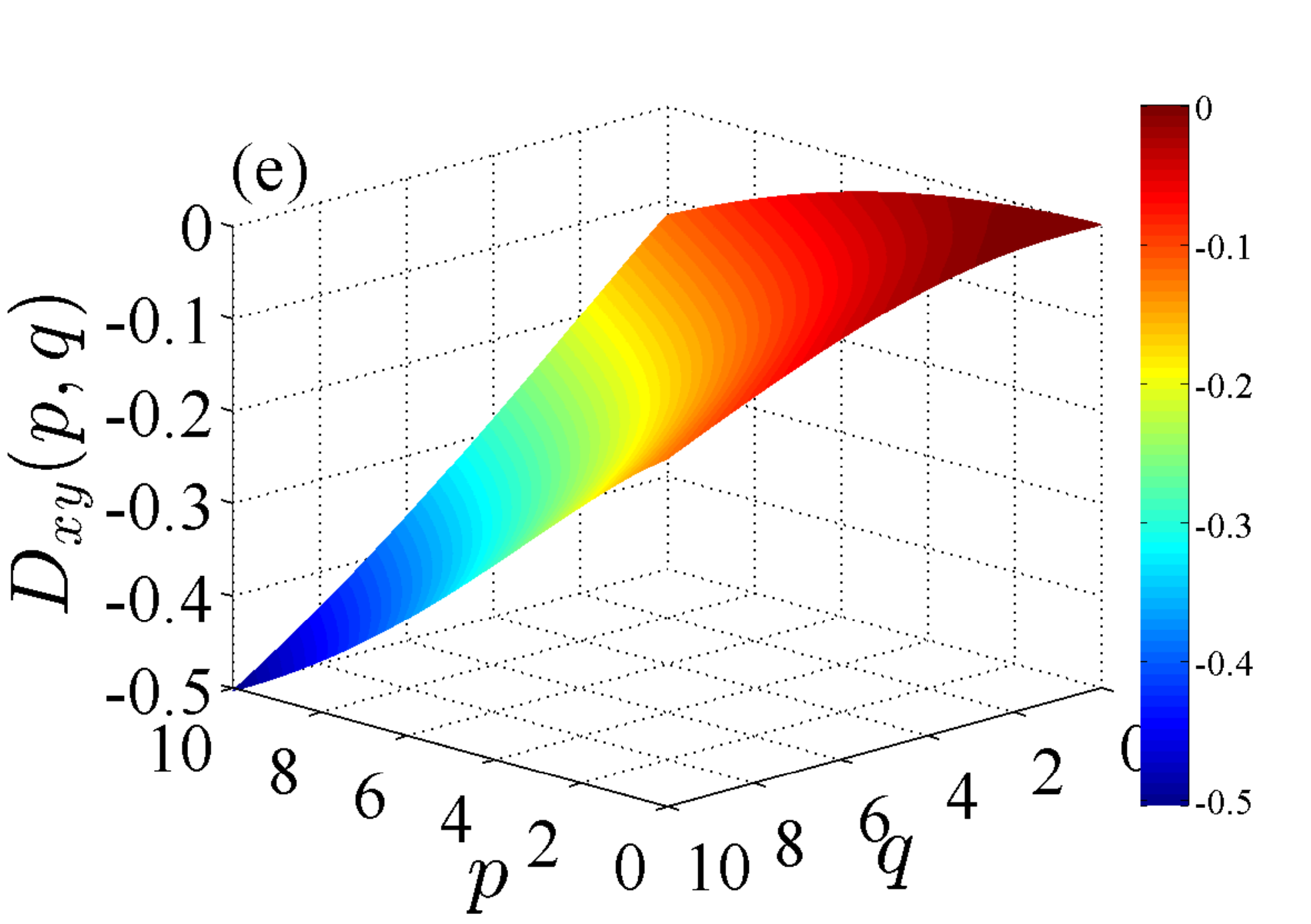}
\includegraphics[width=5.5cm]{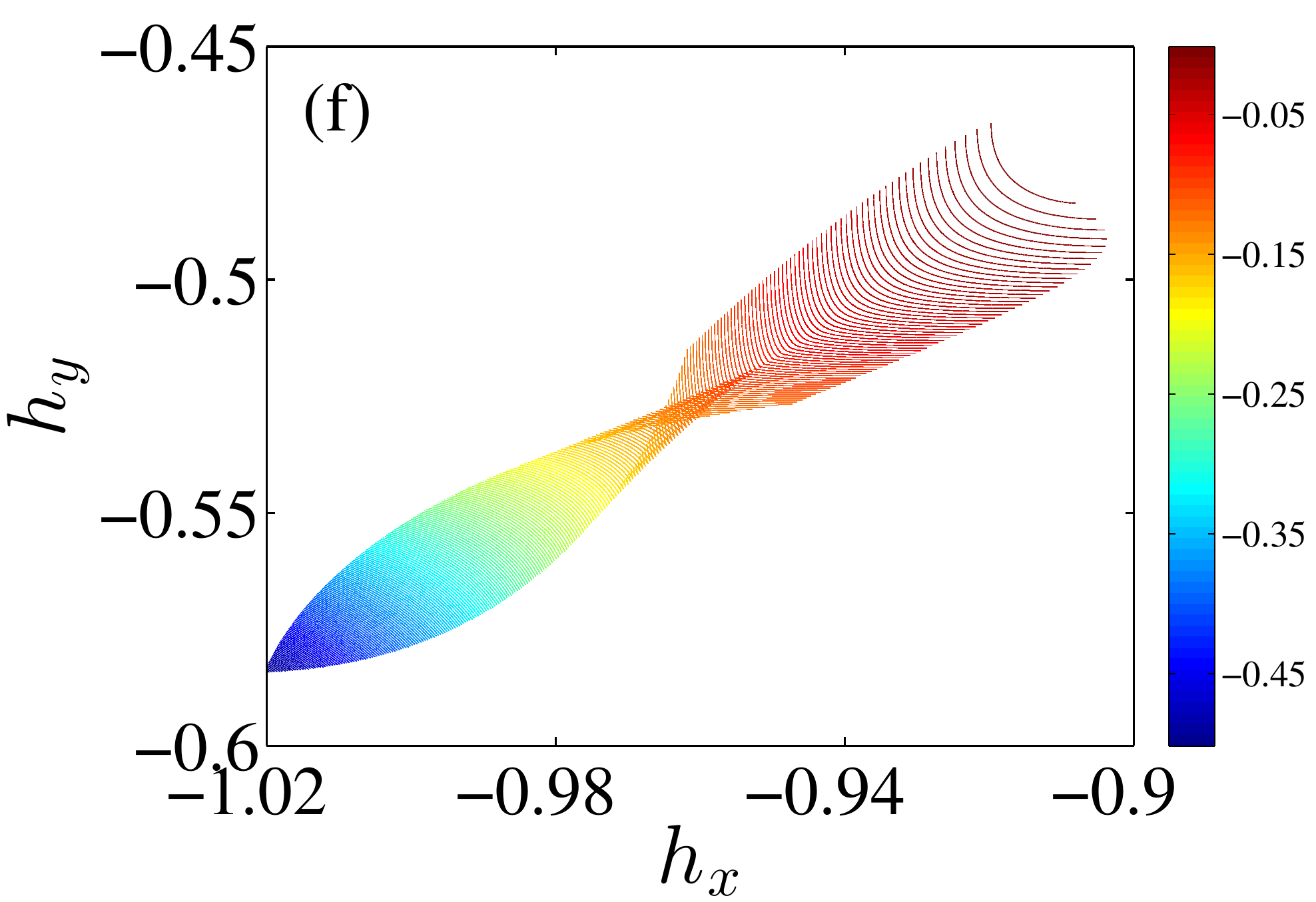}
\caption{\label{Fig:MFXWT:BFBM} (Color online) Multifractal cross
  wavelet analysis of bivariate fractional Brownian motion with $H_{xx}
  = 0.1$, $H_{yy} = 0.5$, and $\rho = 0.5$. (a) Power-law relationship
  between $\chi_{xy}(p, q, s)$ and scale $s$ for fixed $p=2$ and
  different $q$. (b) Joint mass exponent function ${\cal{T}}_{xy}(p,q)$
  obtained from Eq.~(\ref{Eq:MFXWT:Chi:Scaling}). Using the least square
  method, we have ${\cal{T}}_{xy} = -0.485p-0.268q+0.135$. (c) Joint
  singularity function $h_x(p, q)$. (d) Joint singularity function
  $h_y(p, q)$. (e) Joint multifractal spectrum $D_{xy}(p,q)$. (f)
  Contour plots of joint multifractal spectrum $D_{xy}(h_x, h_y)$.}
\end{figure*}

Figure~\ref{Fig:MFXWT:pmodel:PQ}(l) shows that both methods agree, and
this illustrates the joint multifractal spectra $D_{xy} (h_x, h_y)$ of
both methods, as well as the theoretical values (magenta curve)
expressed in Eq.~(\ref{Eq:pModel:fxyalpha}). Xie et al.
\cite{Xie-Jiang-Gu-Xiong-Zhou-2015-NJP} showed that the joint
multifractal spectrum $f_{xy}(\alpha_x, \alpha_y)$ of binomial measures
is a univariate function of $Q$, and thus also of $\alpha_x$ or of
$\alpha_y$ due to Eqs.~(\ref{Eq:pModel:Alphax}) and
(\ref{Eq:pModel:Alphay}), and this means that the joint multifractal
spectrum of MFXWT is also a univariate function of $h_x$ or of $h_y$ due
to Eqs.~(\ref{Eq:pModel:WThx:PFalphax}) and
(\ref{Eq:pModel:WThy:PFalphay}).  Figure~\ref{Fig:MFXWT:pmodel:PQ}(l)
shows that the estimated joint multifractal spectra from both methods
are a curve, not a surface, verifying the univariate function
relationship between $D_{xy}$ and $h_x$ or $h_y$. Both estimated joint
multifractal spectra approximately overlap with the theoretical
multifractal spectrum, suggesting that the MFXWT$(p,q)$ method provides
an accurate joint multifractal analysis of binomial measures.

\subsection{Bivariate fractional Brownian motions (bFBMs)}

A bivariate fractional Brownian motion $[x(t),y(t)]$ with parameters
$\{H_{xx},H_{yy}\}\in(0,1)^2$ is a self-similar Gaussian process with
stationary increments, where $x(t)$ and $y(t)$ are two univariate
fractional Brownian motions with Hurst indices $H_{xx}$ and $H_{yy}$ and
also are the two components of the bFBM
\cite{Lavancier-Philippe-Surgailis-2009-SPL,Coeurjolly-Amblard-Achard-2010-EUSIPCO,Amblard-Coeurjolly-Lavancier-Philippe-2013-BSMF}. The
basic properties of multivariate fractional Brownian motions have been
extensively studied
\cite{Lavancier-Philippe-Surgailis-2009-SPL,Coeurjolly-Amblard-Achard-2010-EUSIPCO,Amblard-Coeurjolly-Lavancier-Philippe-2013-BSMF}. Extensive
numerical experiments of multifractal cross-correlation analysis
algorithms have been performed on bFBMs \cite{Jiang-Zhou-2011-PRE,
  Qian-Liu-Jiang-Podobnik-Zhou-Stanley-2015-PRE,
  Xie-Jiang-Gu-Xiong-Zhou-2015-NJP}. The two Hurst indices $H_{xx}$ and
$H_{yy}$ of the two univariate FBMs and their cross-correlation
coefficient $\rho$ are input arguments in the simulation algorithm. By
using the simulation procedure described in
Ref.~\cite{Coeurjolly-Amblard-Achard-2010-EUSIPCO,
  Amblard-Coeurjolly-Lavancier-Philippe-2013-BSMF}, we generate a
realization of bFBM in which $H_{xx} = 0.1$, $H_{yy} = 0.5$, and
$\rho=0.5$. The length of the bFBM is $2^{16}$.

As described in Ref.~\cite{Xie-Jiang-Gu-Xiong-Zhou-2015-NJP}, if the two
time series is monofractal, the joint singularity strengths $h_x(p,q)$
and $h_y(p,q)$ are constants, and their joint multifractal spectrum is
$D_{xy}(h_x, h_y)=0$. Note that
Eqs.~(\ref{Eq:pModel:WTTau:PFTau}--\ref{Eq:pModel:WTDh:PFfalpha})
obtained from the $p$-model are no longer valid because they are derived
using conservative measures, and the increments of both components
$x(t)$ and $y(t)$ in bFBM are not conservative.

Figure~\ref{Fig:MFXWT:BFBM} shows the results of the joint multifractal
analysis of the bFBM using the MFXWT algorithm.
Figure~\ref{Fig:MFXWT:BFBM}(a) shows how the joint partition functions
$\chi_{xy}(2,q,s)$ of the wavelet coefficients are plotted with respect
to the scale $s$ for fixed $p=2$ and different $q$. Again we see strong
power-law scaling behaviors that allow us to estimate the joint mass
exponents ${\cal{T}}_{xy}$ using the least square estimation
method. Figure~\ref{Fig:MFXWT:BFBM}(b) shows the joint mass exponent
function against different $p$ and $q$. Because of the monofractality of
the bFBMs, we see a plane for ${\cal{T}}(p,q)$. The bivariate regression
yields
\begin{equation}
  {\cal{T}}_{xy}(p,q) = - 0.485 p - 0.268 q + 0.135.
  \label{Eq:MFXWT:Tau:p:q}
\end{equation}
Using Eq.~(\ref{Eq:MFXWT:Dh}), we infer that ${\overline{h}}_x =
-0.970$, ${\overline{h}}_y = -0.536$, and ${\overline{D}}_{xy} = -0.135$
deviate from the theoretical value $D_{xy}(0,0)=0$. When $p=q=0$,
Eq.~(\ref{Eq:MFXWT:Tau:p:q}) gives ${\cal{T}}_{xy}(0,0)=0.135$, which
also deviates from the theoretical value ${\cal{T}}_{xy}(0,0)=0$.

\begin{figure*}[htb]
  \centering
  \includegraphics[width=5.5cm]{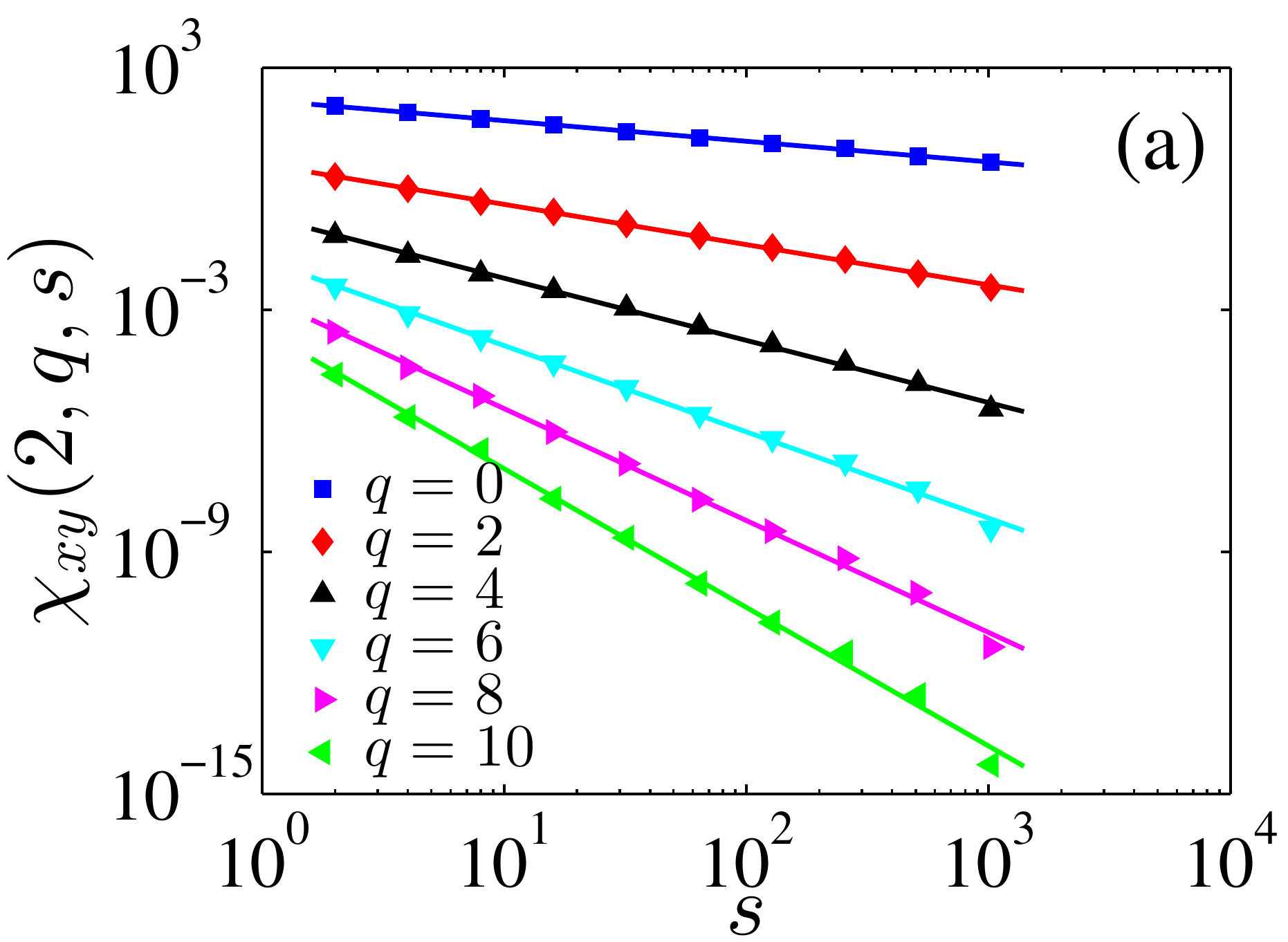}
  \includegraphics[width=5.5cm]{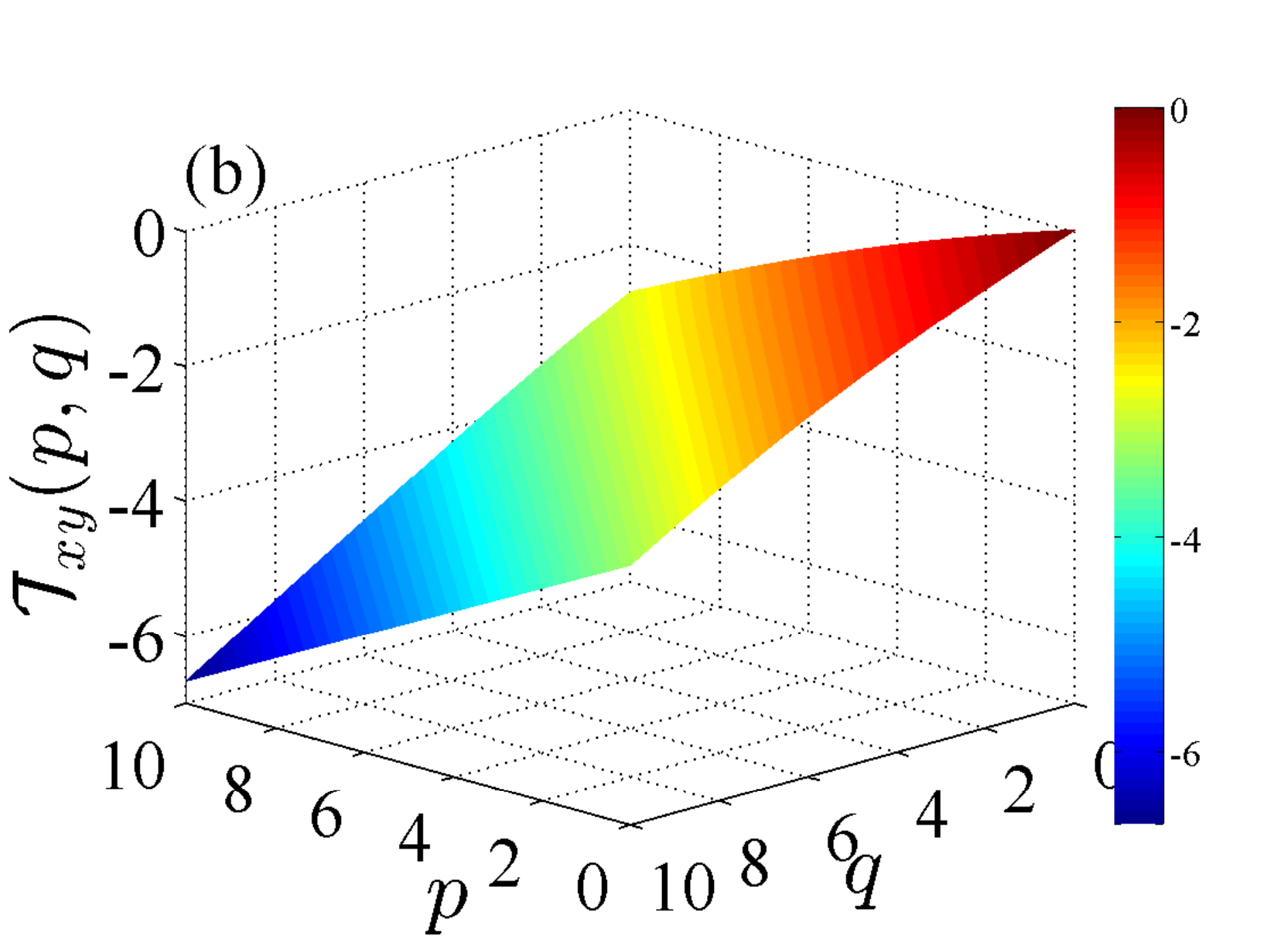}
  \includegraphics[width=5.5cm]{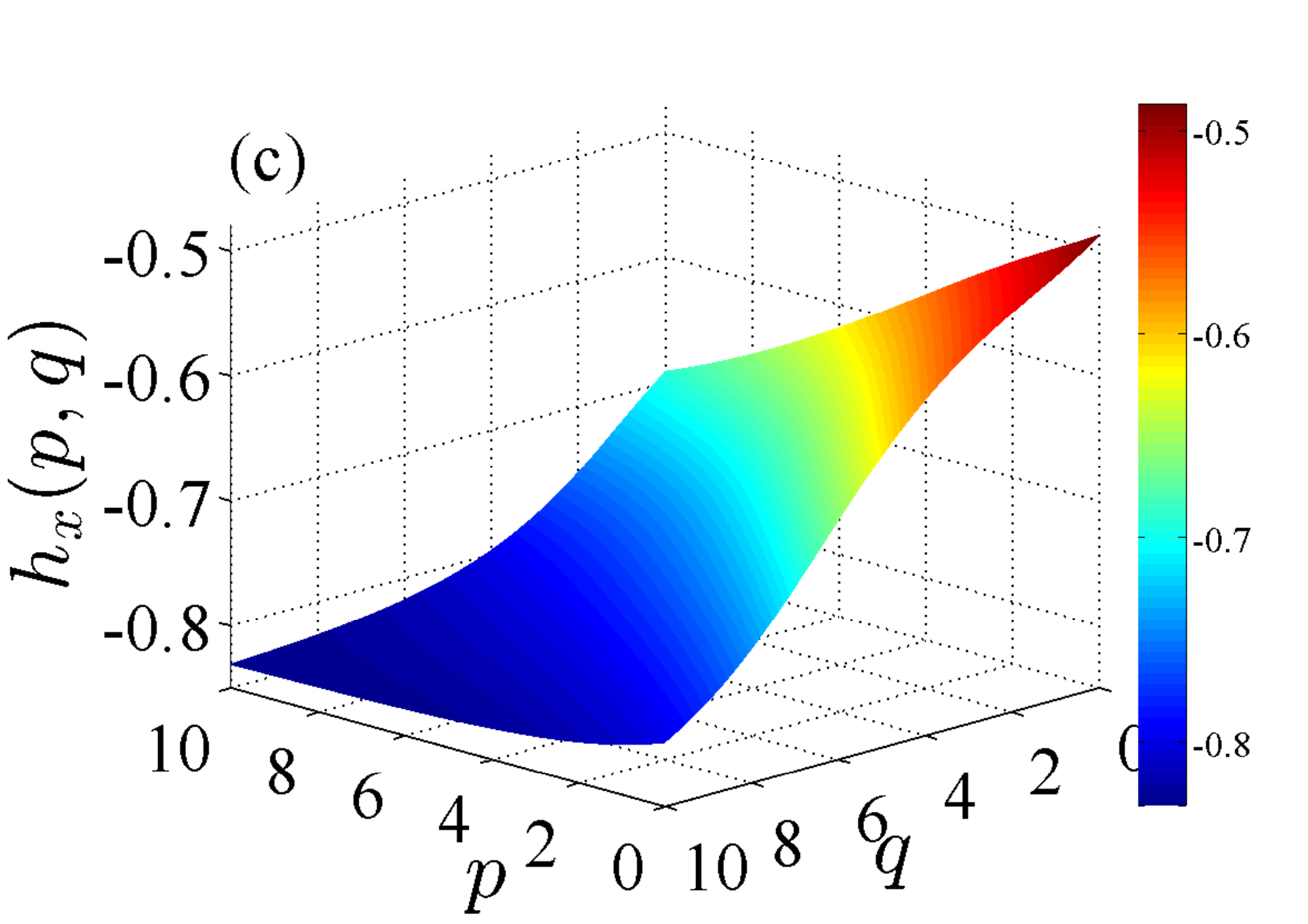}
  \includegraphics[width=5.5cm]{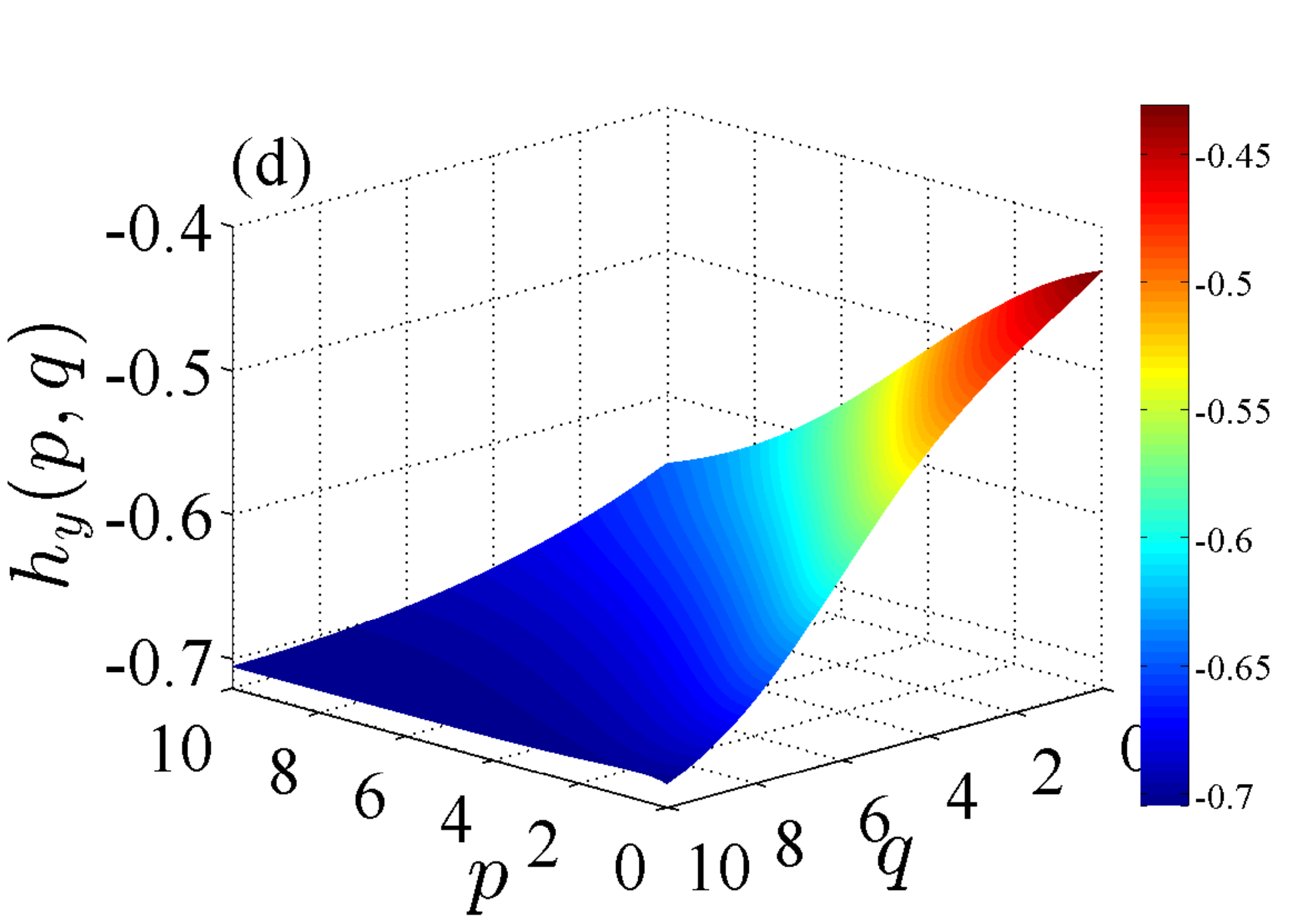}
  \includegraphics[width=5.5cm]{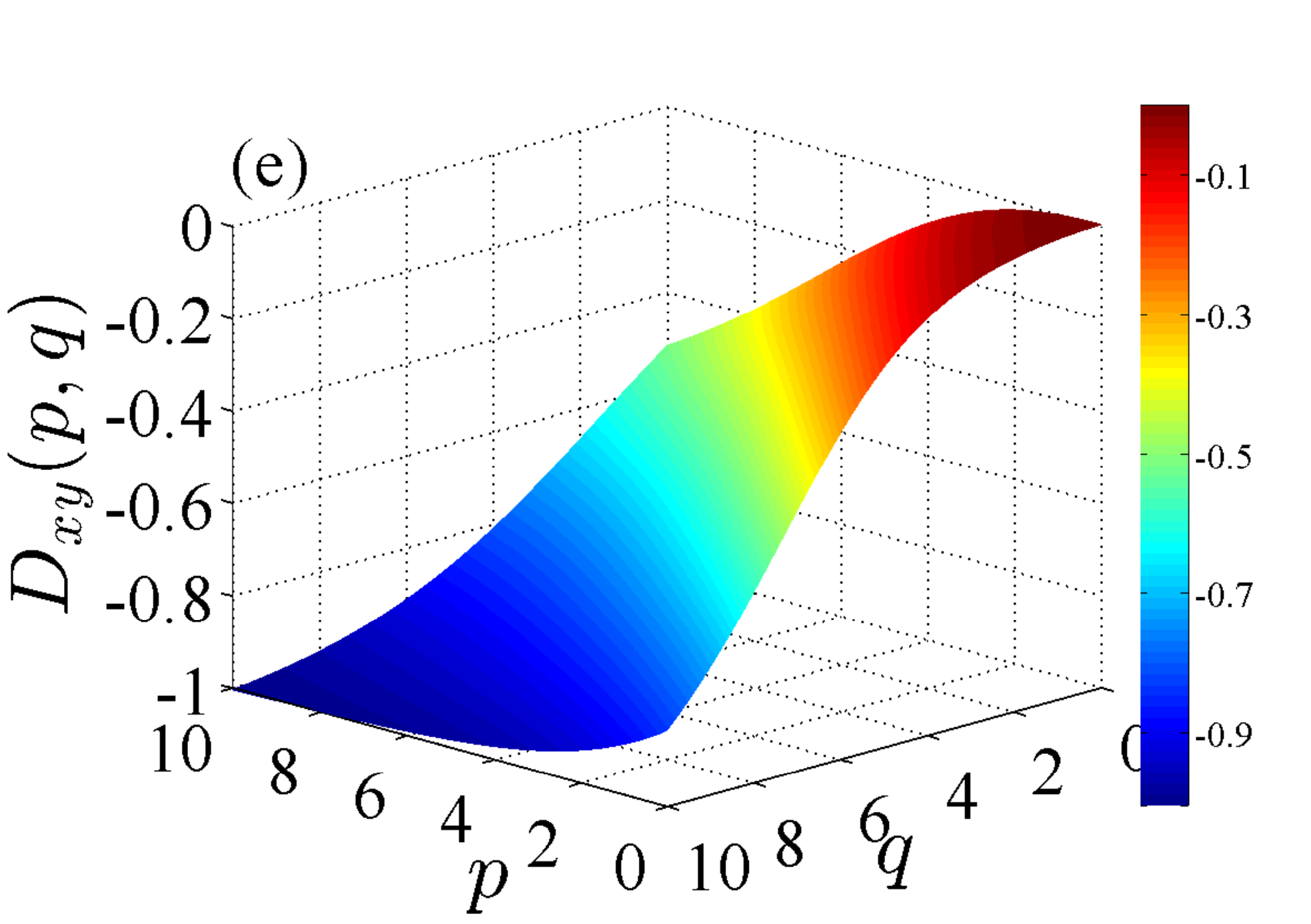}
  \includegraphics[width=5.5cm]{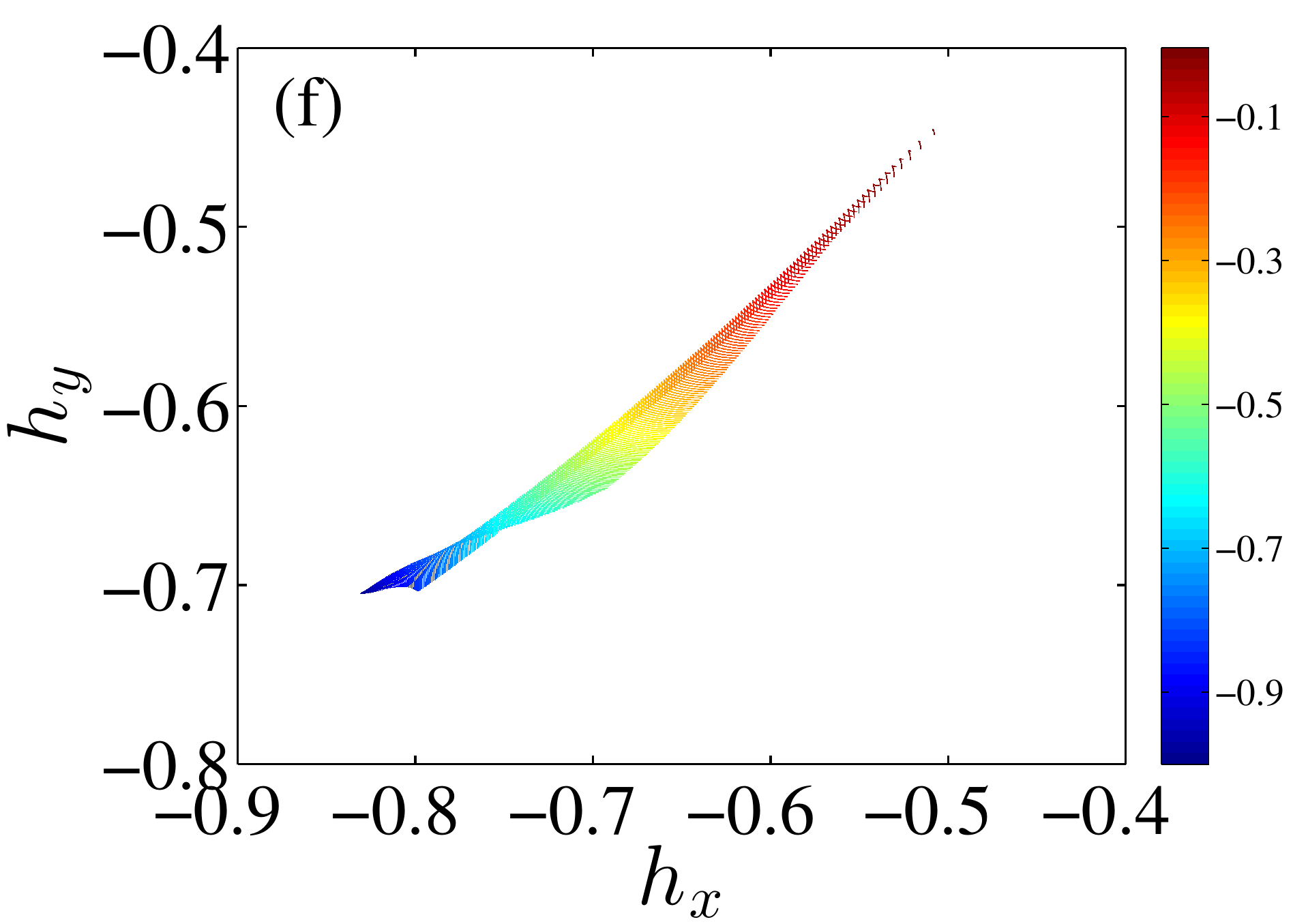}
  \caption{\label{Fig:MFXWT:Returns} (Color online) Multifractal cross
    wavelet analysis of the joint multifractality between the daily
    return series of DJIA index and NASDAQ index using the MFXWT$(p,q)$
    method. (a) Power-law dependence of $\chi_{xy}(p,q,s)$ on scale $s$
    for fixed $p=2$ and different $q$.  (b) Joint mass exponent function
    ${\cal{T}}(p,q)$. (c) Joint singularity strength function
    $h_x(p,q)$. (d) Joint singularity strength function $h_y(p,q)$. (e)
    Joint multifractal function $D_{xy} (p,q)$. (f) Joint multifractal
    singularity spectrum $D_{xy}(h_x, h_y)$.}
\end{figure*}

Alternatively, using Eqs.~(\ref{Eq:MFXWT:hx}) and (\ref{Eq:MFXWT:hy})
and a numerical differentiation of ${\cal{T}}_{xy}$, we can estimate
$h_x(p,q)$ and $h_y(p,q)$. Figures~\ref{Fig:MFXWT:BFBM}(c) and
\ref{Fig:MFXWT:BFBM}(d) plot the estimated joint singularity strength
functions $h_x(p,q)$ and $h_y(p,q)$ obtained from taking the forward
difference of ${\cal{T}}_{xy}(p,q)$. Note that the singularity strength
function $h_x(p,q)$ and $h_y(p,q)$ obtained from the numerical methods
vary across a small range. The corresponding average value is $-0.968$
for $h_x$ and $-0.534$ for $h_y$, which agreea with ${\overline{h}}_x$
and ${\overline{h}}_y$ obtained from the plane equation of
${\cal{T}}_{xy}(p, q)$ in Eq.~(\ref{Eq:MFXWT:Tau:p:q}). Using the double
Legendre transform in Eq.~(\ref{Eq:MFXWT:Dh}), we further obtain the
joint multifractal function $D_{xy}$, which is plotted with respect to
$p$ and $q$ in Fig.~\ref{Fig:MFXWT:BFBM}(e) and with respect to $h_x$
and $h_y$ in Fig.~\ref{Fig:MFXWT:BFBM}(f). The average value of $D_{xy}$
is $-0.178$, also close to ${\overline{D}}_{xy} = -0.135$. However
unlike $h_x$ and $h_y$, which span a narrow range, $D_{xy}$ spans a
relatively wide range from 0 to 0.5. This indicates that the MFXWT
method may indicate a spurious multifractality for bFBM if we determine
the joint multifractality only within the spanning range of
$D_{xy}$. This spurious multifractality often occurs when using the
partition function approach and is usually caused by the finite size
effect \cite{Zhou-2012-CSF}. It suggests that we need to use
bootstrapping to statistically test for the presence of multifractality
\cite{Jiang-Zhou-2007-PA,Jiang-Xie-Zhou-2014-PA}.

\section{Application to stock market indices}
\label{S1:Application}

We now apply the MFXPF$(p,q)$ method to detect joint multifractality in
the daily returns of the dow Jones industrial average (DJIA) and the
National Association of Securities Dealers Automated Quotations (NASDAQ)
index. To compare our results with those in
Ref.~\cite{Xie-Jiang-Gu-Xiong-Zhou-2015-NJP}, we also conduct a similar
analysis on the volatility time series of the two indices. The daily
return is defined as the logarithmic difference of daily closing price,
\begin{equation}
 R(t) = \ln I(t) - \ln I(t-1), \label{Eq:MFXWT:FinIndex:Return}
\end{equation}
where $I(t)$ is the closing price of the DJIA index or the NASDAQ index
on day $t$. Both indices are retrieved from ``Yahoo! Finance.'' The
spanning period of both indices is from 5 February 1971 to 17 June 2016
and contain a total of 11430 data points. The volatilities are
determined by the absolute values of the daily returns.

\subsection{Daily return time series}

We first analyze the joint multifractality of the daily returns of both
indices using the MFXWT$(p,q)$ method. Figure~\ref{Fig:MFXWT:Returns}
shows the results.

Figure~\ref{Fig:MFXWT:Returns}(a) plots the joint partition function
$\chi_{xy}(2,q,s)$ as a function of the scale $s$ for fixed $p=2$ and
varying $q$. We see strong power-law behavior in a scaling range larger
than three orders of magnitude. The results for other $(p,q)$ pairs are
similar. By regressing $\ln \chi_{xy}(p,q,s)$ with respect to $\ln s$
for a given pair of $(p, q)$, we obtain the joint mass exponents
${\cal{T}}_{xy}(p,q)$, which are plotted in
Fig.~\ref{Fig:MFXWT:Returns}(b). Note that the joint mass exponents are
a nonlinear function of $p$ and $q$, indicating the presence of joint
multifractality in the daily returns of the two indices.

\begin{figure*}[htb]
  \centering
\includegraphics[width=5.5cm]{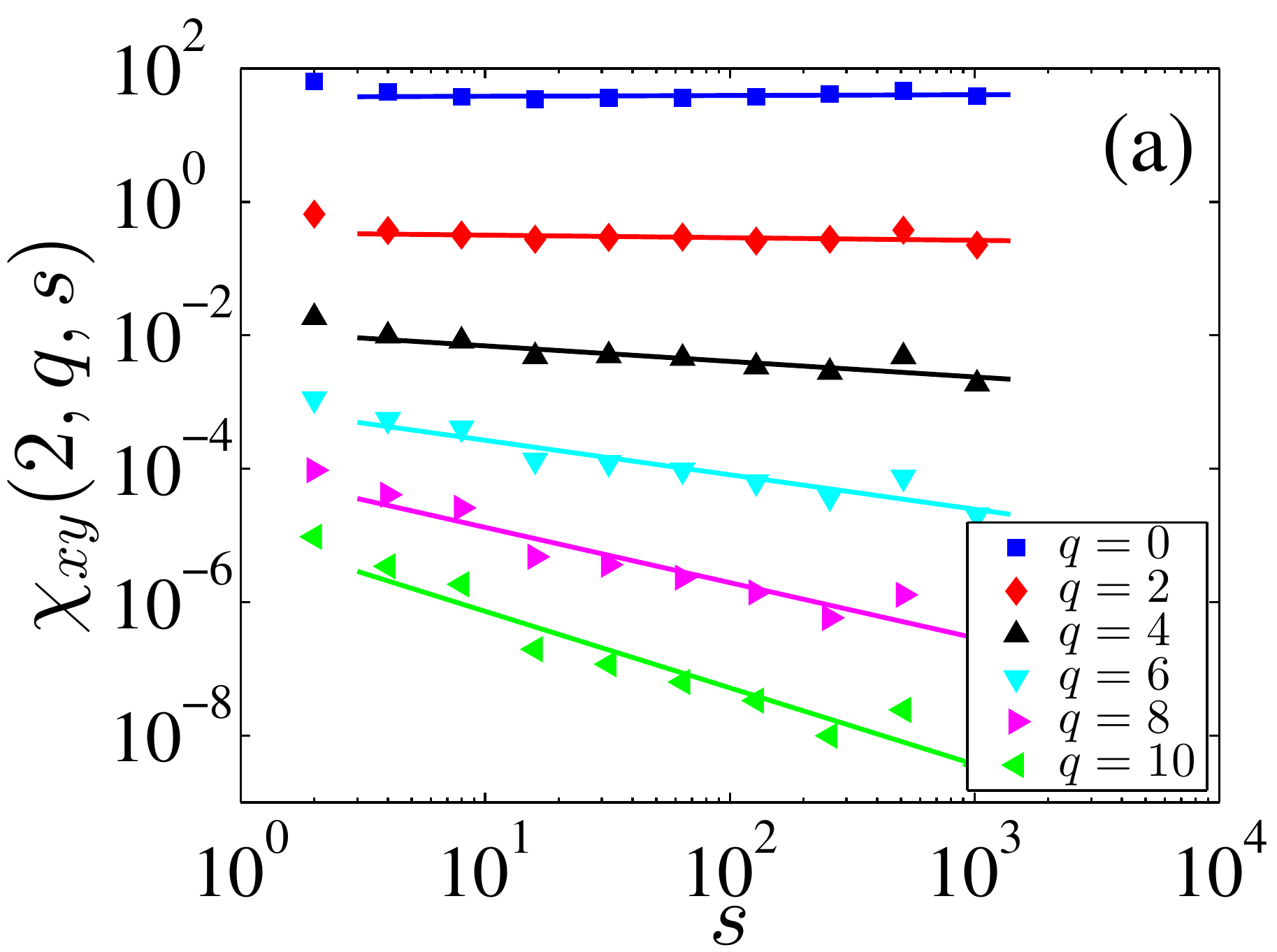}
\includegraphics[width=5.5cm]{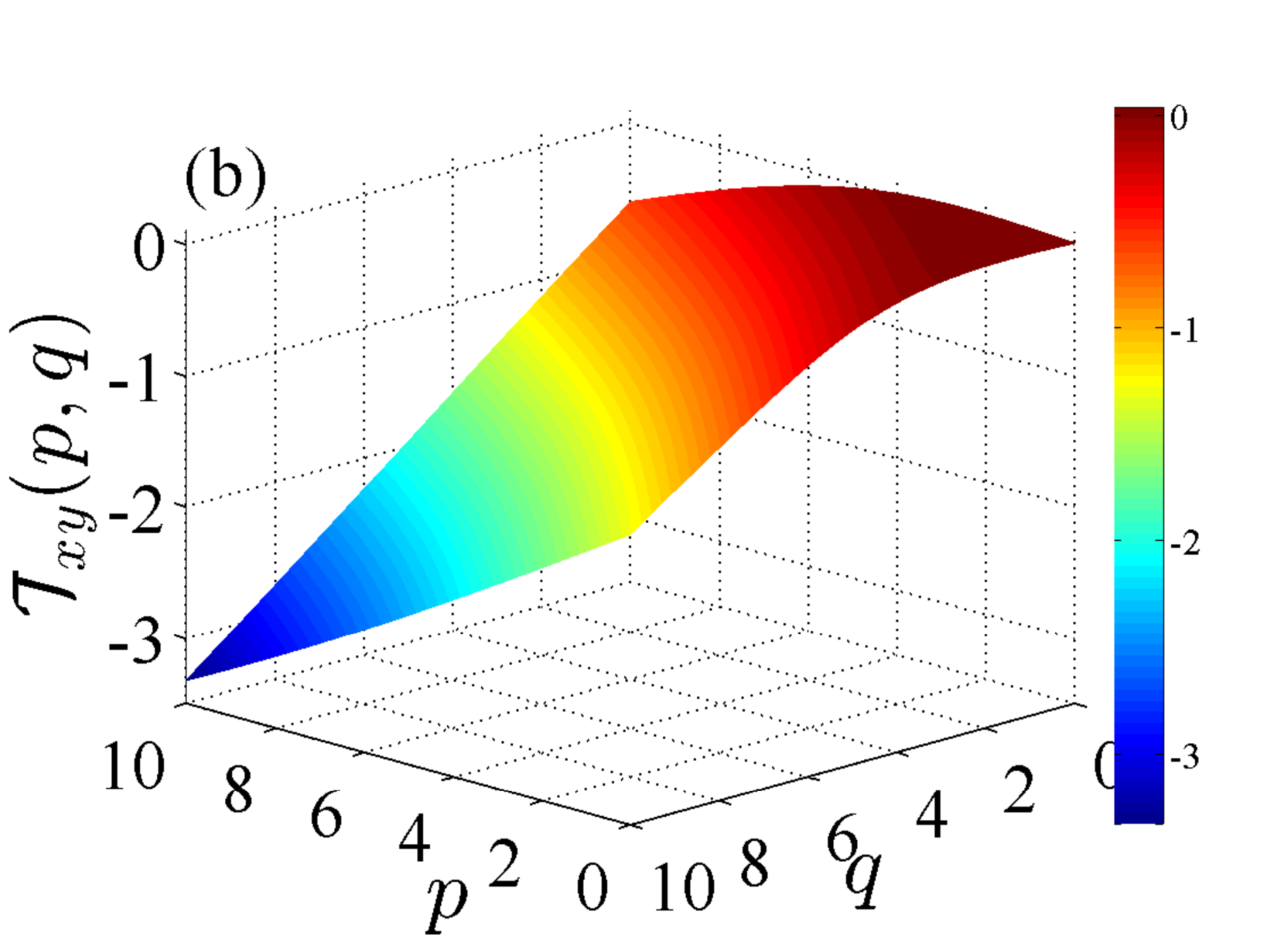}
\includegraphics[width=5.5cm]{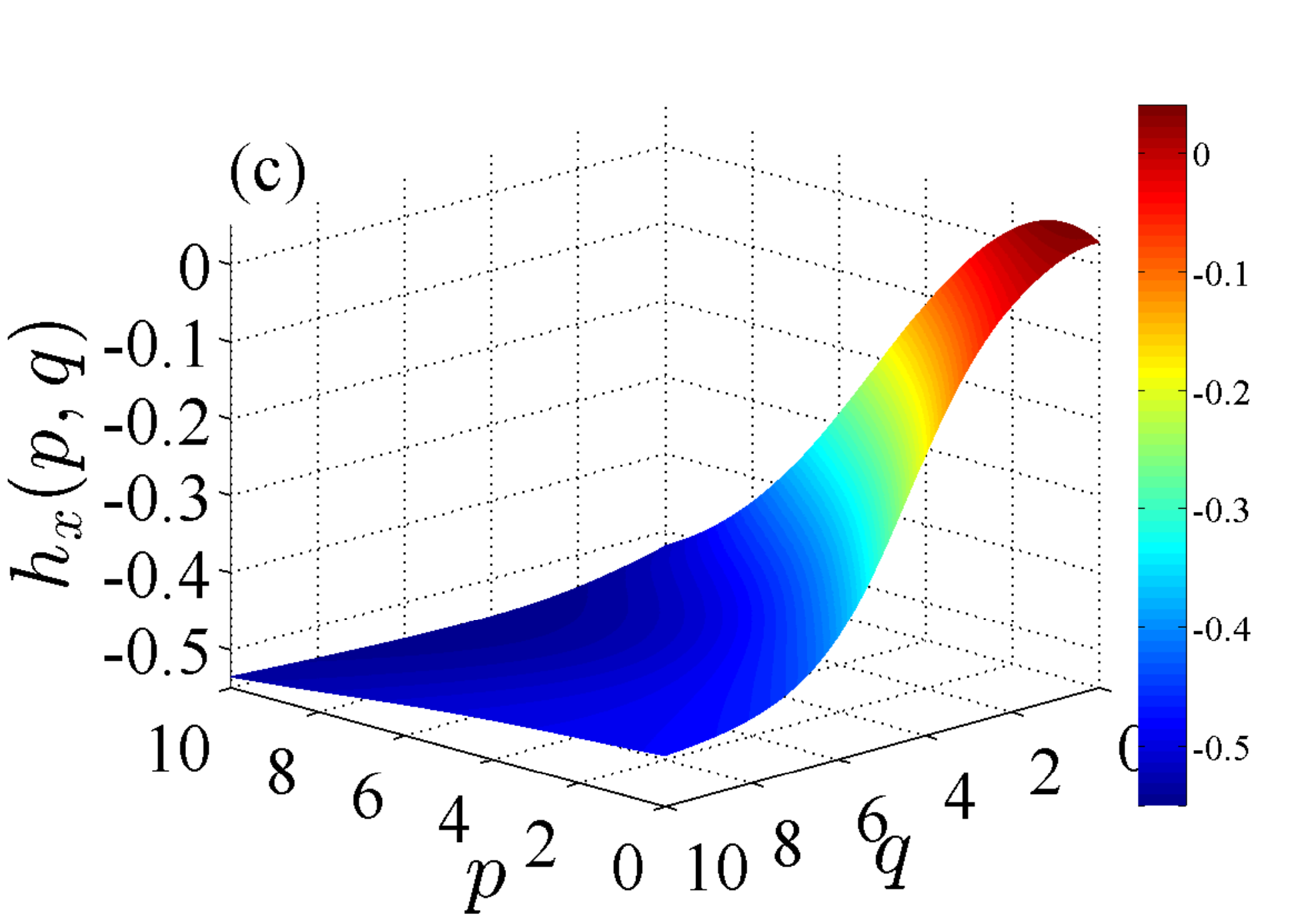}
\includegraphics[width=5.5cm]{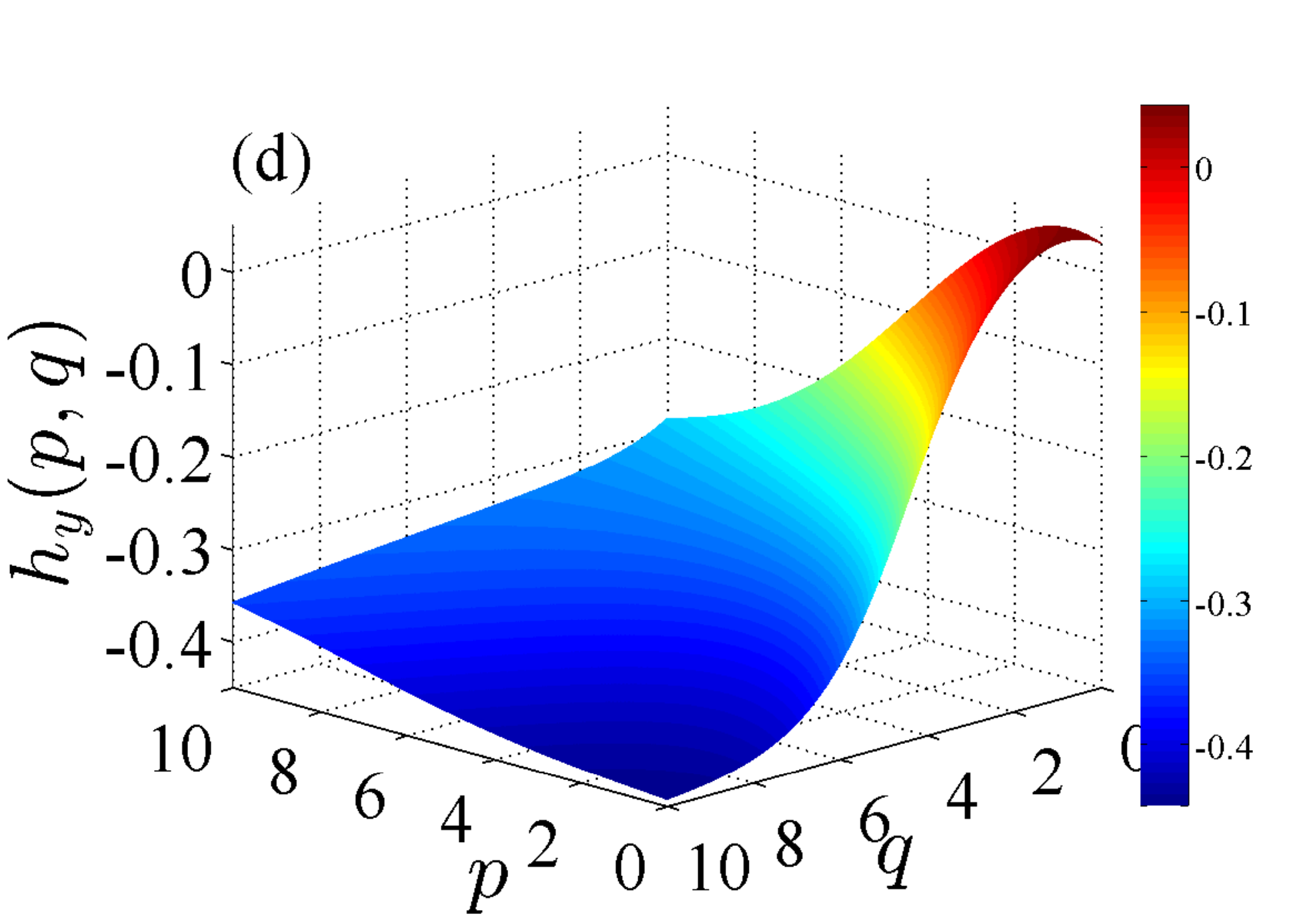}
\includegraphics[width=5.5cm]{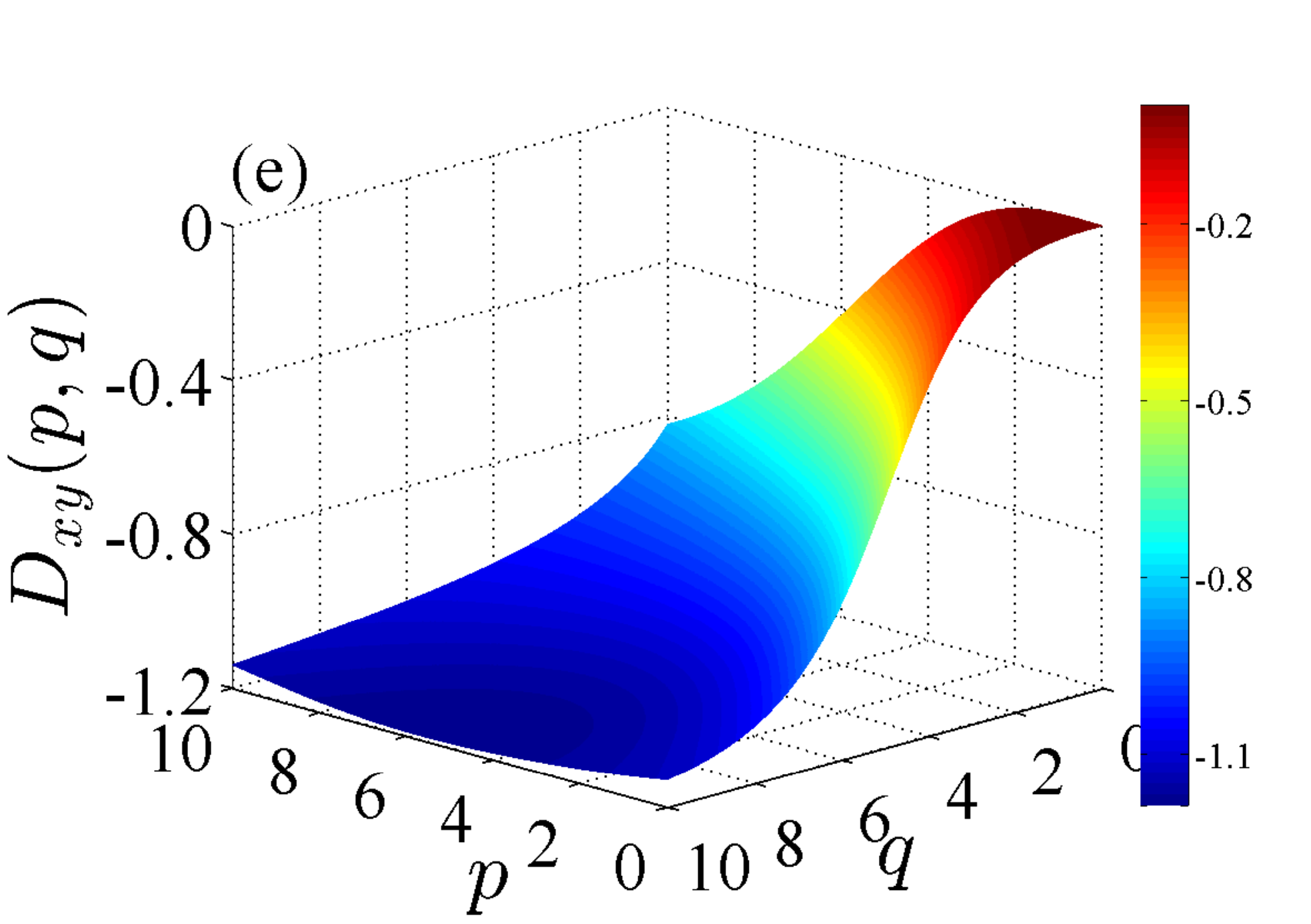}
\includegraphics[width=5.5cm]{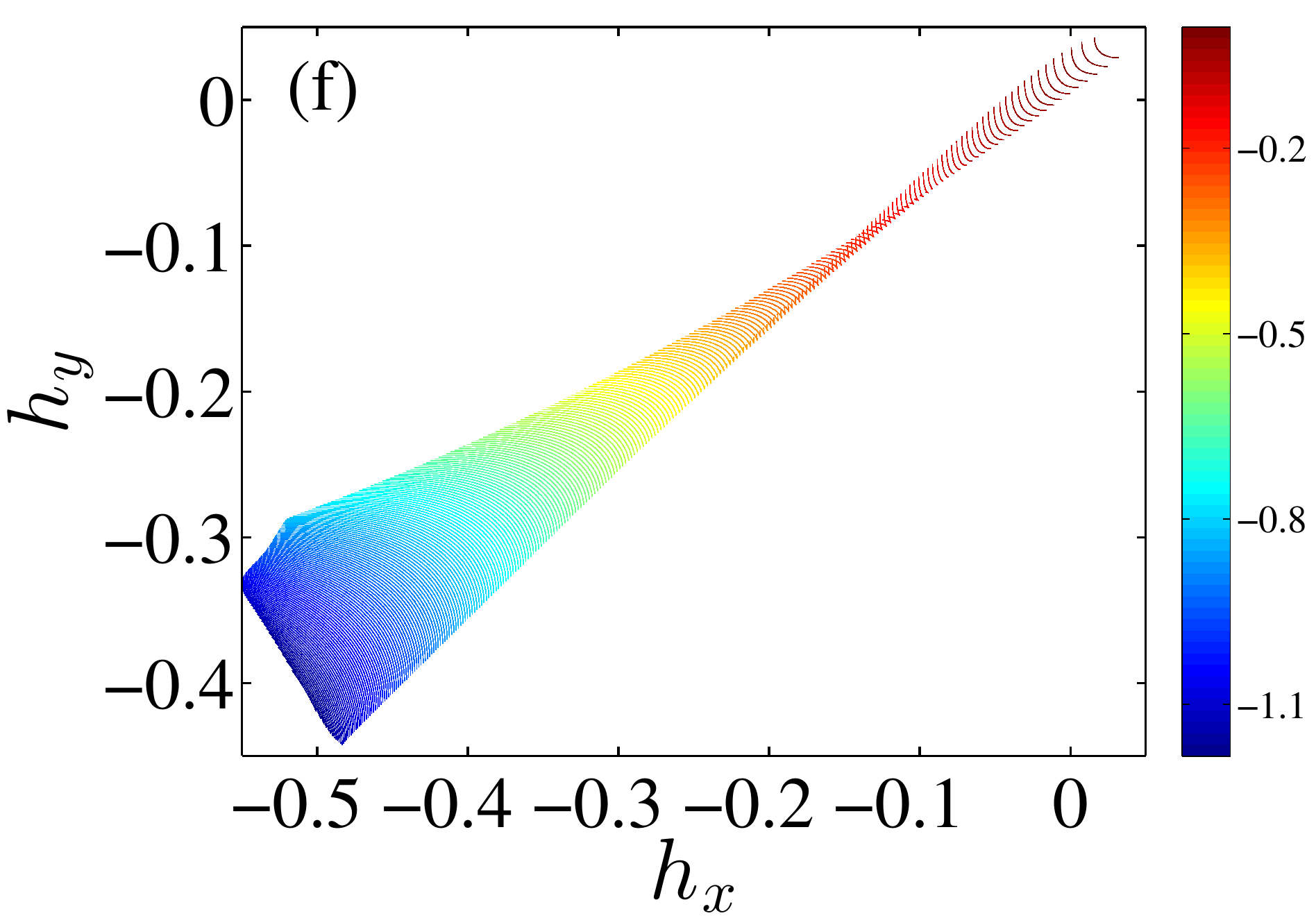}
  \caption{\label{Fig:MFXWT:Volatility} (Color online) Multifractal
    cross wavelet analysis of the joint multifractality between the
    daily volatility series of DJIA index and NASDAQ index using the
    MFXWT$(p,q)$ method. (a) Power-law dependence of $\chi_{xy}(p,q,s)$
    on scale $s$ for fixed $p=2$ and different $q$.  (b) Joint mass
    exponent function ${\cal{T}}(p,q)$. (c) Joint singularity strength
    function $h_x(p,q)$. (d) Joint singularity strength function
    $h_y(p,q)$. (e) Joint multifractal function $D_{xy} (p,q)$. (f)
    Joint multifractal singularity spectrum $D_{xy}(h_x, h_y)$.}
\end{figure*}

Figures~\ref{Fig:MFXWT:Returns}(c) and \ref{Fig:MFXWT:Returns}(d) show
the joint singularity strength functions $h_x(p,q)$ and $h_y(p,q)$,
which are numerically estimated from ${\cal{T}}(p,q)$. We find that both
singularity strength functions are widely dispersed with spanning ranges
greater than 0.3. In addition, the joint singularity strength functions
are monotonic with respect to $p$ and
$q$. Figure~\ref{Fig:MFXWT:Returns}(e) plots the joint multifractal
function $D_{xy}(p,q)$ obtained from the double Legendre transform. Note
that the joint multifractal function is located in the range of
$(-1,0)$. The maximum point of $D_{xy}(p,q)$ is reached at
$(p,q)=(0,0)$. Figure~\ref{Fig:MFXWT:Returns}(f) shows the joint
multifractal spectrum $D_{xy}(h_x, h_y)$, which does not collapse into
the neighbor of a fixed point. Our empirical findings indicate that
there is joint multifractality in the daily returns of the DJIA and the
NASDAQ.

\subsection{Daily volatility time series}

We next use the MFXWT$(p,q)$ method to perform a multifractal cross
wavelet analysis of the daily volatilities of the two indices. The
results are shown in Fig.~\ref{Fig:MFXWT:Volatility}.

Figure~\ref{Fig:MFXWT:Volatility}(a) shows a log-log plot of the
dependence of the joint partition function $\chi_{xy}(2,q,s)$ with
respect to the scale $s$ for fixed $p=2$ and varying $q$. We see strong
power-law behaviors over two orders of
magnitude. Figure~\ref{Fig:MFXWT:Volatility}(b) shows the resulting
joint mass exponents ${\cal{T}}_{xy}(p,q)$, which are monotonically and
nonlinearly increasing, implying that the cross correlations between the
two index volatilities exhibit joint multifractality.

Figures \ref{Fig:MFXWT:Volatility}(c) and \ref{Fig:MFXWT:Volatility}(d)
show the numerical calculations of the joint singularity
strength functions $h_x(p,q)$ and $h_y(p,q)$, respectively. Note that
the widths of both singularity strength functions are significantly
larger than 0, further confirming the existence of joint multifractality
in the cross correlations of the two volatility time
series. Figures~\ref{Fig:MFXWT:Volatility}(e) and
\ref{Fig:MFXWT:Volatility}(f) show the joint multifractal function
$D_{xy}(p,q)$ and the joint multifractal spectrum $D_{xy}(h_x, h_y)$,
respectively, which again affirms the joint multifractal characteristics
in the cross correlations between the two index volatilities. Our
results also show that the joint multifractal volatilities are stronger
than those in the returns, because their widths of joint singularity
strength functions and joint multifractal functions are larger.
Our results are in accordance with the results of the cross multifractal analysis presented in Ref.~\cite{Rak-Drozdz-Kwapien-Oswiecimka-2015-EPL}, because the signs of returns will bring uncorrelated noise in comparison of pure volatilities.

\subsection{Origins of cross multifractality}

The fat-tailed distribution and linear and nonlinear long memory behaviors in financial series are considered as origins of multifractality \cite{Zhou-2009-EPL,Zhou-2012-CSF}. For cross multifractality, the memory behaviors may contain the following two constituents, the auto-correlation  within each series and the cross correlation between series. Thus, it is interesting to investigate how these two types of correlation behaviors affect the cross multifractal nature. To implement the tests, we simply employ the width of multifractal spectrum for $p = q$ to quantitatively measure the degree of cross multifractality, which is defined by
\begin{equation}
\Delta h_{xy} = \max[h_{xy}(p)] - \min[h_{xy}(p)] \label{Eq:MFXWT:OMF:Dhxy}
\end{equation}
Such simplification is reasonable, as the spanning range of the diagonal line with $p=q$ of the ${\cal{T}}_{xy}$ surface approximately equals to the spanning range of the whole surface, evidenced by the surface plots of  ${\cal{T}}_{xy}$ in Figs.~\ref{Fig:MFXWT:pmodel:PQ}-\ref{Fig:MFXWT:Volatility}. We first test the effect of cross correlation behavior on the cross multifractality. Following Ref.~\cite{Oswiecimka-Drozdz-Forczek-Jadach-Kwapien-2014-PRE}, we shift two series from 1 day to 100 days relative to each other to gradually weaken the cross correlation between them without changing the auto-correlation in each series. Such a strategy also allows us to detect the possible time lags or asymmetry effects in cross multifractality.

\begin{figure*}[htb]
  \centering
  \includegraphics[width=4.1cm]{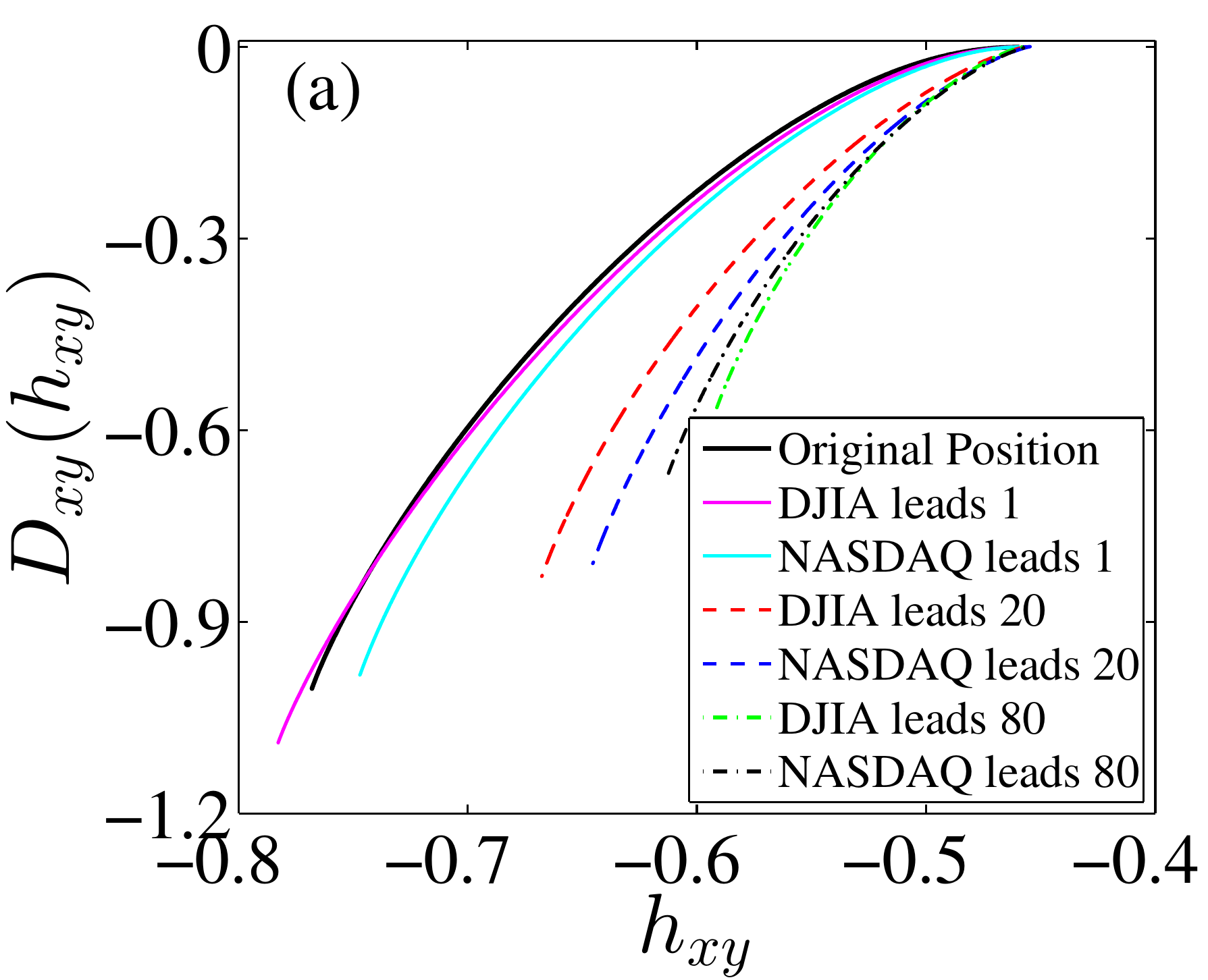}
  \includegraphics[width=4.35cm]{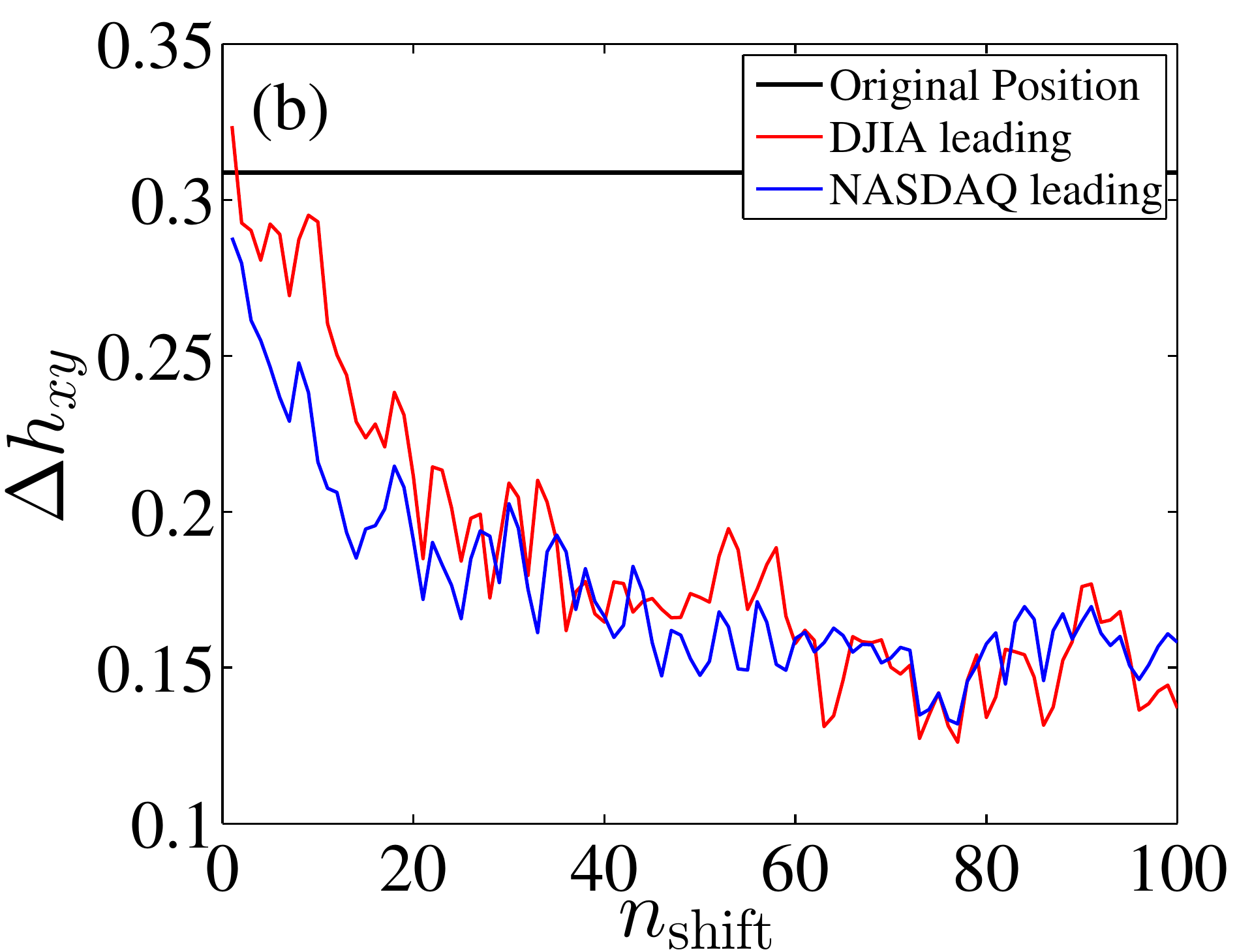}
  \includegraphics[width=4.1cm]{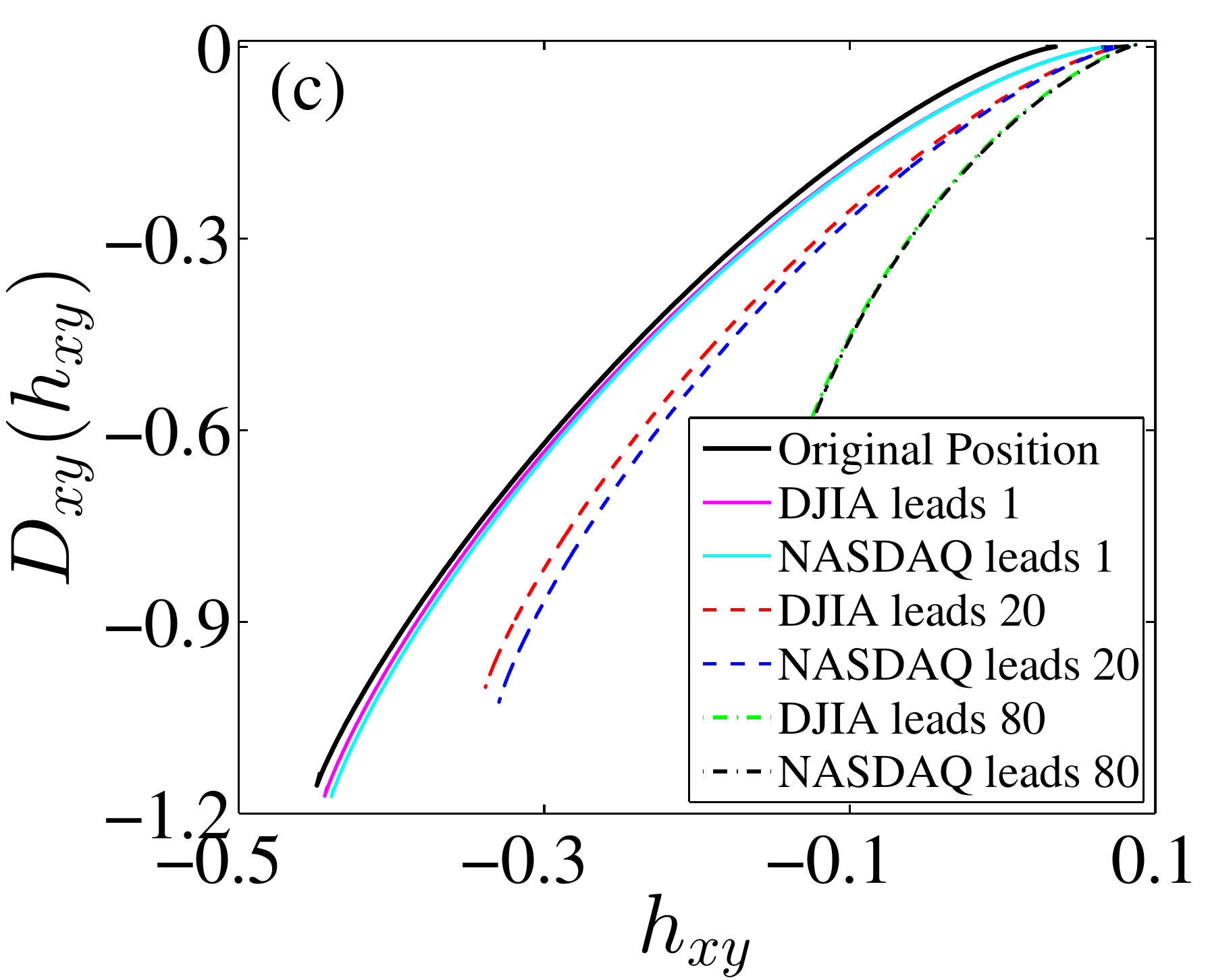}
  \includegraphics[width=4.35cm]{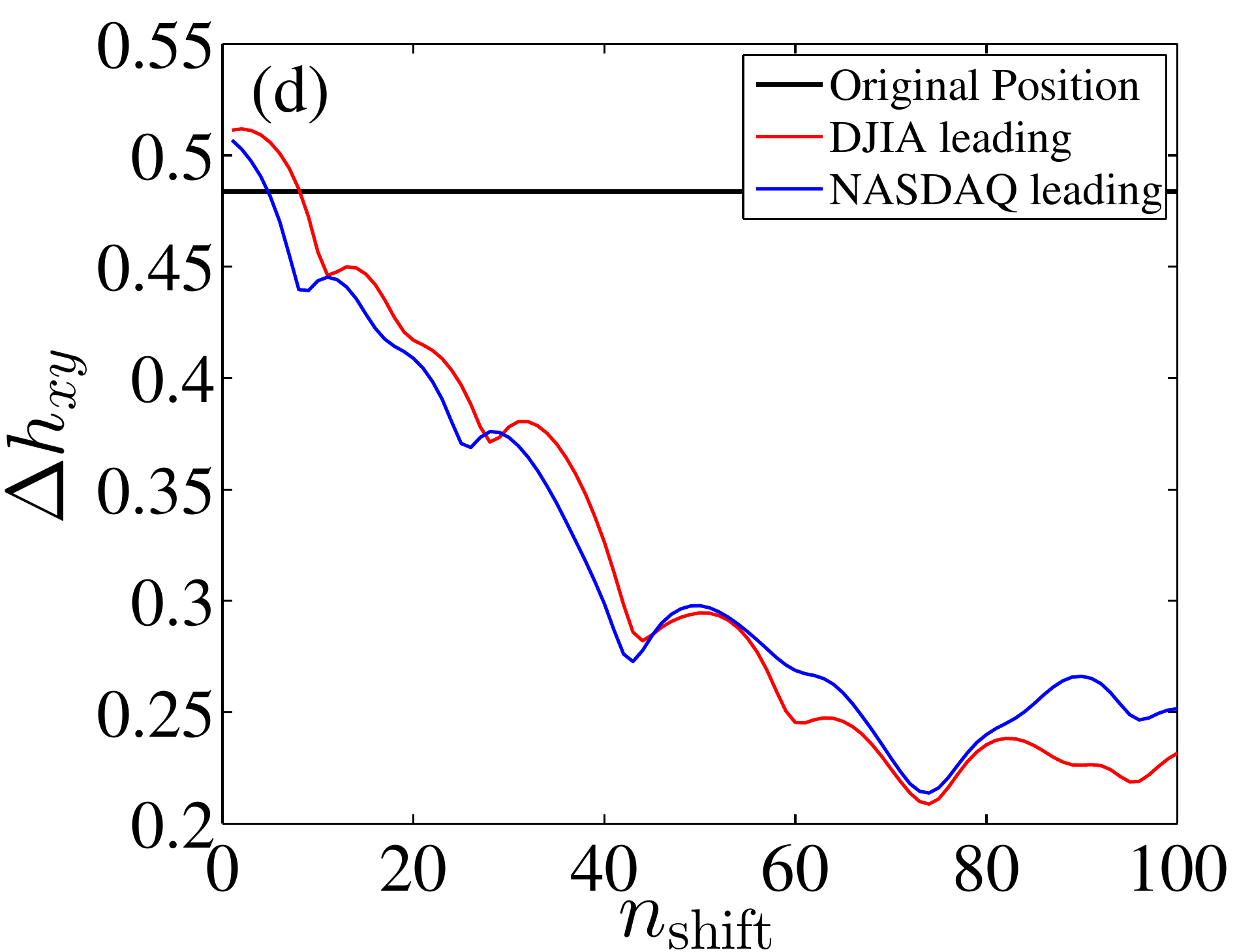}
  \caption{\label{Fig:MFXWT:Shift} (Color online) Results of cross multifractality obtained by gradually decreasing the strength of cross correlations, which is achieved by shifting two series relative to each other. (a) Cross multifractal spectra of the DJIA-NASDAQ returns with the original position, with the leading positions (1 day, 20 days and 80 days) for DJIA, and with the leading positions (1 day, 20 days, and 80 days) for NASDAQ.  (b) Plots of the spectral width as a function of the number of shifted positions for returns. The spectral width of original returns is illustrated as the horizontal line. (c) The same as (a), but for volatilities. (d) The same as (b), but for volatilities. }
\end{figure*}

For the DJIA-NASDAQ returns and volatilities, we estimate the multifractal spectra for three cases of positioning the time series relative to each other. The first case corresponds to no shifts between the two series. The second case is that DJIA is shifted relative to NASDAQ by $n_{\rm{shift}}$ days in advance. The third case is that NASDAQ leads DJIA by $n_{\rm{shift}}$ days. By setting $n_{\rm{shift}}$ = 1, 10, and 80, we plot the obtained multifractal spectra in Fig.~\ref{Fig:MFXWT:Shift} (a) and (c) for returns and volatilities, respectively. We find that the pair of returns in which DJIA leads one day ahead of NASDAQ exhibit the strongest cross multifractality, as well as the pair of volatilities with DJIA leading one day ahead.

We further vary $n_{\rm{shift}}$ from 1 to 100 and estimate the spectral width $\Delta h_{xy}$. The corresponding results of returns and volatilities are illustrated in Fig.~\ref{Fig:MFXWT:Shift} (b) and (d), in which $\Delta h_{xy}$ is plotted with respect to $n_{\rm{shift}}$. We find that $\Delta h_{xy}$ decreases quickly with the increasing of $n_{\rm{shift}}$, indicating the deterioration of the cross multifractality. This is due to that the cross correlation between two series becomes weaker when their lag increases, supporting that the cross correlation can be regarded as the origin of cross multifractality. In panels (b) and (d), another intriguing phenomenon is that for $n_{\rm{shift}} \le 30$ the multifractality of DJIA leading case is stronger than that of NASDAQ leading case, presenting that the influence of DJIA on NASDAQ in next few days is stronger than the influence of NASDAQ on DJIA. Such results also reveal that there is an asymmetry effect in cross correlation between DJIA and NASDAQ returns. We also find that $\Delta h_{xy}$ of volatilities is larger than that of returns, implying that the cross multifractality of volatilities is stronger than that of returns. As mentioned above, the sign of returns will introduce noise that can deteriorate the cross correlations comparing with volatilities \cite{Rak-Drozdz-Kwapien-Oswiecimka-2015-EPL}.

\begin{figure*}[htb]
  \centering
  \includegraphics[width=6cm]{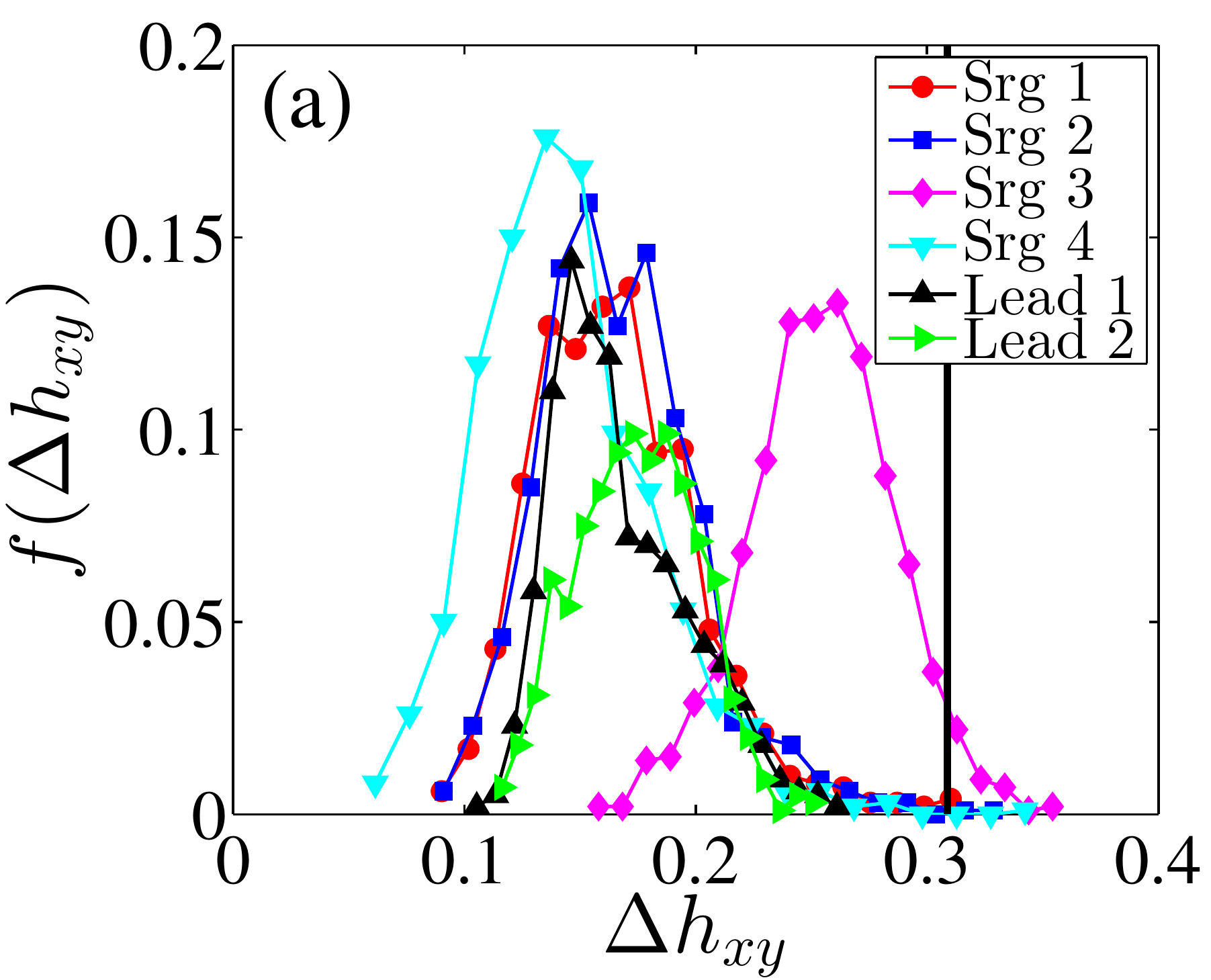}
  \includegraphics[width=6cm]{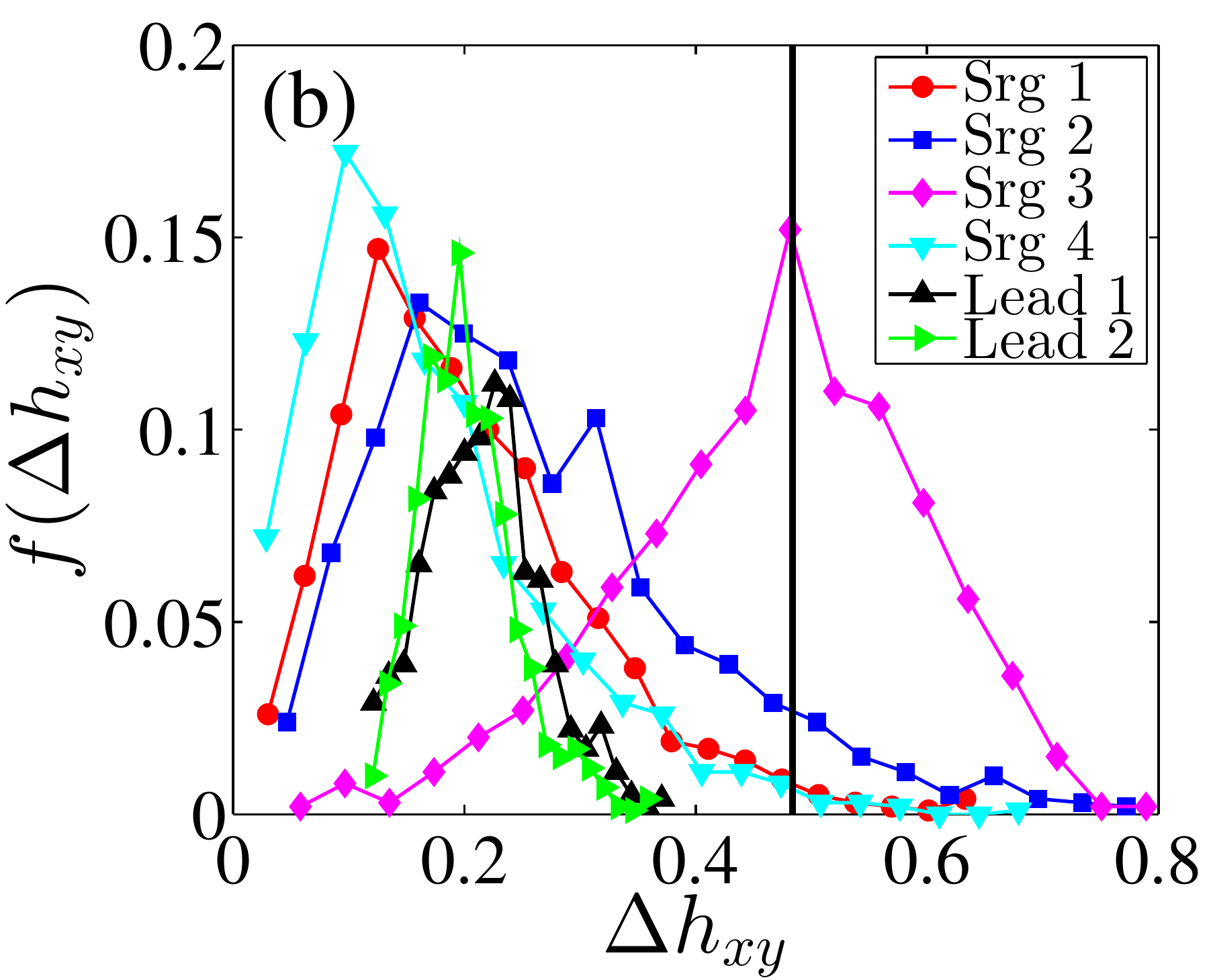}
  \caption{\label{Fig:MFXWT:StatTests} (Color online)  Distribution of the multifractal spectrum width for four surrogate pairs (labeled as Srgs 1--4) and two shifted pairs (labeled as Lead 1 and Lead 2). The spectral width of original pairs are shown in vertical black line. (a) DJIA--NASDAQ Returns. (b) DJIA--NASDAQ Volatilities. }
\end{figure*}

We have shown that destroying the cross correlation between two series can strongly weaken the cross multifractality, which motivates us to further test how the auto-correlation in each series affects the cross multifractality. To conduct the tests, we generate four pairs of surrogate series, including (1) the shuffled DJIA and the original NASDAQ, (2) the original DJIA and the shuffled NASDAQ, (3) the co-shuffled DJIA and NASDAQ in which the data points of the two indexes on the same day are bounded as one single entity in the shuffling procedure, and (4) the shuffled DJIA and the shuffled NASDAQ in which the data points of each index are shuffled independently. The first surrogate pairs (termed as `Srg 1') only preserve the auto-correlation in NASDAQ and destroy the auto-correlation in DJIA and the cross correlation between DJIA and NASDAQ. The second surrogate pairs (termed as `Srg 2') only preserve the auto-correlation in DJIA and remove the auto-correlation in NASDAQ and the cross correlation between DJIA and NASDAQ. The third surrogate pairs (termed as `Srg 3') remove the auto-correlation and cross correlation with non-zero lags, but keep the cross correlation with zero lag unchanged. In the the fourth surrogate pairs (termed as `Srg 4'), all the possible correlations are removed. We generate 1000 synthetic pairs for each surrogate case, and perform the same wavelet cross multifractal analysis on each synthetic pair to estimate its multifractal spectrum width $\Delta h_{xy}$. We also employ two shifted pairs for comparison, One corresponding to DJIA leading NASDAQ (termed as `Lead 1') and the other corresponding to NASDAQ leading DJIA (termed as `Lead 2'). As shown in Fig.~\ref{Fig:MFXWT:Shift} (b) and (d), the width of multifractal spectrum is almost stable when the shifted position is greater than 80, indicating that the cross correlation is completely removed when $n_{\rm{shift}} \ge 80$. Thus, we set $n_{\rm{shift}}$ varying from 101 to 1100 to generate 1000 pairs of synthetic data for each shifted pair. The shifted pairs here represent the surrogate data in which the cross correlation is removed and the auto-correlation remains unchanged.

Fig.~\ref{Fig:MFXWT:StatTests} (a) and (b) illustrate the distributions of $\Delta h_{xy}$ obtained from four surrogate pairs and two shifted pairs for the DJIA--NASDAQ returns and volatilities. The vertical black line in both panels represents the spectral width of original data. Their average multifractal spectrum widths are reported in Table~\ref{Tb:MFXWT:StatTests:AveDhxy}. For DJIA-NASDAQ returns, we have the following inequality,
\begin{equation}
 \langle \Delta h_{xy} \rangle_{\rm{Srg~4}} <  \langle \Delta h_{xy} \rangle_{\rm{Srg~1}} \approx \langle \Delta h_{xy} \rangle_{\rm{Srg~2}} \approx \langle \Delta h_{xy} \rangle_{\rm{Lead~1}} \approx \langle \Delta h_{xy} \rangle_{\rm{Lead~2}} < \langle \Delta h_{xy} \rangle_{\rm{Srg~3}} < \langle \Delta h_{xy} \rangle_{\rm{Original}}. \label{Eq:MFXWT:Returns:StatTests}
\end{equation}
We can observe that the cross multifractal nature is the weakest if we remove both auto-correlations and cross correlations (Srg 4), indicating that both auto-correlation and cross correlation have influences on the cross multifractality. However, the auto-correlation has a very small influence, as the widths of the surrogate pairs, in which one series has the same auto-correlation as the original series (Srg 1 and Srg 2) and both series have the same auto-correlation as the original series (Lead 1 and Lead 2), are only slightly larger than the width of Srg 4. The distribution curves of $\Delta h_{xy}$ for Srg 1,  Srg 2, Lead 1, and Lead 2 almost overlap with each other and their average values are all equal to 0.17, far from the width of the original returns 0.31, implying that the cross correlation plays an crucial role in the origin of cross multifractality. The cross correlation can be further decomposed into cross correlation with zero lag and with nonzero lag. By keeping the zero lag cross correlation unchanged and removing the nonzero lag cross correlation (Srg 3), we find that the obtained cross multifractal nature is the strongest in the surrogate experiments, suggesting that the zero lag cross correlation between returns contributes a great part on the cross multifractality. For DJIA-NASDAQ volatilities, we can obtain the following inequality,
\begin{equation}
 \langle \Delta h_{xy} \rangle_{\rm{Srg~4}} <  \langle \Delta h_{xy} \rangle_{\rm{Srg~1}} \approx \langle \Delta h_{xy} \rangle_{\rm{Srg~2}} \approx \langle \Delta h_{xy} \rangle_{\rm{Lead~1}} \approx \langle \Delta h_{xy} \rangle_{\rm{Lead~2}} < \langle \Delta h_{xy} \rangle_{\rm{Srg~3}} \approx \langle \Delta h_{xy} \rangle_{\rm{Original}}. \label{Eq:MFXWT:Volatilities:StatTests}
\end{equation}
Comparing with Eq.~(\ref{Eq:MFXWT:Returns:StatTests}), the only difference is that $\langle \Delta h_{xy} \rangle_{\rm{Srg~3}} \approx \langle \Delta h_{xy} \rangle_{\rm{Original}}$ in Eq.~(\ref{Eq:MFXWT:Volatilities:StatTests}), suggesting that the zero lag cross correlation between volatilities is the main origin of cross multifractality.

\begin{table}[htp]
\caption{ The average multifractal spectrum width of original pairs, four surrogate pairs, and two shifted pairs for DJIA--NASDAQ returns and volatilities. }
\centering
   \begin{tabular}{cccccccc}
   \hline
    DJIA--NASDAQ & original & Srg 1  & Srg 2 & Srg 3 & Srg 4  & Lead 1 & Lead 2  \\
   \hline
    Returns: $\langle \Delta h_{xy} \rangle$ & 0.31 & 0.17 $\pm$ 0.04 & 0.17 $\pm$ 0.03 & 0.25 $\pm$ 0.03 & 0.14 $\pm$ 0.04 & 0.17 $\pm$ 0.03 & 0.17 $\pm$ 0.03 \\
    Volatilities: $\langle \Delta h_{xy} \rangle$ & 0.48 & 0.20 $\pm$ 0.11 & 0.26 $\pm$ 0.14 & 0.47 $\pm$ 0.12 & 0.17 $\pm$ 0.10 & 0.22 $\pm$ 0.05 & 0.20 $\pm$ 0.04 \\
   \hline
   \end{tabular}
   \label{Tb:MFXWT:StatTests:AveDhxy}
\end{table}

\section{Conclusion and discussion}
\label{S1:conclusion}

We have developed a new method of joint multifractal analysis with two
moment orders based on wavelet transform, which we call
MFXWT$(p,q)$. Because some of the wavelet coefficients approach 0, the
values of $p$ and $q$ must be greater than 0. We check the performance
of the MFXWT$(p,q)$ method using extensive numerical experiments on time
series pairs generated from binomial measures and bivariate fractional
Brownian motions. We also test the ability of this method to detect any
joint multifractality in return pairs and volatility pairs in the US
stock markets.

Using binomial measures from the $p$-model, we derive the theoretical
expressions of the joint multifractality by comparing the scaling
behaviors of the joint partition functions of the MFXWT$(p,q)$ and the
MFXPF$(p,q)$ methods. We find that the joint multifractality
(${\cal{T}}_{xy}$, $h_x$, $h_y$ and the $D_{xy}$) extracted using the
MFXWT method closely agrees with theoretical values. This indicates that
the accuracy of MFXWT$(p,q)$ is sufficient to detect joint
multifractality in binomial measures.

For bFBMs, we find that the joint mass
exponent function ${\cal{T}}_{xy}$ of the cross correlations is linearly
dependent on the orders $p$ and $q$, and this is a hallmark of
monofractality. This clearly indicates the inherent monofractality in
bFMBs. We find that the singularity strength functions $h_x$ and $h_y$
are in an extremely narrow range, which again confirms that bFMBs are
monofractal. On the other hand, we are wary of the multifractal function
$D_{xy}$ indicated by the MFXWT method, because this method yields
biased outcomes for the bFBMs. We are also wary of the multifractality
determined using the multifractal function $D_{xy}$ given by the MFXWT
algorithm because it may indicate spurious multifractality, especially
when we do not know {\textit{a priori}} the underlying fractal
properties. We can compensate for these shortcomings by performing
statistical tests using the bootstrap method.

Unlike the MFXPF$(p,q)$ method, which can be applied only to
conservative measures (volatility), the MFXWT($p,q$) method can analyze
both conservative and non-conservative measures. We thus use it to
analyze joint multifractality in the returns and volatilities of two US
stock market indices. We find joint multifractality both in the returns
and in the volatilities, and find that the joint multifractality in the
volatilities is stronger than in the returns. We also find that the cross correlated behavior, particularly the zero lag cross correlation, is the main origin of cross multifractality.

The well-known shortcoming of the wavelet analysis of multifractals is
that the moment order must be positive due to the presence of small
wavelet coefficients, and thus because all the modulus maxima are
significantly larger than 0 we must use the wavelet transform modulus
maxima (WTMM) method
\cite{Muzy-Bacry-Arneodo-1991-PRL,Bacry-Muzy-Arneodo-1993-JSP,Muzy-Bacry-Arneodo-1993-PRE,Muzy-Bacry-Arneodo-1994-IJBC}.
Unfortunately the WTMM method cannot be generalized to bivariate cases,
because at each scale $s$ the number of modulus maxima of the two time
series usually differ.

\begin{acknowledgments}

We acknowledge financial support from the National Natural Science Foundation of China (11375064, 71532009 and 71571121) and the Fundamental Research Funds for the Central Universities (222201718006).

\end{acknowledgments}


\end{document}